\documentclass[final,11pt,onecolumn,a4]{IEEEtran}

\newcommand{\sinc}{\ensuremath{\mathrm{sinc}}}

\newtheorem{proposition}{Proposition}[section]
\newtheorem{theorem}{Theorem}[section]
\newcommand{\E}{\ensuremath{\mathrm{E}}}
\newcommand{\var}{\ensuremath{\mathrm{var}}}
\newcommand{\rel}{\ensuremath{\mathrm{rel}}}
\newcommand{\cov}{\ensuremath{\mathrm{cov}}}

\usepackage{graphics,graphicx,bm,amsmath,amssymb}
\usepackage{pst-all,amstext,amsfonts,epsf,epsfig,pifont}
\usepackage{subfigure}
\begin{document}
%
\title{Estimation of Ambiguity Functions With Limited Spread}
\author{Heidi Hindberg \& Sofia~C.~Olhede
\thanks{H. Hindberg thanks the Research Council of Norway for financial support under project 162831/V00.
}
\thanks{H. Hindberg is with the Department of Physics
and Technology,
University of Troms\o , NO-9037 Troms\o, Norway (heidih@phys.uit.no)
Tel:+4777645118, Fax+4777645580.}
\thanks{S. C. Olhede is with the Department of Statistical Science,
University College London, Gower Street,
London, WC1 E6BT, UK (s.olhede@ucl.ac.uk). Tel:
+44 (0)20 7679 8321, Fax: +44 (0)20 7383 4703.}}

\markboth{Department of Statistical Science Research Report 293}{Hindberg \& Olhede: Estimation of the Ambiguity Functions with Limited Spread}
\maketitle

\begin{abstract}
This paper proposes a new estimation procedure for the ambiguity function of a non-stationary time series. The stochastic properties of the empirical
ambiguity function calculated from a single sample in time are derived.
Different thresholding procedures are introduced for the estimation of the ambiguity function. Such estimation methods are suitable if the ambiguity function is only non-negligible in a limited region of the ambiguity plane. The thresholds of the procedures are formally derived for each point in the
plane, and methods for the estimation of nuisance parameters that the thresholds depend on are proposed. The estimation method is tested on several signals, and reductions in mean square error when
estimating the ambiguity function by factors of over a hundred are obtained. An estimator of the spread of the ambiguity function is proposed.
\end{abstract}

\small{
}

\section{Introduction \label{Introduction}}
We propose a new estimation procedure for the Ambiguity Function
(AF) of a non-stationary Gaussian process. The suitable definition of a `locally stationary process', and the development of inference methods for a given sample from a non-stationary process is still an open question, see \cite{Donoho98}. The AF forms an essential component to this task, as it
provides a characterisation
of dependency between a given time series and its translates in
time and frequency.
The AF of a non-stationary process can often be assumed to be mainly limited in support to a small region of the ambiguity plane, and this corresponds to correlation in the process of interest being limited in support. If the support is additionally centred at time and frequency lags of zero, then the process is underspread, see Matz and Hlawatsch \cite{Matz06}.  An important class of non-stationary processes, namely semi-stationary processes \cite{Priestly65}, exhibit limited essential spread in the ambiguity plane. Non-stationary processes can also be constructed from the viewpoint of time-variant linear filtering.
Pfander and Walnut \cite{Pfander} have shown that if the spreading of a filtering operator is sufficiently limited then the operator may be identified from its measurements. It is not necessary in this case
to assume that the spread is centered at the origin \cite{Pfander}.
The key to the estimation or characterization of the generating mechanism of non-stationary processes, in all of these cases, is the compression of the AF of the process.

The AF has also been popularly used in
radar and sonar applications \cite{Blahut1991ed}, where the echo
of a known signal is recorded.
The delay and frequency shift of the echo can be determined from the AF of the signal \cite{Flandrin99}.
In such applications the emitted signal is deterministic,
even if the echo has been contaminated by noise. By using the compression of the emitted signal in the AF domain, the properties of the process of reflection can be determined, see Ma and Goh \cite{Ma2006}. The estimation of the AF of a non-zero mean signal immersed in noise is of interest. Unlike Ma and Goh, we shall not assume the compression of the AF is known, but propose an estimator of the AF, suitable if the AF satisfies some (unknown) compression constraints.

Given the importance of the AF it is surprising to find that existing methods for its estimation are quite naive. Methods are usually based on calculating the EMpirical AF (EMAF), or some smoothed version of the EMAF. It is clearly unpalatable to implement a smoothing procedure on the EMAF, as the compression of the EMAF should be preserved, or even preferably increased, by any proposed estimation procedure. In this spirit Jachan {\it et al.}~\cite{Jachan07} have used shrinkage methods to estimate the EMAF with an assumed knowledge of its support, using a multiplier that attenuates larger local time and frequency lags. Often the support of the AF is not known, and arbitrary shrinkage
at larger lags will not be an admissible estimation procedure.

Given the compression or sparsity of the AF, shrinkage or threshold estimators are a natural choice. In this paper we propose a set of threshold estimators for the AF. Thresholding has been implemented before in the global time-frequency plane for wavelet spectra, or evolutionary spectra, see Fryzlewicz and Nason \cite{Frylewicz06}, or von Sachs and Schneider \cite{vSachs96}. The estimation problem in this case is quite different to estimating the AF, as we cannot assume the raw wavelet/evolutionary spectra exhibit compression. Hedges and Suter \cite{Hedges2002}
calculated the spread of the AF, based on exceedances  of
the magnitude of the EMAF over a (fixed and user tuned) threshold in a given
direction in the local time-frequency plane. The stochastic properties of the EMAF were not treated by Hedges and Suter, and instead their investigation focused on some effects from different boundary treatments. 

To be able to select a suitable threshold procedure at each local time-frequency point we must establish the distribution of the EMAF, and this requires modelling the observed process. We discuss the classes of non-stationary processes that we intend to treat in Section \ref{modelobs}, and initial method of moments estimators of second order structure in Section \ref{initialest}. In Section \ref{prop} we determine the first and second order properties of the EMAF. For strictly underspread processes satisfying constraints on their degree of uniformity we show that the EMAF is asymptotically Gaussian proper, if calculated for local frequency and time lags outside the support of the AF. 

We introduce the proposed estimation procedure in Section \ref{Estimation} for deterministic signals immersed in noise and stochastic processes. In the first estimation procedure we threshold the EMAF based on a variance estimated from the entire plane, this yielding the Thresholded EMAF (TEAF). For stochastic processes, we also introduce a Local TEAF (LTEAF), where we estimate the variance locally in regions of the ambiguity plane. 
Such procedures are suitable as smooth time-frequency spectra will correspond to decaying ambiguity functions and are a natural extension to treating wavelet coefficients differently depending on the level of the wavelet transform, so called level-dependent thresholding, see {\em e.g.} Johnstone and Silverman \cite{Johnstone97}.
The EMAF of deterministic signals immersed in analytic white noise will be biased. We propose a method to remove this bias prior to local thresholding, thus obtaining the Bias-corrected LTEAF (BLTEAF). We define an estimator of the total spread of a signal in the ambiguity plane in Section \ref{estspread}, to numerically measure the support of the signal in the ambiguity plane. The proposed TEAFs will produce sparse approximations to processes that are not strictly underspread. Whenever the EMAF is small in comparison to the variance of the EMAF, the AF will be estimated as zero. The AF can provide essential information as to the natural bandwidth of a process via the spread of the TEAF. 
 
In Section \ref{examples} we provide simulation studies of the proposed estimators. The Mean Square Errors (MSE) of our procedures reduce in many cases to around a hundredth of the MSE of the EMAF. Plots of single realisations show the accuracy of the threshold estimators, based on a single sample. The proposed methodology thus introduces a new method of characterising the features of structured non-stationary processes with high accuracy.

\section{Second Order Structure Descriptions}
\subsection{Modelling the Observations \label{modelobs}}
We assume that the process under consideration, $X[t]$, is an analytic, zero-mean, discrete-time, Gaussian harmonizable random process. Note that $[\cdot]$ is used to indicate a discrete argument, and $(\cdot)$ is used for a continuous argument.
As $X[t]$ is harmonizable it admits the spectral representation~\cite{Cramer40} of
\begin{equation}
X[t]=\int_{0}^{\frac{1}{2}} e^{j2\pi t f}\,d\widetilde{X}(f),
\quad t\in{\mathbb{Z}},
\end{equation}
where $\left\{d\widetilde{X}(f)\right\}$ is a Gaussian increment process, and $X[t]\in{\mathbb{C}}$. We note that $\left\{d\widetilde{X}(f)\right\}$ has a correlation structure specified by the dual-frequency spectrum $S_{X X^*}(\nu,f)\,d\nu\,df=\E\left\{d\widetilde{X}(\nu+f)d\widetilde{X}^*(f) \right\}$. The AF of $X[t]$ is defined as a Fourier Transform (FT) pair with the dual-frequency spectrum in variable $f$. The AF forms an FT pair with the dual-time second moment of the process, or 
$M_{X X^*}[t,\tau]=\E\left\{X[t]X^*[t-\tau]\right\}$, for $t$ and $\tau \in{\mathbb{Z}}$, now in variable $t$. We refer to \cite{Kozek,Matz} for a more detailed exposition of the AF and forms of
AFs of common processes.

Estimation of the AF of a real-valued process is known to display artifacts from aliasing and interference from negative frequencies \cite{Jeong92}. For this reason we consider analytic processes solely. Note that since we are working with the analytic signal of a non-stationary process, it may (in general) be improper~\cite{Neeser93} and thus may exhibit complementary correlation~\cite{Schreier03}. We leave the estimation of the complimentary AF outside the scope of this paper, noting that our proposed methods can in a straight forward fashion be extended to include the estimation of such objects. Furthermore, we will in this paper assume that we are working with proper processes, and any terms that will only be non-zero for improper processes will be omitted. 
\subsection{Initial Estimation}\label{initialest}
We observe $x[t]$, a finite sample of a realization of $X[t]$. The Sample AF (SAF) of a sampled deterministic signal, $\left\{g[t]\right\}_{t=0}^{N-1}$, is given by
\begin{equation}
A_{gg^*}(\nu,\tau]=\sum_{t=\max(0,\tau)}^{N-1+\min(0,\tau)} g[t]g^*[t-\tau]\;e^{-j2\pi
\nu t}, \nu\in \left[-\frac{1}{2},\frac{1}{2}\right],\quad \tau=[-(N-1),\dots,(N-1)].
\end{equation}
The properties of $A_{gg^*}(\nu,\tau]$ follow directly from properties of AFs of deterministic sequences,  see
\cite{Auslander1984,Auslander1985}. The effects of discretization on deterministic structure, i.e. trying to infer properties of $g(t)$ from $g[t]\equiv g(t\Delta t)$ for some appropriate sampling period $\Delta t>0$, will also be left outside the scope of this article~\cite{Bekir1993,Tolimieri1995}. The Empirical Second Moment (ESM) of a stochastic process based on the realization $x[t], t=0,\dots,N-1$ is given by $\widehat{M}_{XX^*}[t,\tau]=x[t]x^*[t-\tau]$, which has an expectation $\E\left\{\widehat{M}_{XX^*}[t,\tau]\right\}=M_{XX^*}[t,\tau]$ and a variance $\var\left\{\widehat{M}_{XX^*}[t,\tau]\right\}=M_{XX^*}[t,0]M_{XX^*}^*[t-\tau,0]$. Note that the ESM, its expected value and variance is defined as above for $0\leq t\leq N-1$ and $t-(N-1)\leq \tau \leq t$, and they are zero otherwise. Clearly the variance of $\widehat{M}_{XX^*}[t,\tau]$ is very large and the estimator should not be put to direct use. We form the EMAF
of $X[t]$
by
{\small \begin{equation}
\label{eq:eafdef}
\begin{split}
\widehat{A}_{X X^*}(\nu,\tau]
&=\sum\limits_{t=\max(0,\tau)}^{N-1+\min(0,\tau)}\widehat{M}_{XX^*}[t,\tau]  e^{-j2\pi \nu t}
=\sum\limits_{t=\max(0,\tau)}^{N-1+\min(0,\tau)}x[t] x^*[t-\tau]  e^{-j2\pi \nu t}
\end{split}.
\end{equation}}
The EMAF corresponds to an {\em estimator} of the AF of $X[t]$. However, it is {\em not} a suitable estimator of $A_{XX^*}(\nu,\tau]$, as its variance will be extremely large. When implementing (\ref{eq:eafdef}) we have chosen zero-padding as our choice of extending the estimated covariance matrix. We found that periodic (circular) extension created unpalatable mixing effects of a fundamentally non-stationary object. Edge-effects are discussed more carefully in \cite{Hedges2002}. The best choice of edge treatment will depend on the statistical and deterministic properties of the chosen EMAF. 

\section{Properties of the Empirical Ambiguity Function \label{prop}}
\subsection{Representation and Moments of the EMAF}
We shall construct estimators of the AF, based on the EMAF. 
The EMAF of an analytic zero-mean harmonizable process $X[t]$ admits the representation of
\begin{eqnarray}\label{repofEAF}
\nonumber
\widehat{A}_{X X^*}(\nu,\tau]=\int_{0}^{1/2}\int_{0}^{1/2} e^{j\pi (\nu'-\nu) (N+\tau-1)}e^{j2\pi f\tau}D_{N-|\tau|}(\nu'-\nu)d\widetilde{X}(\nu'+f)d\widetilde{X}^*(f),
\end{eqnarray}
where
$D_{N}(f)=\sin(\pi f N)/\sin(\pi f)$. We respectively denote $\mu_{A}(\nu,\tau]$, $\sigma^2_{A}(\nu,\tau]$ and $r_{A}(\nu,\tau]$ as the mean, variance and relation of the EMAF. We denote an analytic process corresponding to a real-valued stationary white process as an analytic white process, but keep in mind that this process is not actually white. Thus, an analytic white process is a stationary process that has a power spectral density that is constant for nonnegative frequencies, and zero for negative frequencies. 
\begin{proposition}{Moments of the EMAF for noisy Deterministic Signals \label{edet}}\\
For a process that is the sum of a deterministic analytic signal and analytic white noise with variance $\sigma_W^2$, $X[t]=g[t]+W[t]$, the expected value, variance and relation of the EMAF take the form
{\small\begin{alignat}{1}
\mu_{A}(\nu,\tau]&=\widehat{A}_{gg^*}(\nu,\tau]+\frac{\sigma^2_W}{2}e^{-j\pi\nu(N+\tau-1)}D_{N-|\tau|}(\nu) e^{j\pi\tau/2}\sinc(\tau/2)\\
\sigma^2_{A}(\nu,\tau]&=\sigma^2_{W}h(\nu,\tau]+\sigma^2_{W}h(-\nu,-\tau]+\sigma^4_{W}(N-|\tau|)(1/2-|\nu|)+O\left(1\right)\\
\nonumber
r_{A}(\nu,\tau]&=-\sigma^4_{W}|\nu|(N-|\tau|) L\left(N-\tau,\nu\right)e^{-2j\pi \nu(N-1)}
\sinc(2|\nu| \tau)+2\sigma^2_{W}h'(\nu,\tau]+O\left(1 \right)\\
\nonumber
L\left(N-|\tau|,\nu\right)&=\int_{-\infty}^{\infty} \sinc(f)\sinc(f+2(N-|\tau|)\nu) \,df,\quad
h(\nu,\tau]=\int_{\max(-\nu,0)}^{1/2-\max(0,\nu)} \left| G(f,\tau]\right|^2 df\\
\nonumber
h'(\nu,\tau]&=\int_{\max(0,-\nu)}^{1/2+\min(0,-\nu)} G^*(f,\tau]G(f+2\nu,-\tau]e^{j2\pi(f-\text{sign}(\tau)\nu)\tau}df
\end{alignat}}
where $\sinc(x)=\sin(\pi x)/(\pi x)$ and $G(f,\tau]=\sum\limits_{t=0}^{N-1-|\tau|} g[t-|\tau|I(\tau < 0)]e^{-j2\pi ft}$. The indicator function is defined as $I(\tau<0)=1$ if $\tau<0$ and $I(\tau<0)=0$ otherwise. 
\end{proposition}
\begin{proof}
See appendix \ref{prop1}.
\end{proof}
\begin{proposition}{Moments of the EMAF for a Stationary Process \label{estat}}\\
For a zero-mean, analytic stationary stochastic process $X[t]$ the expected value, variance and relation
of the EMAF take the form
\begin{eqnarray}
\label{estat2}
\mu_{A}(\nu,\tau]
&=& D_{N-|\tau|}(\nu)e^{-j\pi\nu(N+\tau-1)}\widetilde{M}_{X X^*}[\tau]\\
\sigma^2_{A}(\nu,\tau]
&=&(N-|\tau|)(1/2-|\nu|)\overline{A}_{X X^*}(-\nu,0]+O\left(1\right)\\
r_{A}(\nu,\tau] 
&=&e^{-2j\pi \nu (N+\tau-1)}(N-|\tau|)(1/2-|\nu|)L(N-|\tau|,\nu)\overline{A}_{X X^*}(\nu,\tau]+O\left(1 \right)
\end{eqnarray}
where $\widetilde{M}_{X X^*}[\tau]$ is the autocorrelation sequence of the process, $\widetilde{S}_{X X^*}(f)$ is the spectral density of the process, and
\begin{equation}
\overline{A}_{X X^*}(\nu,\tau]=\int_{\max(0,-\nu)}^{1/2+\min(\nu,0)}\widetilde{S}_{X X^*}(f-\nu) \widetilde{S}_{X X^*}(f)e^{j4\pi f\tau}\;df/(1/2-|\nu|).
\end{equation}
\end{proposition}
\begin{proof}
See appendix \ref{prop4}.
\end{proof}
The AF of a stationary process will be nonzero only on the stationary manifold $\nu=0$, and takes the form $A_{X X^*}(\nu,\tau]=\widetilde{M}_{X X^*}[\tau]\delta(\nu)$. For a {\em fixed} value of $N$, $D_{N-|\tau|}(\nu)$ does {\em not} correspond to a delta function in $\nu$. Ideally however the estimated EMAF should not exhibit spreading in $\nu$. 

\begin{proposition}{Moments of the EMAF for a Uniformly Modulated White Noise Process \label{eunif}}\\
For a zero-mean, analytic process $X[t]$ corresponding to a real-valued uniformly modulated white noise process, the expected value, variance and relation of the EMAF take the form
\begin{eqnarray}
\label{eunif2}
\mu_{A}(\nu,\tau]  &=&(1/2-|\nu|)\Sigma_{X X^*}(\nu) e^{j\pi (1/2-|\nu|)\tau}\sinc((1/2-|\nu|)\tau)+O\left(1\right)\\
\sigma^2_{A}(\nu,\tau]&=&(1/2-|\nu|)(N-|\tau|)\int_{-1/2+|\nu|}^{1/2-|\nu|} \frac{\left|\Sigma_{X X^*}(f) \right|^2}{N-|\tau|}e^{j2\pi f\tau}df+O\left(1\right)\\
\nonumber
r_{A}(\nu,\tau]&=&e^{-4j\pi\nu\tau}\int_{\max(0,\nu)}^{1/2+\min(0,\nu)}\int_{\max(0,\nu)}^{1/2+\min(0,\nu)}\Sigma_{X X^*}(f-\alpha+\nu)\Sigma_{X X^*}^*(f-\nu-\alpha) e^{j2\pi(f+\alpha)\tau}
\,df d\alpha+O\left(1\right).
\end{eqnarray}
where $\Sigma_{XX^*}(\nu)$ is the DFT of the time-varying variance of $X[t]$, $\sigma^2_X[t]$.
\end{proposition}
\begin{proof}
See appendix \ref{prop7}.
\end{proof}
We note that the result of the integral in the expression for the variance decreases with $|\nu|$ and also with $|\tau|$.

\subsection{Distribution of the EMAF of an Underspread Process}
To determine suitable estimation procedures for the AF we need to derive the distribution of the EMAF.
\begin{theorem}{Distribution of the EMAF of an underspread process
\label{distEAF}}\\
Assume $X[t]$ is the analytic process corresponding to a harmonizable real-valued zero-mean Gaussian
process whose EMAF is strictly underspread. A strictly underspread process has an ambiguity function that is only non-zero for $(\nu,\tau]\in{\cal D}$, where there exists some finite non-negative $T$ and $\Omega$ such that ${\cal D}\subset [-\Omega,\Omega]\times [-T,T]$. Then the EMAF of $X[t]$ evaluated at $(\nu,\tau]$ has a mean, variance and relation of
{\small\begin{eqnarray}
\label{meanunder2}
\mu_{A}(\nu,\tau]&=& \int\limits_{0}^{1/2}\int\limits_{-f}^{1/2-f} S_{XX^*}(\nu',f)e^{j2\pi f\tau}
e^{j\pi(N+\tau-1)(\nu'-\nu)}D_{N-|\tau|}(\nu'-\nu)d\nu'df\\
\label{theovar}
\sigma^2_{A}(\nu,\tau]&=&
\sum_{t=\max(0,\tau)}^{N-1+\min(0,\tau)}\sum_{\tau'=-(T-1)}^{T-1}
e^{-j2\pi \nu \tau'}M_{XX^*}[t,\tau']M_{XX^*}^*[t-\tau,\tau']+O\left(\log\left[\frac{N}{T}\right] \right)\\
r_{A,N}(\nu,\tau]&=&
\begin{cases}
O\left(\log\left[\frac{N}{T}\right] \right)\quad {\mathrm{if}} \quad & \left|\tau\right|>T\\
0  \quad {\mathrm{if}} \quad & \left|\nu\right|>\Omega,
\end{cases}
\label{relunder2}
\end{eqnarray}}
while if $|\nu|<\Omega$ and $|\tau|<T$ then the relation in noted in \eqref{rely2}.
Fix $\left(\nu,\tau\right)\notin {\cal D}$. Let $\tau\ge 0$ and take
\begin{equation}
Q_{N-|\tau|}[t]=e^{-j2\pi\nu(t+\tau)}\left(X[t+\tau]X^*[t]-M_{XX^*}[t+\tau,\tau] \right),\quad t=0,\,\dots,N-1-\tau.
\end{equation}
Assume that the triangular array $\left\{Q_{N-|\tau|}[t],\quad
t=0,\dots,N-1-\tau\right\}$ is strongly mixing and satisfies for any $\epsilon>0$
\begin{equation}
\frac{1}{\sigma^2_{A}(\nu,\tau]}\sum_{t=0}^{N-\tau-1}\var\{Q_{N-|\tau|}[t]\}
I\left(|Q_{N-|\tau|}[t]|>\epsilon \sigma_{A}(\nu,\tau]\right)\longrightarrow 0,
\end{equation}
as $N-|\tau|\rightarrow \infty$, as well as the condition:
\begin{alignat}{1}&
\label{haga}
\sup_{N}\frac{1}{\sigma^2_{A}(\nu,\tau]}
\sum_{t=0}^{N-\tau-1}M_{XX^*}[t+\tau,0]M_{XX^*}[t,0]<\infty.
\end{alignat}
Then, with $\overset{\cal L}{\longrightarrow}$ denoting convergence in law
\cite{Ferguson96},
\begin{equation}
\frac{\widehat{A}_{X X^*}(\nu,\tau]-\mu_{A}(\nu,\tau]}
{\sigma_{A}(\nu,\tau]}\overset{\cal L}{\longrightarrow}{\cal N}^C\left(0,
1,0\right),\quad (\nu,\tau)\notin {\cal D},\quad t\ge 0,
\end{equation}
where ${\cal N}^C\left(0,1,0\right)$ denotes a complex Gaussian distribution with mean 0, variance 1 and relation 0. 
The result follows {\em mutatis mutandis} for $\tau<0$.
\end{theorem}
\begin{proof}
See appendix \ref{theorem13}.
\end{proof}
It is tempting to attempt to deduce the distributional results directly from
the first and second order structure established in the first part of the
theorem. In fact, if $X[t]$ corresponds to a stationary process, where $T$ and $\tau$ are both even, the result follows
directly. For more general classes of processes the result is slightly more involved,
and a condition on the variance needs to be combined with a suitable
mixing assumption, as above. Note that the CLT theorem that we use does not
provide rates of convergence. Also, in some degenerate cases the joint distribution of  $\left\{\widehat{A}_{X X^*}(\nu,\tau]\right\}_{(\nu,\tau)}$ may not be asymptotically multivariate Gaussian. For ease of exposition, such cases are not treated here.
Finally, whilst (asymptotically) retrieving the Gaussian distribution for the EMAF by Theorem \ref{distEAF}, we fail to obtain simple interpretable forms for $\sigma^2_{A}(\nu,\tau]$ and $r_{A}(\nu,\tau]$. For this reason, it is convenient to derive the first and second order structure directly from the modelling assumptions of some commonly used processes, as done in Section~\ref{prop}. 
\section{Estimation Procedure \label{Estimation}}
To determine the approximate distribution of the EMAF of a deterministic signal immersed in analytic white noise, we need to note the distributions of the four components of~(\ref{sum}). Here, $A_{gg^*}(\nu,\tau]$ is constant, whilst $A_{WW^*}(\nu,\tau]$
is asymptotically Gaussian, and its form is determined by Theorem \ref{distEAF}.
We note that since $W[t]$ is Gaussian analytic white noise, and so from~(\ref{grej}) and~(\ref{nyekva1})
{\small 
\begin{alignat}{1}
\nonumber
A_{Wg^*}(\nu,\tau]+A_{gW^*}(\nu,\tau]&\overset{d}{=}
{\cal N}^C\left(0,\sigma^2_{W}\left(h(\nu,\tau]+h(-\nu,-\tau]\right),2\sigma^2_{W}h'(\nu,\tau]\right).
\end{alignat}}
From~(\ref{ae4}) and~(\ref{eq:3er}) we can note that $A_{Wg^*}(\nu,\tau]+A_{gW^*}(\nu,\tau]$ is independent of $A_{WW^*}(\nu,\tau]$.
Combining~(\ref{sum}) and Theorem~\ref{distEAF}, as well
as using proposition \ref{edet}, it follows for $(\nu,\tau)\neq (0,0)$
\begin{eqnarray}
&&\frac{\widehat{A}_{X X^*}(\nu,\tau]-\mu_{A}(\nu,\tau]}{\sigma_{A}(\nu,\tau]}\overset{\cal
L}{\longrightarrow}{\cal N}^C
\left(0, 1,r_{A}(\nu,\tau]/\sigma^2_{A}(\nu,\tau]\right)\\
\end{eqnarray}
Hence we may note $|\mu_{A}(\nu,\tau]|^2\ll \sigma^2_{A}(\nu,\tau]$ in most of the ambiguity
plane. Furthermore it follows that
{\small\begin{eqnarray}
\frac{\left|\widehat{A}_{X X^*}(\nu,\tau]-\mu_{A}(\nu,\tau]\right|^2}
{\sigma^2_{A}(\nu,\tau]}&
\overset{d}{=}&
\frac{1}{2}\left[1+\frac{|r_{A}(\nu,\tau]|}{\sigma^2_{A}(\nu,\tau]}\right]U_1^2+
\frac{1}{2}\left[1-\frac{|r_{A}(\nu,\tau]|}{\sigma^2_{A}(\nu,\tau]}\right]U_2^2+o\left(1\right),
\end{eqnarray}}
where $U_1$ and $U_2$ are independent standard Gaussian random variables.
Thus for such points in the ambiguity plane it follows that $\left|\widehat{A}_{X X^*}(\nu,\tau]-\mu_{A}(\nu,\tau]\right|^2$ is a weighted sum of $\chi^2_1$'s. These are equally weighted if $r_{A}(\nu,\tau]= 0$. For most points in the
ambiguity plane this statement will be roughly valid as apparent from proposition
\ref{edet}.
Then if $(\nu,\tau]\notin{\cal D}$, where ${\cal D}$ denotes the support region of the ambiguity function of the real part of $X[t]$, we may note that
\begin{eqnarray}
\frac{\left|\widehat{A}_{X X^*}(\nu,\tau]-\mu_{A}(\nu,\tau]\right|^2}{\sigma^2_{A}(\nu,\tau]}&
\overset{d}{=}&
\left[\frac{1}{2}\chi^2_2+
O\left(\frac{\left|\log(N) \right|^2}{N-|\tau|}\right)\right]+o\left(1\right).
\end{eqnarray}
A suitable estimation procedure for such random variables with $\sigma^2_{A}(\nu,\tau]$ known a priori, is then the following thresholding procedure, for some given
threshold $\lambda^2$,
\begin{eqnarray}
\label{troskelvarde}
\widehat{A}_{XX^*}^{({\mathrm{ht}})}(\nu,\tau]&=&
\left\{\begin{array}{lcr}
\widehat{A}_{XX^*}(\nu,\tau] &{\mathrm{if}} & \left|\widehat{A}_{XX^*}(\nu,\tau]\right|^2>
\lambda^2\sigma^2_{A}(\nu,\tau]\\
0 &{\mathrm{if}} & \left|\widehat{A}_{XX^*}(\nu,\tau]\right|^2\le
\lambda^2\sigma^2_{A}(\nu,\tau]\end{array}\right. .
\end{eqnarray}
If we normalize $\left\{\left|\widehat{A}_{X X^*}(\nu,\tau]\right|^2\right\}$
via dividing the sequence, element by element, by $\sigma_{A}^2(\nu,\tau]$,
then we retrieve a set of {\em correlated} positive random variables. 
For any such collection, we may note from \cite{Olhede07}, that using a threshold $\lambda^2_{N_X}(C)=2\log(N_X[\log(N_X)]^C)$ for $C\ge 1$ is suitable. In our example $N_X$ is chosen to be twice the number of observations, as we threshold the ambiguity function frequency by frequency, across the total collection of all time lags. Olhede~\cite{Olhede07}[p.~1529] has calculated the risk of this non-linear estimator for sums of unequally weighted $\chi^2_1$'s.

Of course $\sigma^2_{A}(\nu,\tau]$ is {\em not} known. We standardize the EMAF
by
\begin{equation}
\label{eq:2}
\widehat{A}_{XX^*}^{(S)}(\nu,\tau]=\frac{\widehat{A}_{XX^*}(\nu,\tau]}
{\sqrt{(N-|\tau|)(1/2-|\nu|)}},
\end{equation}
and note that
\begin{eqnarray}
\label{stuff1}
\var\left\{\widehat{A}_{XX^*}^{(S)}(\nu,\tau]\right\}&=& \sigma^4_{W}+
\sigma^2_{W}\frac{h(\nu,\tau]+h(-\nu,-\tau]}{(N-|\tau|)(1/2-|\nu|)}+
O\left(\frac{1}{N-|\tau|} \right).
\end{eqnarray}
We have assumed that $g[t]\in \ell_2$, and so $\left|h(\nu,\tau]\right|=O\left(1\right)$. Thus $\sigma^2_{W}\frac{h(\nu,\tau]+h(-\nu,-\tau]}{(N-|\tau|)(1/2-|\nu|)}=o\left(1\right)$, where the form of $h(\nu,\tau]$ at high values of $|\nu|$ and $|\tau|$ ensure that the statement is still true for such values. We shall use the Median Absolute Deviation (MAD) estimator of the variance. MAD has been used for estimating the scale of correlated data before, see
\cite{Johnstone97}. The MAD estimator will need an adjustment factor that is different for $\frac{1}{2}\chi^2_2$ random variables compared to $\chi^2_1$ random variables. We note the median of a $\frac{1}{2}\chi^2_2$ is $\ln[2]$.
We define an estimator of the analytic white noise variance for any region
$\left\{(\nu,\tau] \right\}={\cal M}\subset\left[-1/2,1/2\right]
\times  \left\{-(N-1),\dots,(N-1)\right\}$ by
\begin{equation}
\widehat{\sigma}^4_{W}({\cal M})
=\frac{{\mathrm{median}}\left\{\left|\widehat{A}_{XX^*}^{(S)}(\nu,\tau]
\right|^2
\right\}_{(\nu,\tau)\in{\cal M}}}{\ln[2]}.
\label{varest}
\end{equation}
The imprecision of this procedure will depend on the lack of compression of the representation of $\left\{X[t]\right\}$ in the ambiguity domain. We note that MAD has a breakdown point
of 50 \%, and so with quite severe contamination the estimator will still
be useful, if somewhat inefficient. We then take
\begin{equation}
\label{eq:sdfwes}
\widehat{\sigma}^2_{A}(\nu,\tau|{\cal M}]=\widehat{\sigma}^4_{W}({\cal M})(N-|\tau|)(1/2-|\nu|).
\end{equation}
A suitable threshold procedure (now with an unknown $\sigma^2_{A}(\nu,\tau]$)
simply corresponds to using~(\ref{troskelvarde}) with the variance replaced by its estimated value from~(\ref{eq:sdfwes}). If the variance is estimated from the entire plane, i.e., ${\cal M}=\left\{-N/(2N),\dots,N/(2N)\right\}
\times \left\{-(N-1),\dots,(N-1)\right\}$, we denote $\widehat{A}_{XX^*}^{({\mathrm{ht}})}(\nu,\tau]$ the Thresholded EMAF (TEAF). 

\begin{figure}
    \begin{center}
   \includegraphics[scale=0.4]{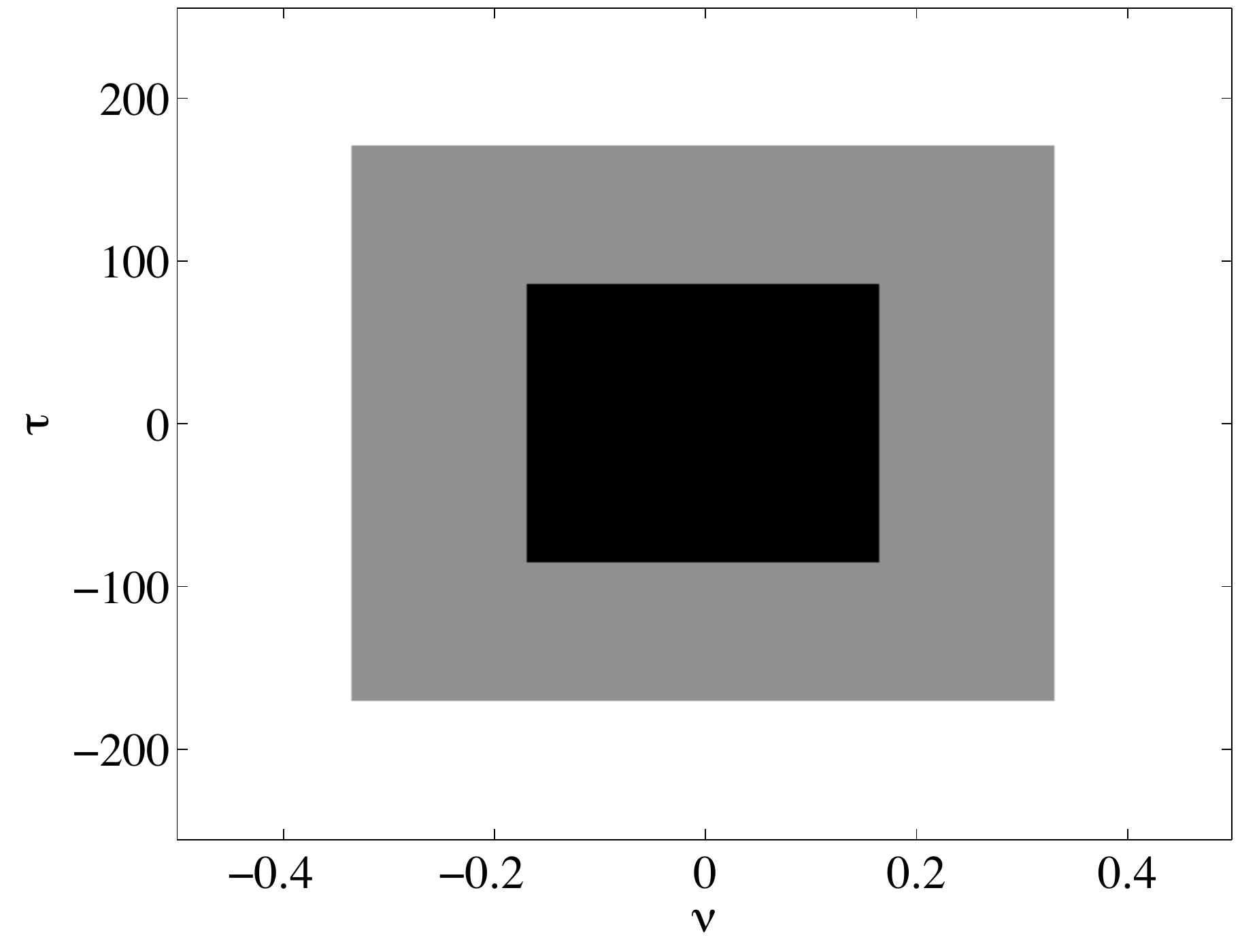}
   \includegraphics[scale=0.4]{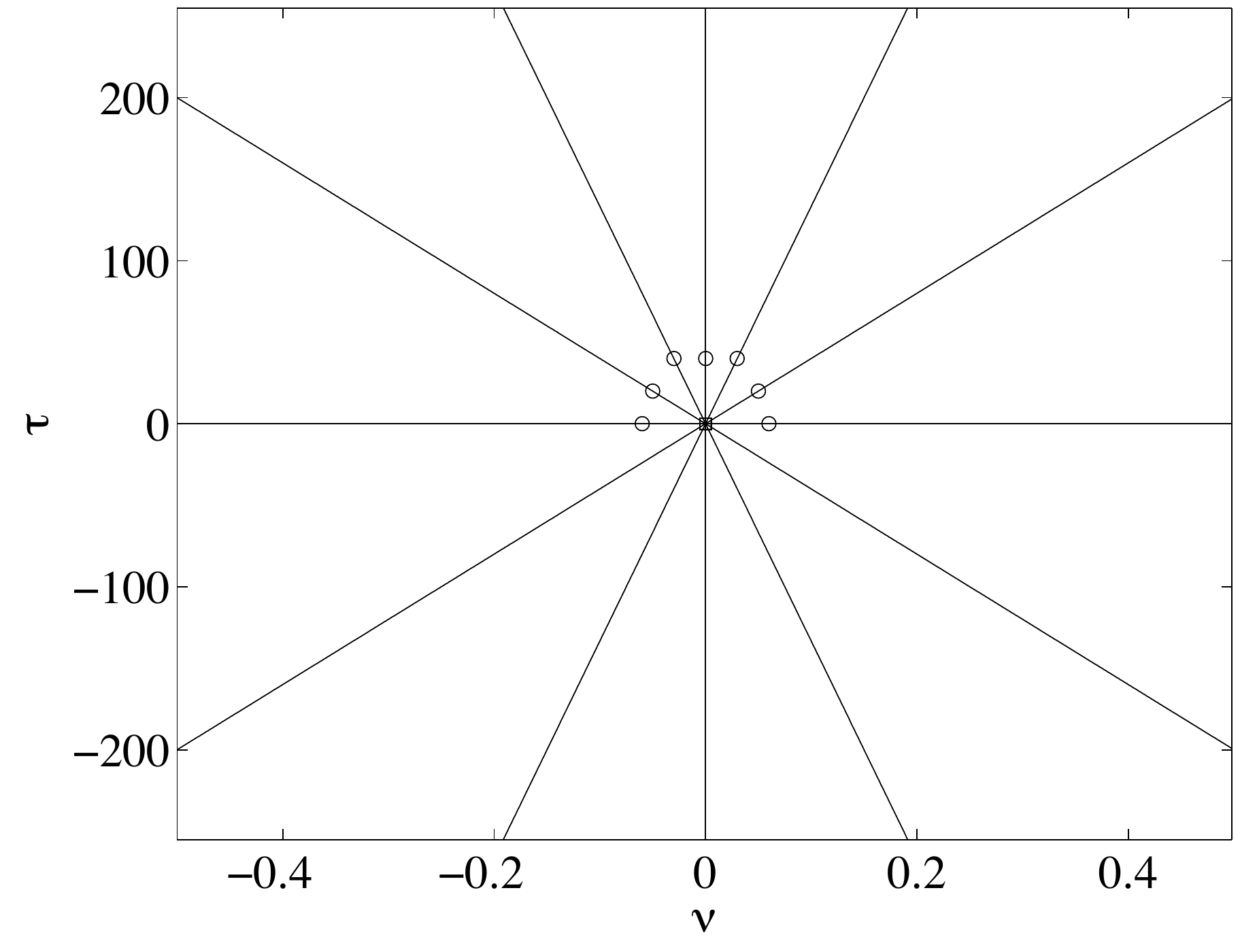}
     \end{center}
     \caption{(a) Partitioning of the ambiguity plane. (b) Radial lines over
     which the integrand in (\ref{lalavar}) exhibits stationary phase.}
     \label{fig:annuli}
\end{figure}

The EMAF of a deterministic signal immersed in analytic white noise is biased, see proposition \ref{edet}. Given an estimator of $\sigma^2_{W}$, we can remove this bias prior to thresholding.
We define the bias-corrected EMAF by
\begin{equation}
\widehat{A}^{(B)}_{XX^*}(\nu,\tau]=
\widehat{A}_{XX^*}(\nu,\tau]-\frac{1}{2}\widehat{\sigma}^2_{W}({\cal M}_1)e^{-j\pi \nu(N+\tau-1)}D_{N-|\tau|}\left(
\nu\right) e^{j\pi \tau/2}\sinc\left(\tau/2 \right).
\end{equation}
The noise variance is estimated from some given region ${\cal M}_1$, which is chosen to only include the rims of the ambiguity plane. We normalize $\widehat{A}^{(B)}_{XX^*}(\nu,\tau]$ as in~(\ref{eq:2}). For smaller sample sizes, it may be unreasonable to assume that the contributions of $h(\nu,\tau]$ are negligible compared to $\sqrt{N-|\tau|}$. We define $\overline{\sigma}_X^4(\nu,\tau]=
\sigma^2_{A}(\nu,\tau]/[(N-|\tau|)(1/2-|\nu|)]$. This is explicitly subscripted by $X$, as it is only non-constant from contributions due to $h(\nu,\tau]$.
We propose to segment the plane into regions, and implement the proposed procedure after
taking a separate estimate of $\overline{\sigma}^4_X(\nu,\tau]$ in each region. The optimal choice of region is given by (\ref{stuff1}). Since $h(\nu,\tau]$ is unknown, we propose to use a centre square around $(0,0]$,
and square annuli, see Fig.~\ref{fig:annuli}(a), to estimate the local variance. This is based on assuming $\sigma^2_{A}(\nu,\tau]$ smooth and decaying from $(0,0)$. In general
we separate the ambiguity plane into $K$ regions
$\left\{{\cal M}_k\right\}$, so that $\overline{\sigma}^4_X(\nu,\tau]\approx
\overline{\sigma}^4_X({\cal M}_k]$, a constant for all $(\nu,\tau]\in {\cal M}_k$. We estimate the variance in each region as
\begin{equation}
\widehat{\overline{\sigma}}^4_{X}({\cal M}_k)
=\frac{{\mathrm{median}}\left\{\left|\widehat{A}_{XX^*}^{(S)}(\nu,\tau]
\right|^2
\right\}_{(\nu,\tau)\in {\cal M}_k}}{\ln[2]}.
\end{equation}
We threshold $\widehat{A}^{(B)}_{XX^*}(\nu,\tau]$ according to~(\ref{troskelvarde}), but now using $\widehat{\overline{\sigma}}^4_{X}({\cal M}_k)$ in each defined region instead of $\widehat{\sigma}^4_{W}({\cal M})$ in~(\ref{eq:sdfwes}), this yielding $
\widehat{A}_{XX^*}^{({\mathrm{lbht}})}(\nu,\tau]$, or the Local Bias-corrected TEAF (LBTEAF). 

Next, we derive results for estimating the EMAF of an underspread zero-mean stochastic process. This is strictly speaking a correct treatment for processes satisfying the constraints of Theorem~\ref{distEAF}, but we expect that the distributional result is valid under less constrained scenarios. The conditions are sufficient, but by no means necessary, for the distributional result to hold.
We note from Theorem~\ref{distEAF} that for most points in the ambiguity plane
\begin{equation}
\frac{\left|\widehat{A}_{X X^*}(\nu,\tau]-\mu_{A}(\nu,\tau]\right|^2}
{\sigma^2_{A}(\nu,\tau]}
\overset{d}{=}
\frac{1}{2}\chi^2_2+
O\left(\frac{\left|\log(N) \right|^2}{N-|\tau|}\right)+o\left(1\right).
\end{equation}
The question then arises of yet again estimating $\sigma^2_{A}(\nu,\tau]$
appropriately. We note from Theorem~\ref{distEAF} that $\sigma^2_{A}(\nu,\tau)$ takes different forms depending on the spreading of the process.
The more $X[t]$ is like an analytic white noise process, the less variation will $\sigma^2_{A}(\nu,\tau]$ exhibit from the form of the variance of the EMAF of an analytic white noise process. We again define $\overline{\sigma}^4_X(\nu,\tau]=\sigma^2_{A}(\nu,\tau]/[(N-|\tau|)(1/2-|\nu|)]$.
Then 
\begin{eqnarray}
\frac{\left|\widehat{A}_{X X^*}(\nu,\tau]-\mu_{A}(\nu,\tau]\right|^2}{(N-|\tau|)(1/2-|\nu|)}
\overset{d}{=}\overline{\sigma}^4_X(\nu,\tau]
\left[ \frac{1}{2}\chi^2_2+
O\left(\frac{\left|\log(N) \right|^2}{N-|\tau|}\right) 
+o\left(1\right)\right].
\end{eqnarray}
Using inverse FTs we note from Theorem~\ref{distEAF} that the variance of the EMAF of a strictly underspread process can in fact be rewritten for $\tau>0$ 
\begin{equation}
\label{lalavar}
\begin{split}
&\sigma^2_{A}(\nu,\tau]=
\sum_{\tau'=-(T-1)}^{T-1} \int_{-\Omega}^{\Omega} \int_{-\Omega}^{\Omega}
e^{-j2\pi(\nu\tau'-\nu'\tau)}A_{XX^*}(\nu',\tau']\\
& A_{XX^*}^*(\nu'',\tau']e^{j\pi(N-\tau-1)(\nu'-\nu'')}
D_{N-\tau}(\nu'-\nu'')\;d\nu'\;d\nu''+O\left(\log\left[\frac{N}{T}\right] \right)\\
&=\sum_{\tau'=-(T-1)}^{T-1} \int_{-\Omega}^{\Omega}
e^{-j2\pi(\nu\tau'-\nu'\tau)}|A_{XX^*}(\nu',\tau']|^2\;d\nu'+O\left(\log\left[\frac{N}{T}\right] \right)+O(1).
\end{split}
\end{equation}
Deriving this requires that $A_{XX^*}(\nu,\tau]$ is sufficiently smooth in $\nu$. If this is not the case, use~(\ref{theovar}). The variance of the EMAF corresponds to aggregating the total magnitude squared of the AF over the plane and implementing phase shifts. If the function is smooth in $\nu$
with a stationary phase approximation to the integral we mainly pick up contributions
on the line $\nu'=\nu \tau'/\tau$, which we later sum over $\tau'$. This is exemplified for some choices of $\nu$ and $\tau$ in Figure~\ref{fig:annuli}(b),
where we plot the lines summed over.
Despite this geometrical intuition, the form in (\ref{lalavar}) is not extremely
informative. Equation~(\ref{lalavar}) does
indicate that $\sigma^2_{A}(\nu,\tau]$ should be a smooth function of
$\nu$ and $\tau$, as we only integrate over ${\cal D}$. We rely on the forms of propositions \ref{estat} and \ref{eunif} to give more precise understanding of the variance of the EMAF in these special
cases. We note from these set of results that the variance is smooth in $\nu$ and $\tau$, and often exhibits a decay in $(\nu,\tau)$ that resembles that of the variance of analytic white noise.

If $\overline{\sigma}^4_X(\nu,\tau]$ is exactly constant across the ambiguity plane
we can propose
an estimator of $\overline{\sigma}^4_X(\nu,\tau]$ given by equating $
\widehat{\overline{\sigma}}^4_{X}(\nu,\tau]$ to the estimator from~(\ref{varest}). Propositions~\ref{estat} and \ref{eunif} argue that in regions of the ambiguity plane $\overline{\sigma}^4_X(\nu,\tau]$ is smooth, and can be approximated as constant over a given region, again using~(\ref{varest}). Noting that as the process is underspread $|\mu_{A}(\nu,\tau]|\ll \sigma_{A}(\nu,\tau]$ for most $(\nu,\tau]$, and so thresholding is an admissible estimation procedure. We therefore propose a Local TEAF (LTEAF), by estimating the variance in given regions, just like the LBTEAF, but without removing the bias. We will divide the ambiguity plane into regions according to Fig.~\ref{fig:annuli}(a) to obtain both the LTEAF and LBTEAF. Using separate regions in the local time-frequency domain to estimate the variance of the EMAF is a natural procedure also for stochastic processes. It is reasonable to assume that the time-varying spectrum is smooth in global frequency and global time. If the time-varying spectrum is a member of a Sobolev space of order $k$, then the modulus of the Fourier transform must decay faster than $\nu^{-(k+1)}$ in local frequency, and $\tau^{-(k+1)}$ in local time, see for example \cite{Wasserman07}. It would therefore be natural to use regions defined in terms of distance from $(\nu,\tau)=(0,0)$ as ${\cal M}_k$, but to more naturally meld with digital sampling we propose to use the set of square annuli, similar to sampling discussed in \cite{Averbuch2006}.
Johnstone and Silverman in a similar spirit \cite{Johnstone97} 
proposed level dependent thresholding using a different variance for each wavelet level, in 1-D.  As a final note the proposed estimators
may not correspond to valid AFs. Only AFs that are the FTs of valid autocovariance
sequences are valid. We do not think this corresponds to a major flaw
of the procedure, as we are mainly concerned with functions whose support
will inform us of the resolution of the autocovariance. 

\subsection{Estimated Spread \label{estspread}}
Using thresholding methods, we have produced an estimate of the AF in the entire ambiguity
plane. We also propose an estimator of the spread of the estimated AF, see
\cite{Matz06,Hedges2002}. Note that for processes that are not strictly underspread, our thresholding procedure will identify regions where the mean of the EMAF
is sufficiently distinct in magnitude from the variance of the EMAF. This in essence corresponds to comparing the magnitude of the AF at a point, with the magnitude of the AF at other points, see (\ref{lalavar}).
We additionally note from (\ref{lalavar}) that
if the AF is mainly centered at other $(\nu',\tau']$ then the variance of the EMAF will be large compared to the mean of the EMAF. In this instance the contribution at $(\nu,\tau]$ is not significantly contributing to the structure of $X[t]$. We define the ambiguity indicator cell to be
\begin{equation}
\zeta_X\left(\nu,\tau\right]=I\left(\left|A_{X X^*}\left(\nu,\tau\right]\right|\neq
0\right),\quad \nu\in\left(-\frac{1}{2},\frac{1}{2}\right),\quad \tau \in{\mathbb{Z}}.
\end{equation}
The total spread of the AF of $X[t]$ over a time-frequency
region ${\cal S}$ is given by
\begin{equation}
0\le \xi_X\left({\cal S}\right)=\frac{\int \sum_{\cal S} \zeta_Z\left(\nu,\tau\right]\;d\nu }{\int \sum_{\cal S} \;d\nu}\le 1.
\end{equation}
A process $X[t]$ is strictly AF Compressible  if $\xi_X\left({\cal S}\right)<<1$.
We define
\begin{equation}
\widehat{\zeta}_X\left(\nu,\tau\right]=I\left(\left|\widehat{A}_{X X^*}^{({\mathrm{ht}})}
\left(\nu,\tau\right]\right|\neq
0\right),\quad \nu\in\left(-\frac{1}{2},
\frac{1}{2}\right),\tau\in\left[-(N-1),(N-1)\right],
\end{equation}
and thus
\begin{equation}
\label{eq:totspread}
\widehat{\xi}_X\left({\cal S}\right)=\frac{\sum \sum_{\cal S} \widehat{\zeta}_X\left(\nu,\tau\right] }{\sum \sum_{\cal S} 1}
\end{equation}
is an estimator of the total spread of the process.
Matz and Hlawatsch define extended underspread processes, as those whose spread is essentially limited.
Such processes will be approximated by the TEAF to the region of their essential support, i.e., where their magnitude is non-negligible in comparison to the rest of the ambiguity plane. 
This permits the quantification of the degree of variability of the time series at each lag, and the band of variability supported at a given lag $\tau_0$, can be determined by $\widehat{\xi}_X\left(\left\{(\nu,\tau]
:\;\tau=\tau_0\right\}\right)$.

\section{Examples \label{examples}}
We estimate the mean squared error (MSE) by comparing the estimated AFs to a theoretically based quantity that is the expected value of the EMAF, for a finite value of $N$, limited to the support $\rho$ of the AF of the process of interest, $\mu_{A}(\nu,\tau]I\left((\nu,\tau]\in \rho\right)$, and refer to this as the $N$-AF. If the process is the analytic process constructed from a real-valued process, then $\rho$ is the support of the AF of the real-valued process.
Note that $A_{X X^*}(\nu,\tau]$ is
based on the {\em infinite} sequence $\left\{M_{X X^*}[t,\tau]\right\}_{t,\tau\in{\mathbb{Z}}}$ and does {\em
not} exhibit any finite sample issues with spreading in local frequency and time lag due to finite sample
effects. We can never observe such values in a finite digital sample,
because the maximum concentration of energy will behave like $N$ (our number
of sample points), and thus will be finite for finite $N$. Therefore, we ideally insist on the concentration of the AF to where $A_{X X^*}(\nu,\tau]$ is supported, but only letting the $N$-AF take finite sample length realizable values.

\subsection{Linear chirp immersed in white noise.}
\begin{figure}
    \begin{center}
    \subfigure[]{\includegraphics[scale=0.27]{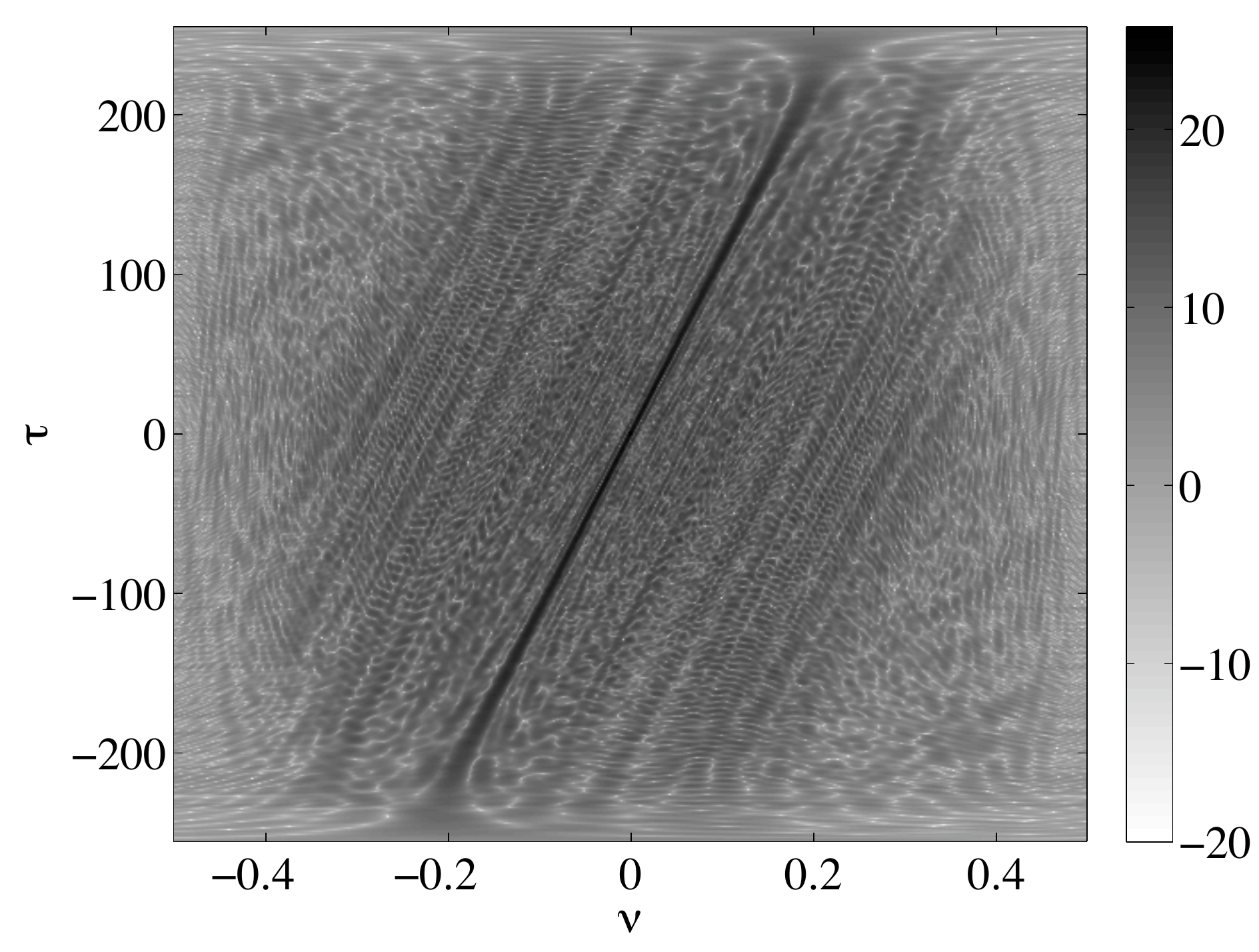}}
    \hspace{0.5cm}
     \subfigure[]{\includegraphics[scale=0.27]{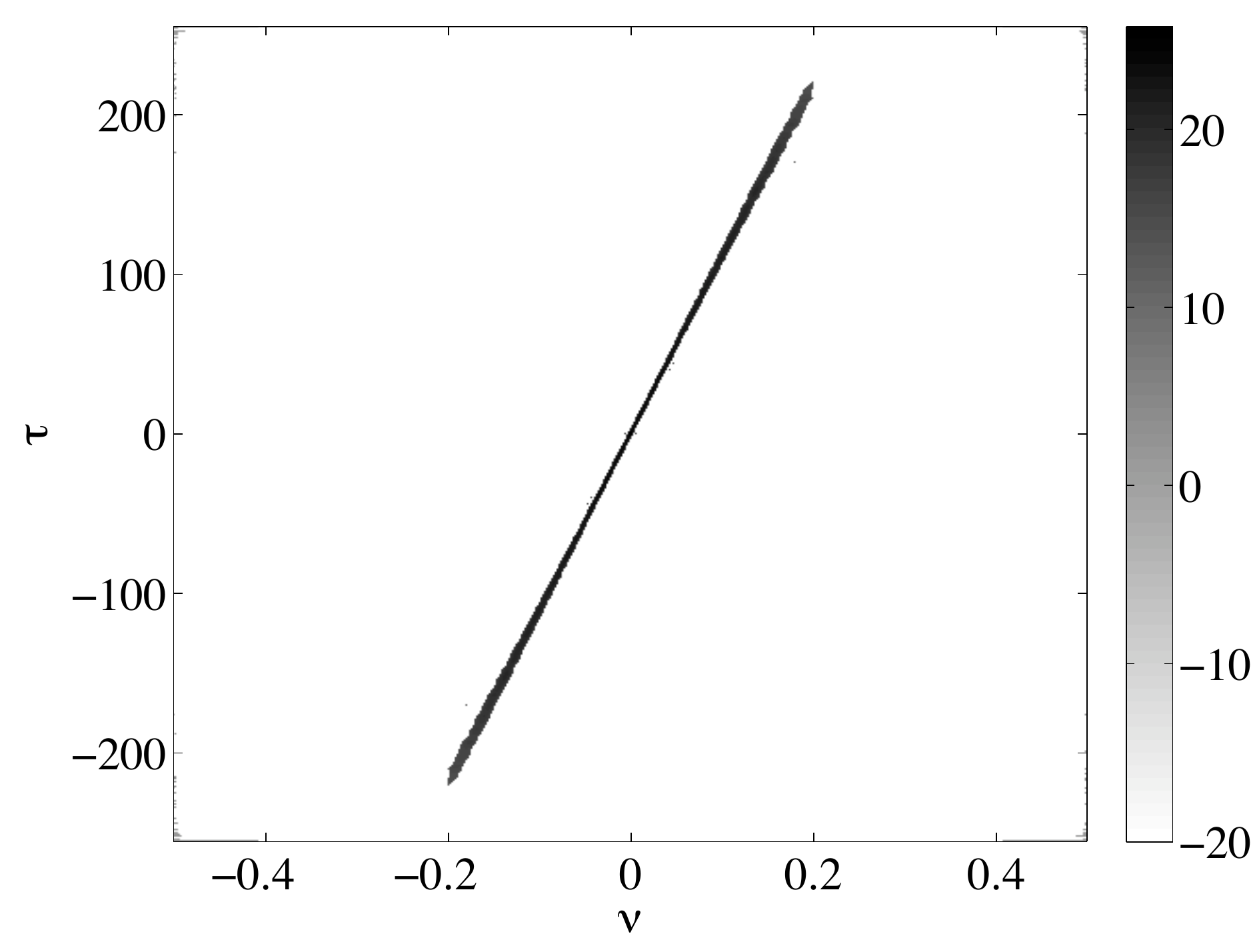}}\\
     \subfigure[]{\includegraphics[scale=0.27]{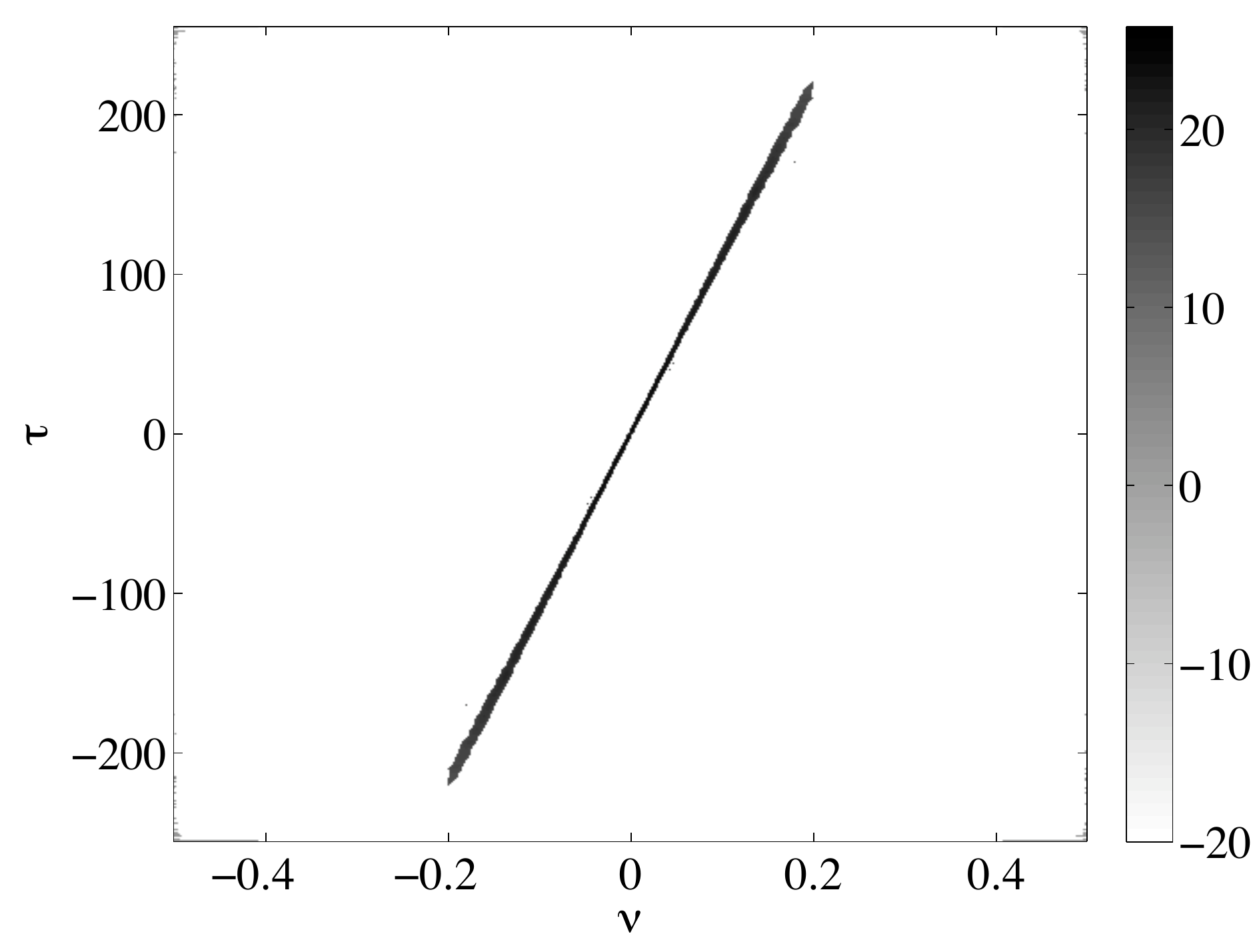}}
         \hspace{0.5cm}
          \subfigure[]{\includegraphics[scale=0.27]{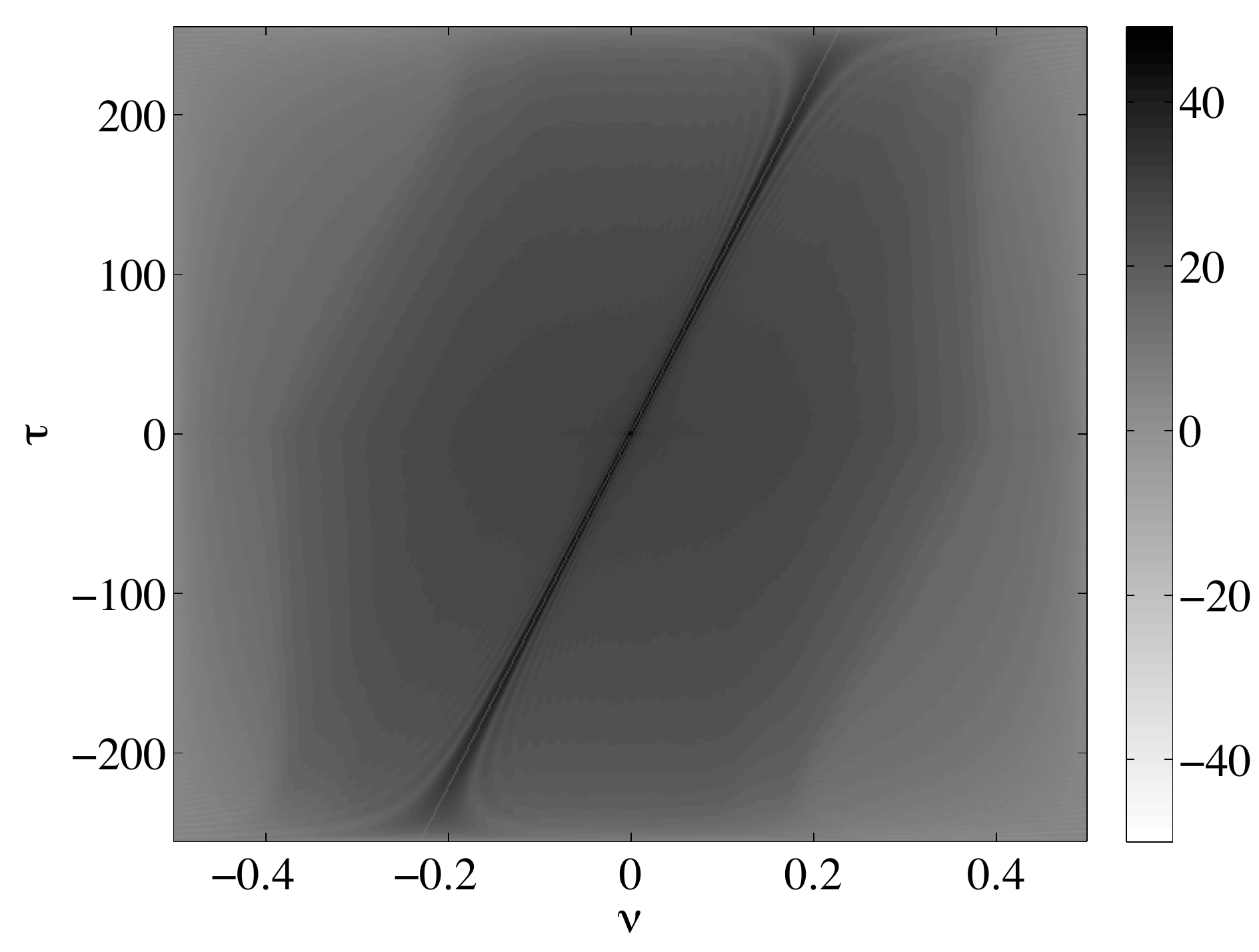}}\\
     \subfigure[]{\includegraphics[scale=0.27]{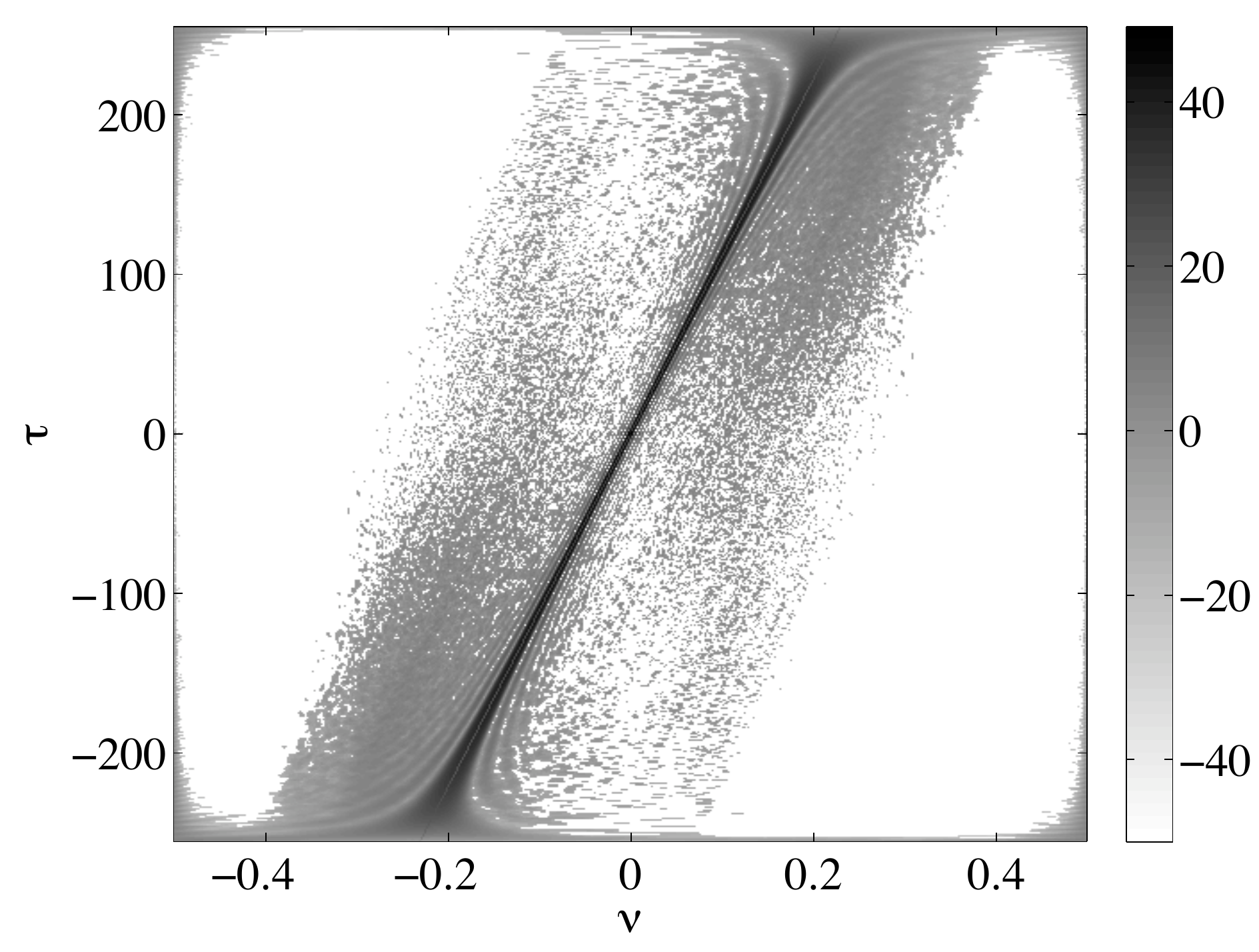}}
              \hspace{0.5cm}
     \subfigure[]{\includegraphics[scale=0.27]{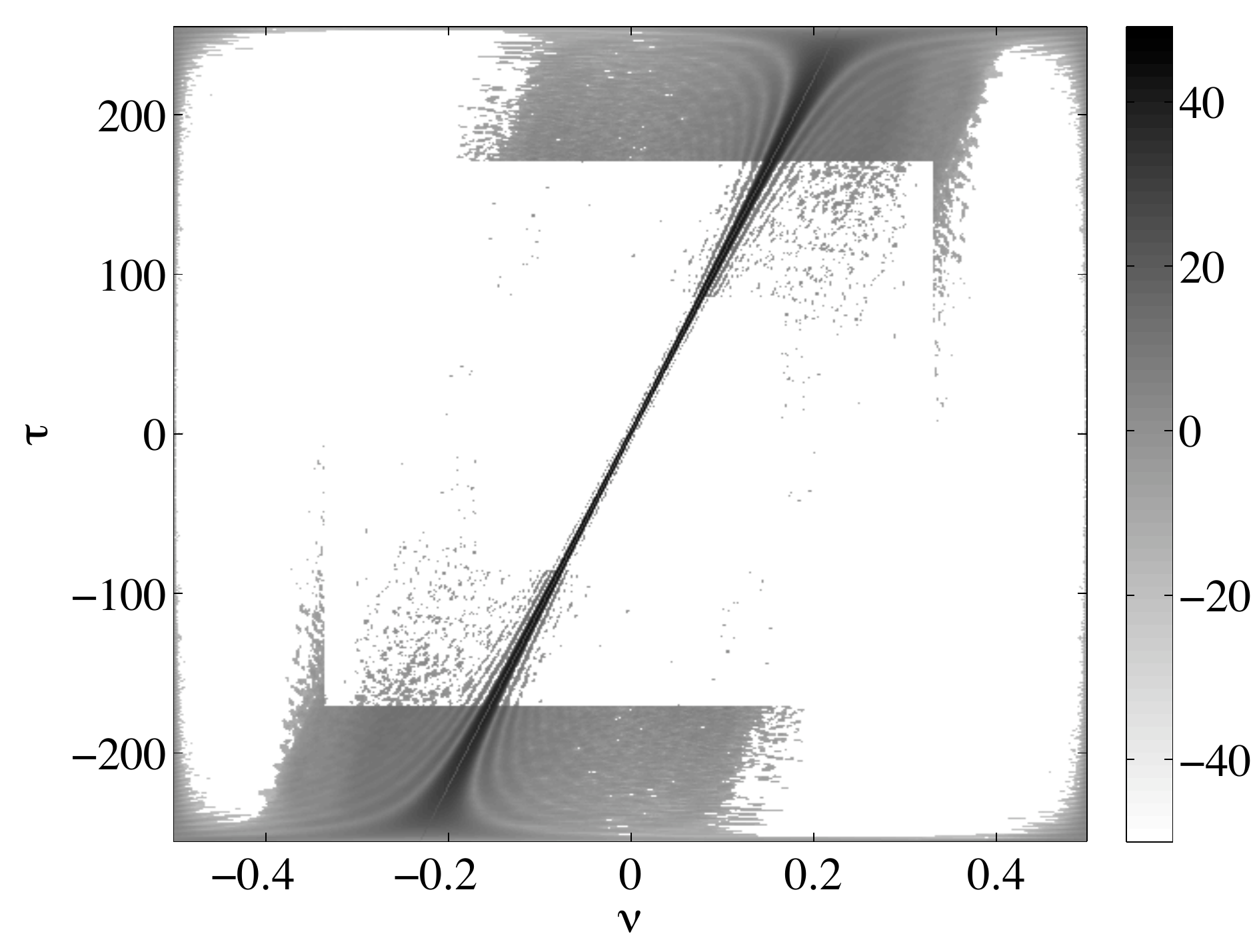}}
     \end{center}
     \caption{Chirp immersed in white noise. (a) EMAF (b) TEAF (c) LBTEAF for one realisation. Estimation of the MSE for (d) EMAF (e) TEAF (f) LBTEAF. All are on a dB scale.}
     \label{fig:exchirp912}
\end{figure}

We first consider the case of a deterministic linear chirp, $g[t]=\exp[j\pi(2\alpha t+\beta t^2)]$, immersed in analytic white noise. Chirps are commonly characterised in the ambiguity plane~\cite{Ma2006}. The chirp has a starting frequency $\alpha=0.1$, with chirp rate $\beta=9.0196\cdot 10^{-4}$, and the noise has variance $\sigma_{W}^2=0.6$. We generate data sets of length $N=256$ from this process. We find the $N$-AF of the chirp 
\begin{equation}
\label{eq:sdfwe}
\widehat{\cal{A}}_{XX^*}^{(N)}(\nu,\tau]=
\exp\left\{j\pi\left[2\alpha\tau-\beta\tau^2+(\beta\tau-\nu)(N+|\tau|-1)\right]\right\}D_{N-|\tau|}(\beta\tau-\nu)
\end{equation}
for $\beta\tau=\nu$ and zero otherwise. We have not inserted $\beta\tau=\nu$ in this equation, since we are dealing with discrete values and $\beta\tau$ will not be identically equal to $\nu$ anywhere. Instead, we will for each $\tau_i$ find $\nu_i=\min|\beta\tau_i-\nu|$. Thus, the $N$-AF will be zero except for at the points $(\nu_i,\tau_i)$, where it is defined by~(\ref{eq:sdfwe}). Fig.~\ref{fig:exchirp912} (a)--(c) shows the EMAF, TEAF and LBTEAF for one realisation of the process.
The EMAF is substantially corrupted by noise, but both thresholding methods have removed most of the noise. It should be noted that the line has exhibited some increasing thickness especially towards high $|\tau|$. This will explain why the MSE results are not quite as good as might be anticipated from this figure.
To quantify the performance of the estimators, we will estimate the MSE by Monte Carlo simulation, where we compare with the $N$-AF. We generate $K=5,000$ realisations of the process. The estimated MSE is shown in Fig.~\ref{fig:exchirp912}(d)--(f). Thresholding has reduced the MSE a great deal. To observe differences between the two
thresholding procedures, note that
the TEAF suffers from estimating a single noise variance, in that the regions where $h(\nu,\tau]$ inflates the variance, noise remains in the estimation (see for example the band around the chirp in Fig.
~\ref{fig:exchirp912}(e)). The LBTEAF looks like a considerable improvement, as it removes more noise, but does suffer from difficulties in estimating the noise around the rim of the domain. This leaves a ``badly cleaned window'' effect.

To quantify our visual impression of these procedures, we look at the total MSE of each estimator, which is obtained by summing the MSE over the $\nu$-$\tau$-plane, averaged over the five thousand realisations. The resulting total MSEs and their standard deviations are shown in Table~\ref{table:tabl1}. We see that the local bias corrected thresholding yields the lowest MSE, but both thresholding methods have quite significantly reduced the MSE of the EMAF. The reduction, is as mentioned, not quite as large as might have been anticipated, as the chirp has been broadened near its true support. But the reduction in MSE is respectable corresponding to a factor of almost four. 

We estimate the total spread for the chirp immersed in analytic white noise by Monte Carlo simulation. The estimated total spread and the standard deviation of the total spread is given in Table~\ref{table:tabl2}. The EMAF will not be zero anywhere, so we do not need to estimate the total spread for this estimate, but equate this to $1$. 
Theoretically, the AF of the chirp should be nonzero for only 511 of $512\times 511$ cells, which gives us a spread of $0.002$. Our estimated spread is larger than this number, but reflects the sparsity of the TEAF. 
\begin{table}
\begin{center}
\begin{tabular}{|c||c|c|c|c|}
\hline  
& Chirp & MA & UM & TVMA\\ \hline
Total MSE of EMAF& $7.5382\cdot 10^{7}$ & $2.0337\cdot 10^{8}$ & $3.3225\cdot 10^{7}$& $5.0842\cdot 10^{7}$ \\ \hline
Total MSE of TEAF&  $1.9622\cdot 10^{7}$& $9.7961\cdot 10^{5}$& $1.7409\cdot 10^{5}$& $3.1209\cdot 10^{5}$\\ \hline
Total MSE of LTEAF& - & $9.5598\cdot 10^{5}$ & $1.7076\cdot 10^{5}$ & $2.951\cdot 10^{5}$ \\ \hline
Total MSE of LBTEAF& $1.8398\cdot 10^{7}$ & - & - & -\\ 
\hline
STD of total MSE of EMAF& $9.9790\cdot 10^{6}$ & $4.6983\cdot 10^{7}$ & $7.3221\cdot 10^{6}$&$1.3809\cdot 10^{7}$ \\ \hline
STD of total MSE of TEAF&  $4.0799\cdot 10^{6}$& $4.9168\cdot 10^{5}$& $5.1692\cdot 10^{4}$& $1.7156\cdot 10^{5}$\\ \hline
STD of total MSE of LTEAF& - & $4.1841\cdot 10^{5}$ & $4.6430\cdot 10^{4}$ & $1.1522\cdot 10^{5}$\\ \hline
STD of total MSE of LBTEAF& $4.0407\cdot 10^{6}$ & - & - & -\\ \hline
\end{tabular}
\end{center}
\caption{Average total MSE and the standard deviation (STD) of total MSE.}
\label{table:tabl1}
\end{table}
\begin{table}
\begin{center}
\begin{tabular}{|c||c|c|c|c|}
\hline
& Chirp & MA & UM& TVMA \\ \hline
Total spread of TEAF&  $0.013 $& $0.0013$& $6.5513\cdot 10^{-4}$&$0.0012$\\ \hline
Total spread of LTEAF& - & $0.001$ & $6.3735\cdot 10^{-4}$ & $0.001$\\ \hline
Total spread of LBTEAF& $0.0156 $ & - & -&- \\ \hline
STD of total spread of TEAF&  $0.0028$& $0.0016$& $8.9984\cdot 10^{-4}$& $0.0018$\\ \hline
STD of total spread of LTEAF & - & $0.0012$ & $8.1374\cdot 10^{-4}$& $0.0014$ \\ \hline
STD of total spread of LBTEAF& $0.0039$ & - & - & -\\ \hline
\end{tabular}
\end{center}
\caption{Average total spread and STD of total spread.}
\label{table:tabl2}
\end{table}
 \begin{figure}
    \begin{center}
    \subfigure[]{\includegraphics[scale=0.27]{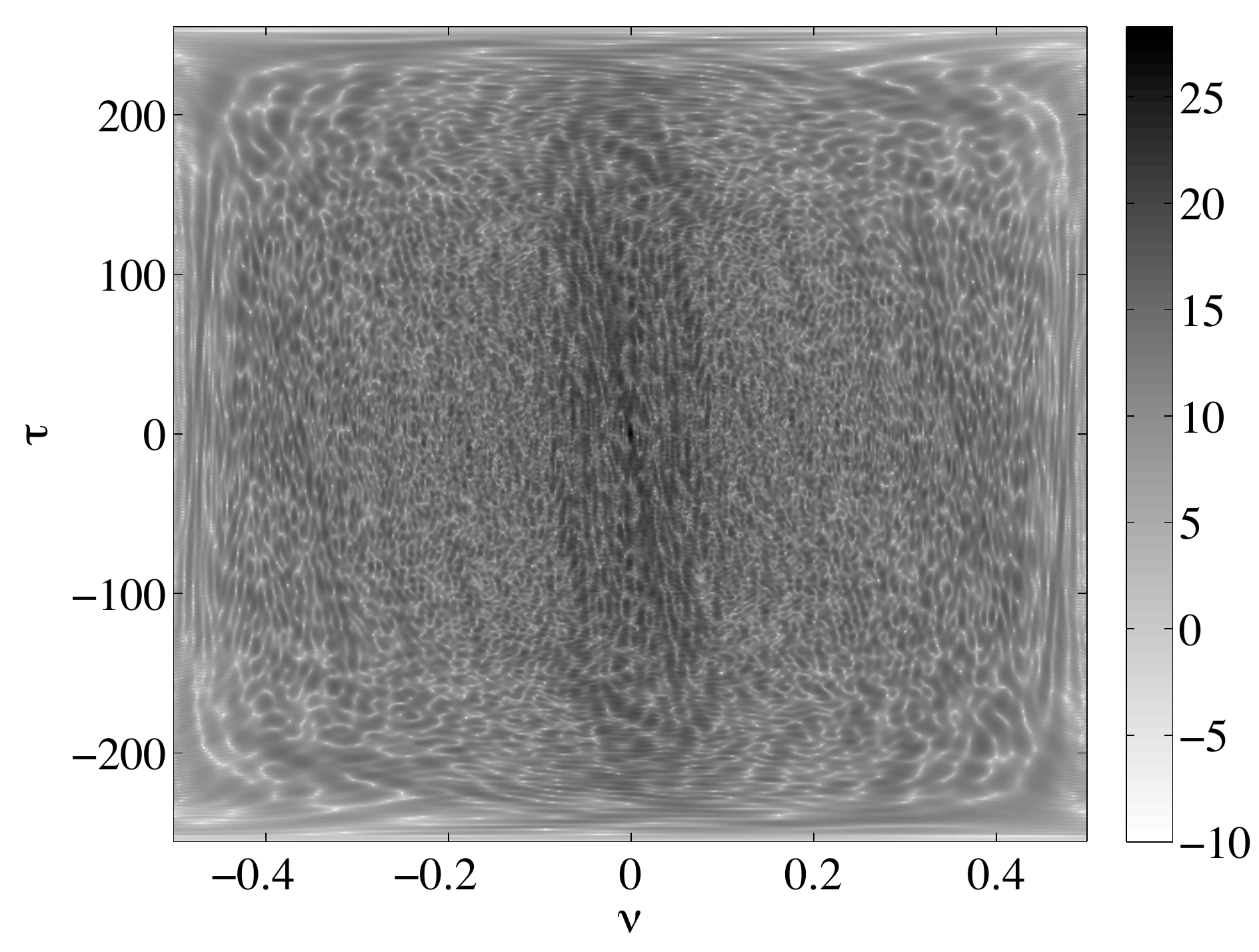}}
    \hspace{0.5cm}
     \subfigure[]{\includegraphics[scale=0.27]{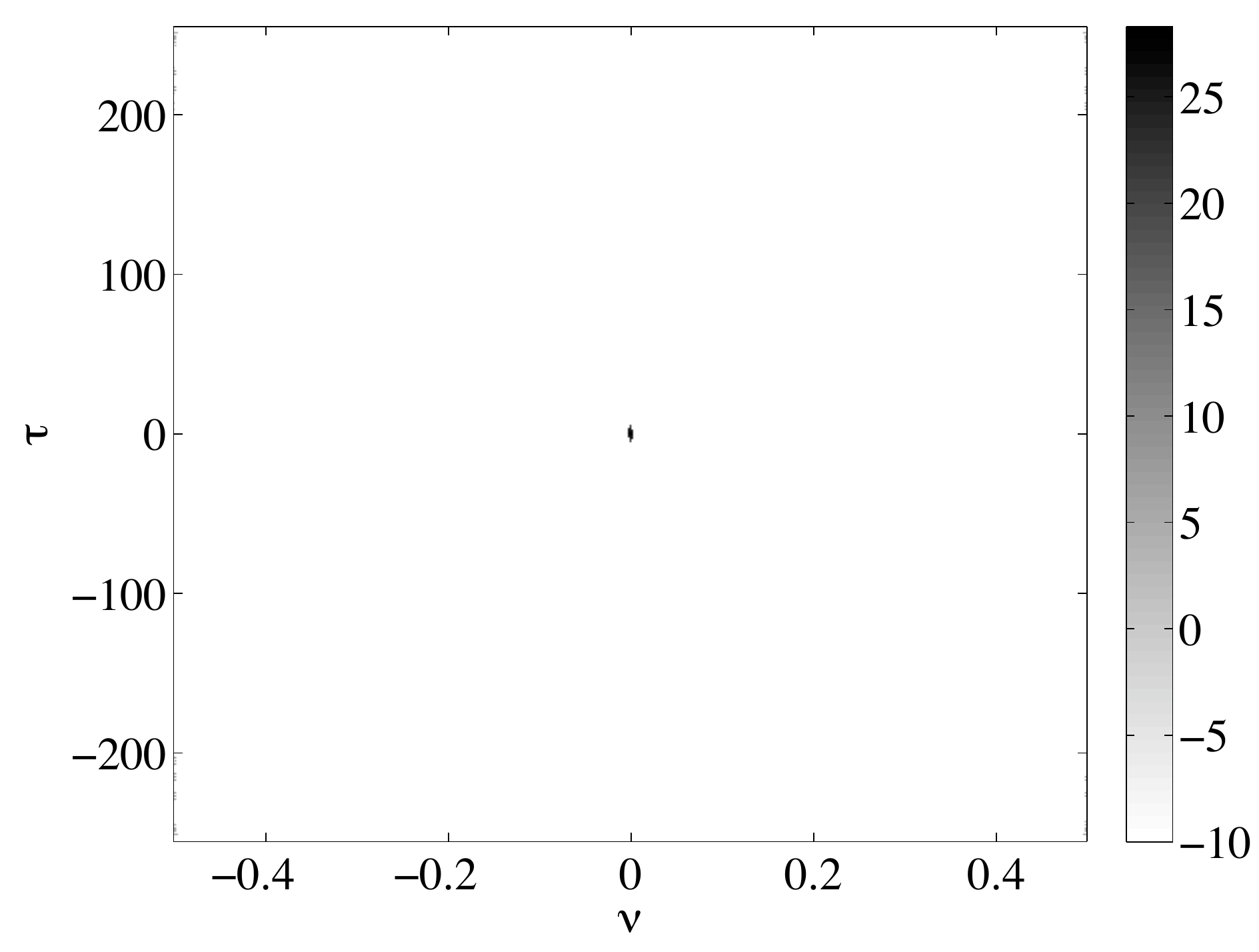}}\\
      \subfigure[]{\includegraphics[scale=0.27]{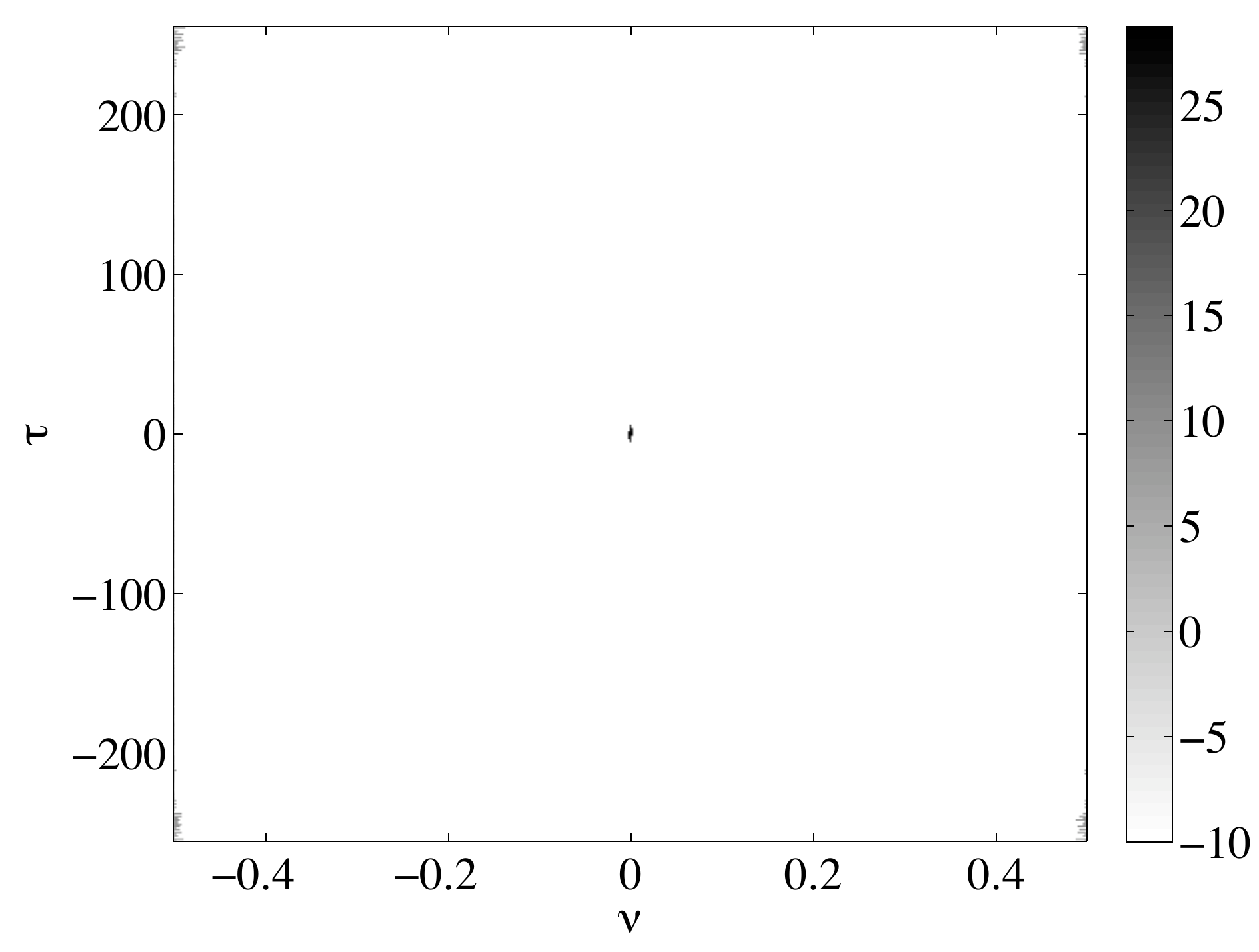}}
          \hspace{0.5cm}
           \subfigure[]{\includegraphics[scale=0.27]{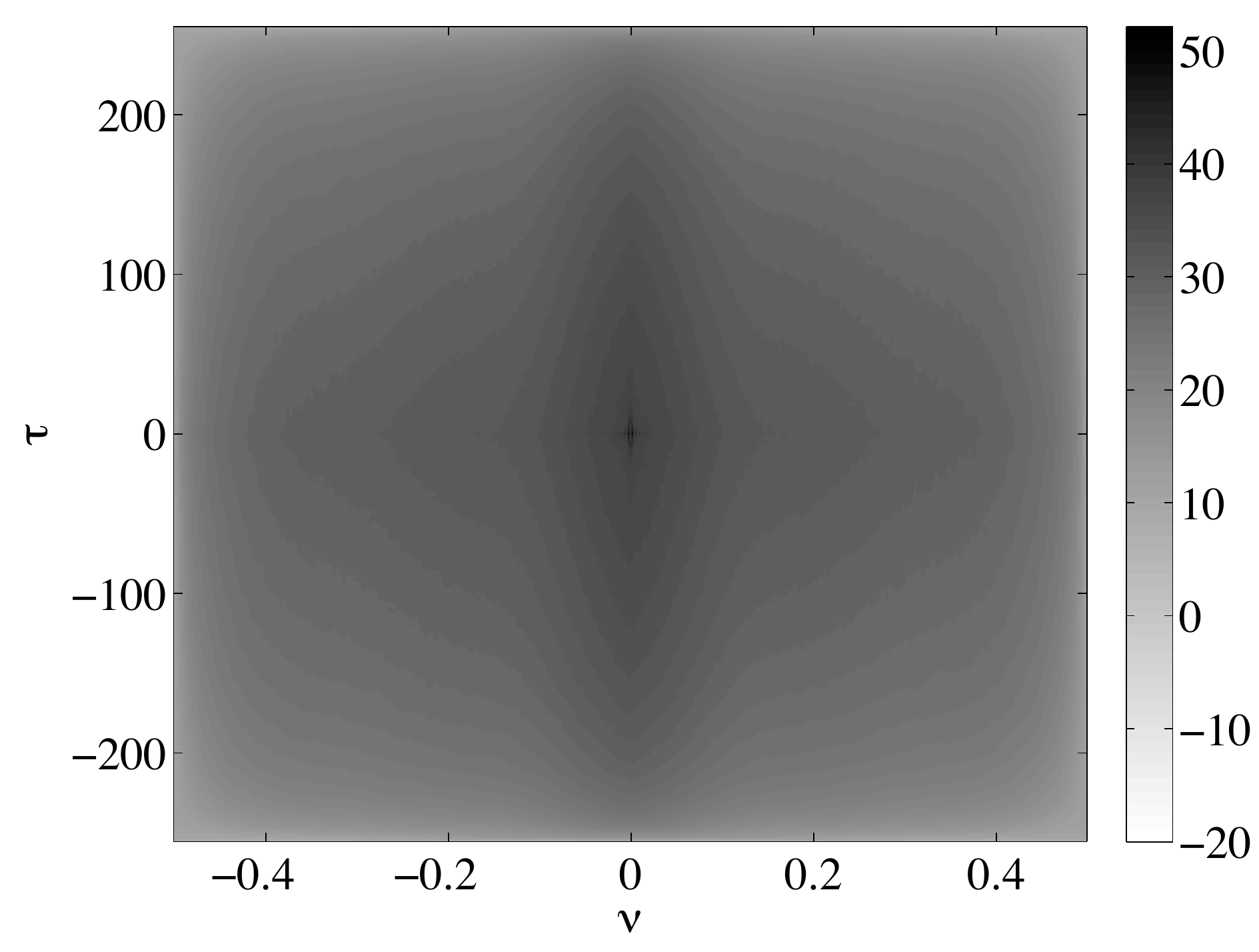}}\\
     \subfigure[]{\includegraphics[scale=0.27]{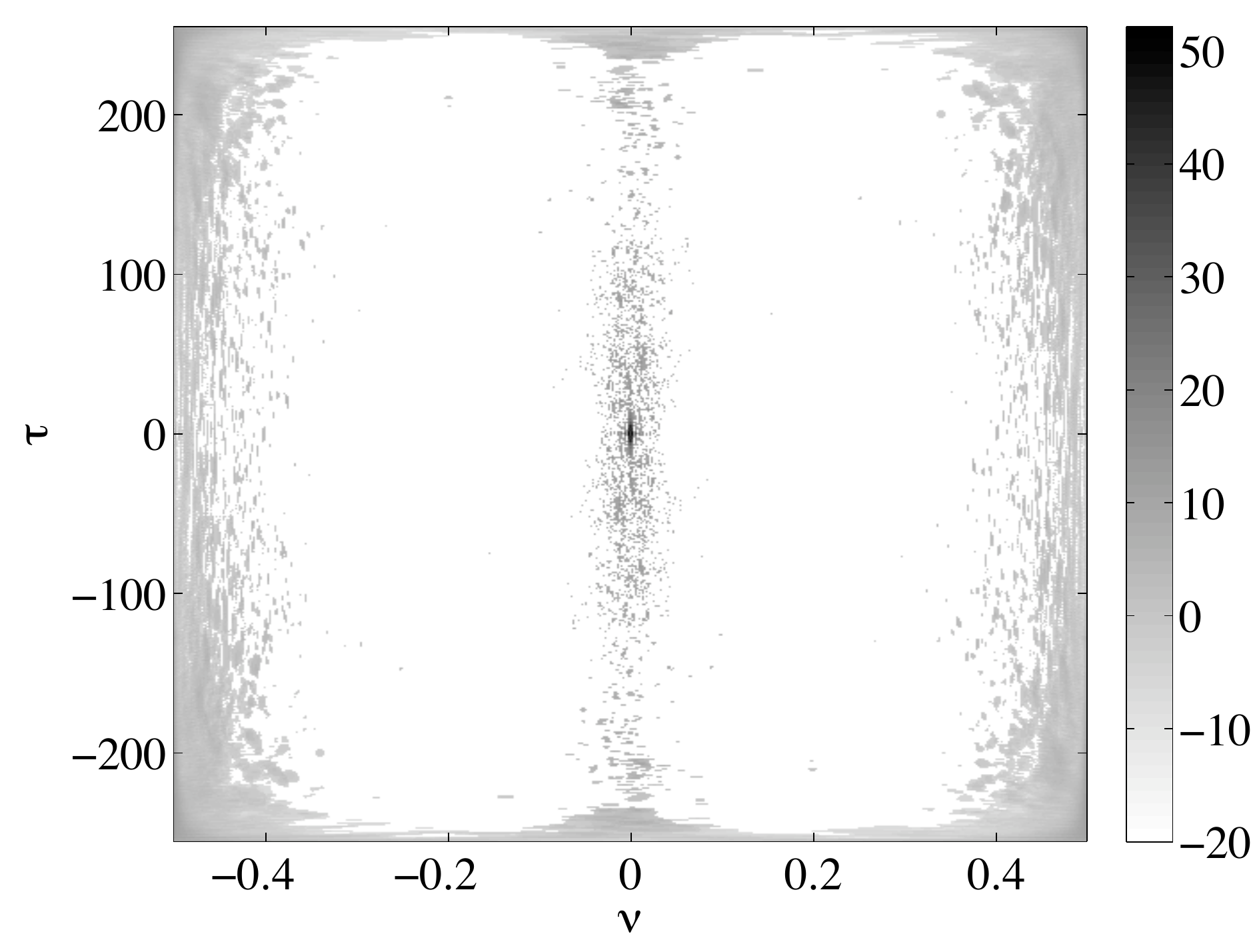}}
               \hspace{0.5cm}
     \subfigure[]{\includegraphics[scale=0.27]{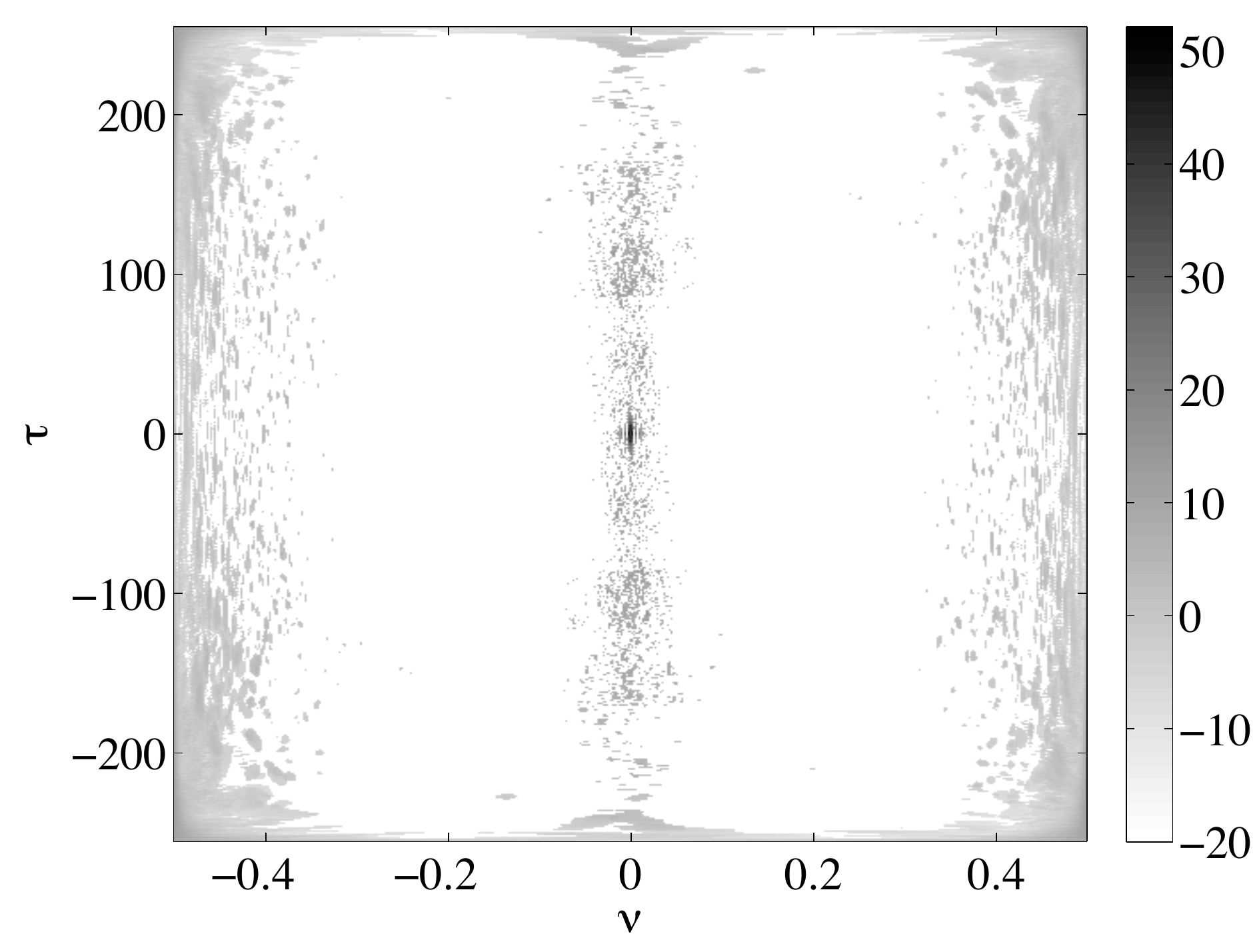}}
     \end{center}
     \caption{MA process. (a) EMAF (b) TEAF (c) LTEAF for one realisation. Estimation of the MSE for (d) EMAF (e) TEAF (f) LTEAF. All are on a dB scale. }
     \label{fig:exstat78}
\end{figure}

\subsection{Stationary process}\label{subsecma}
\begin{figure}
    \begin{center}
      \includegraphics[scale=0.3]{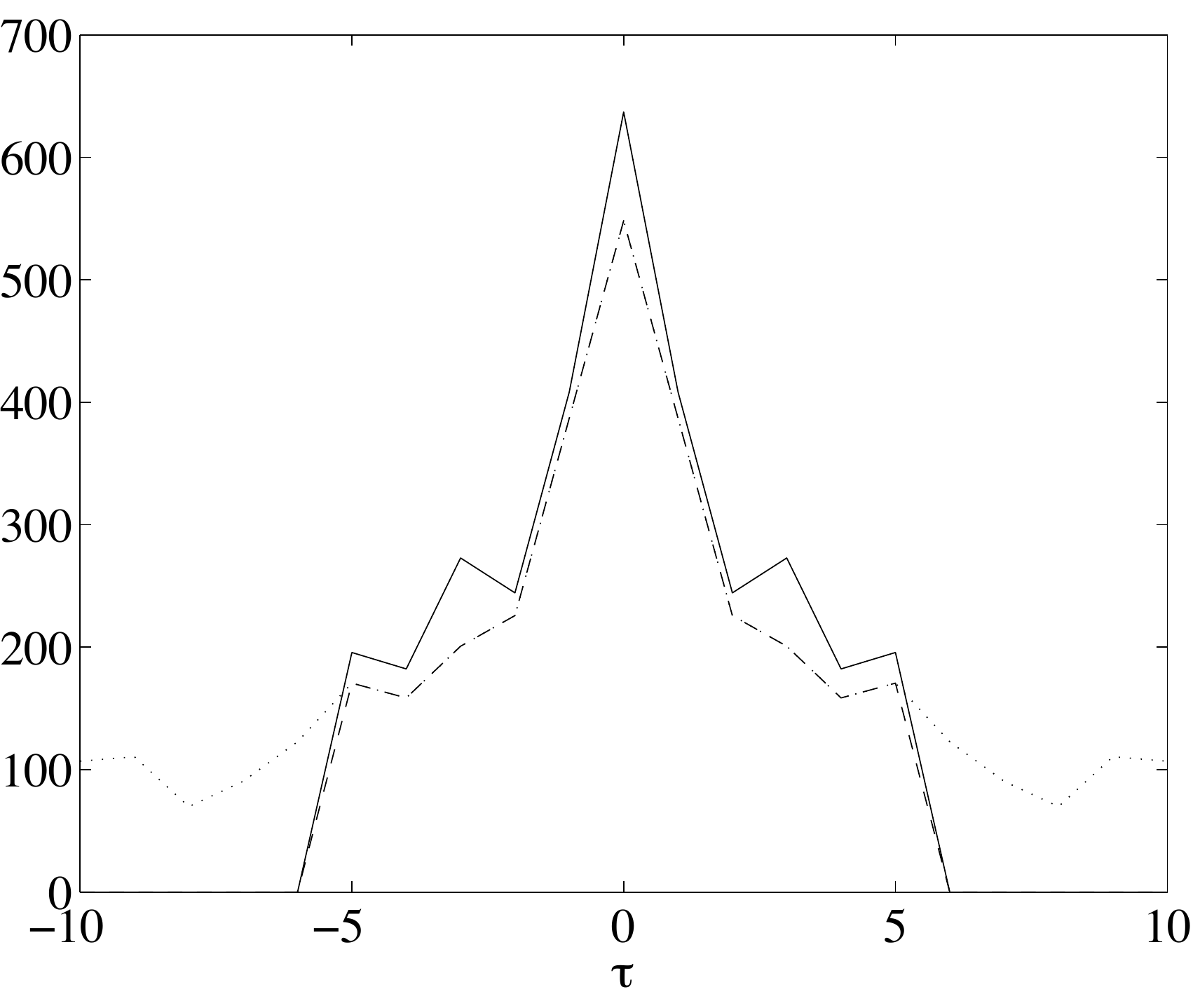}
      \hspace{0.5cm}
      \includegraphics[scale=0.3]{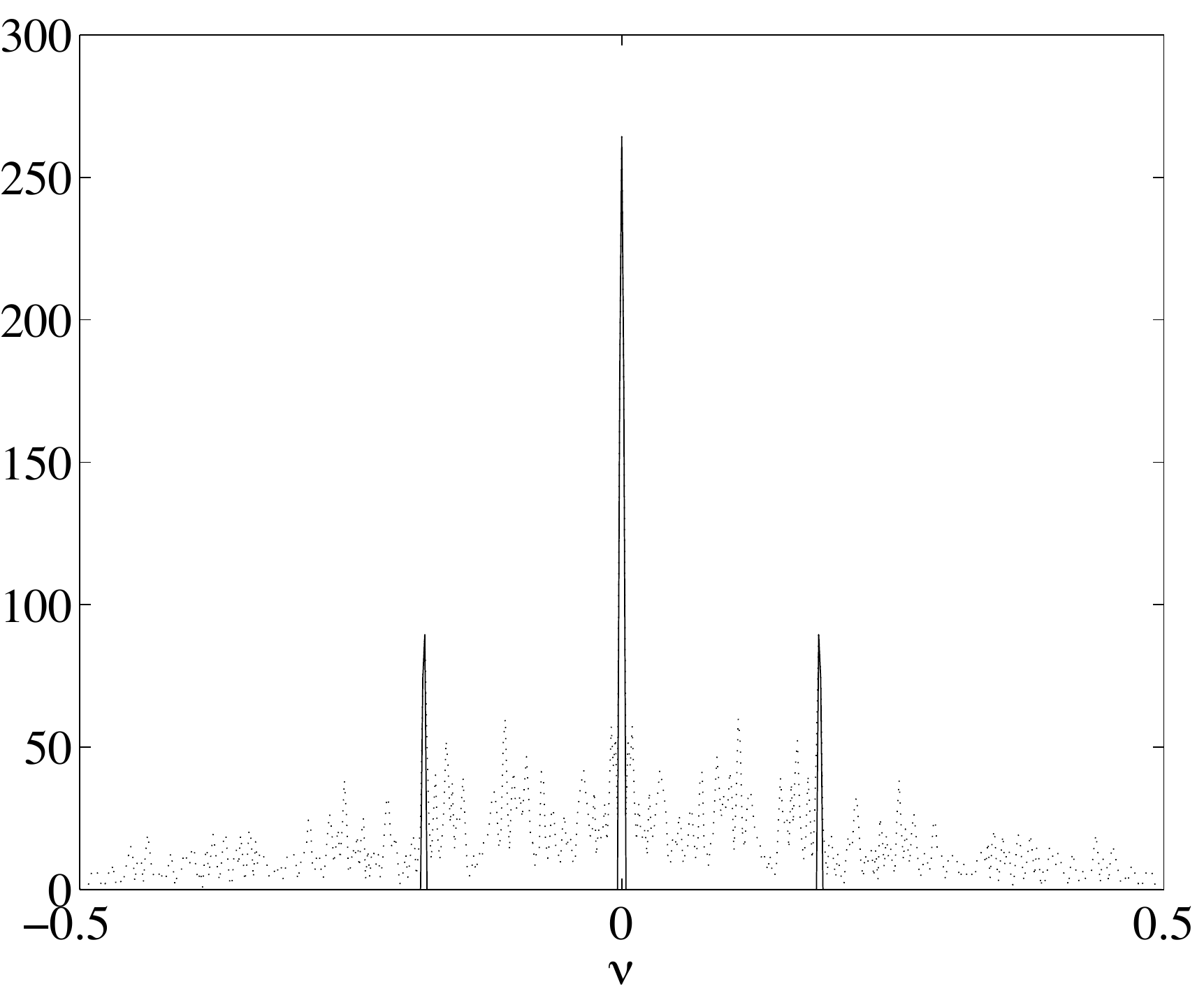}\\
            \includegraphics[scale=0.3]{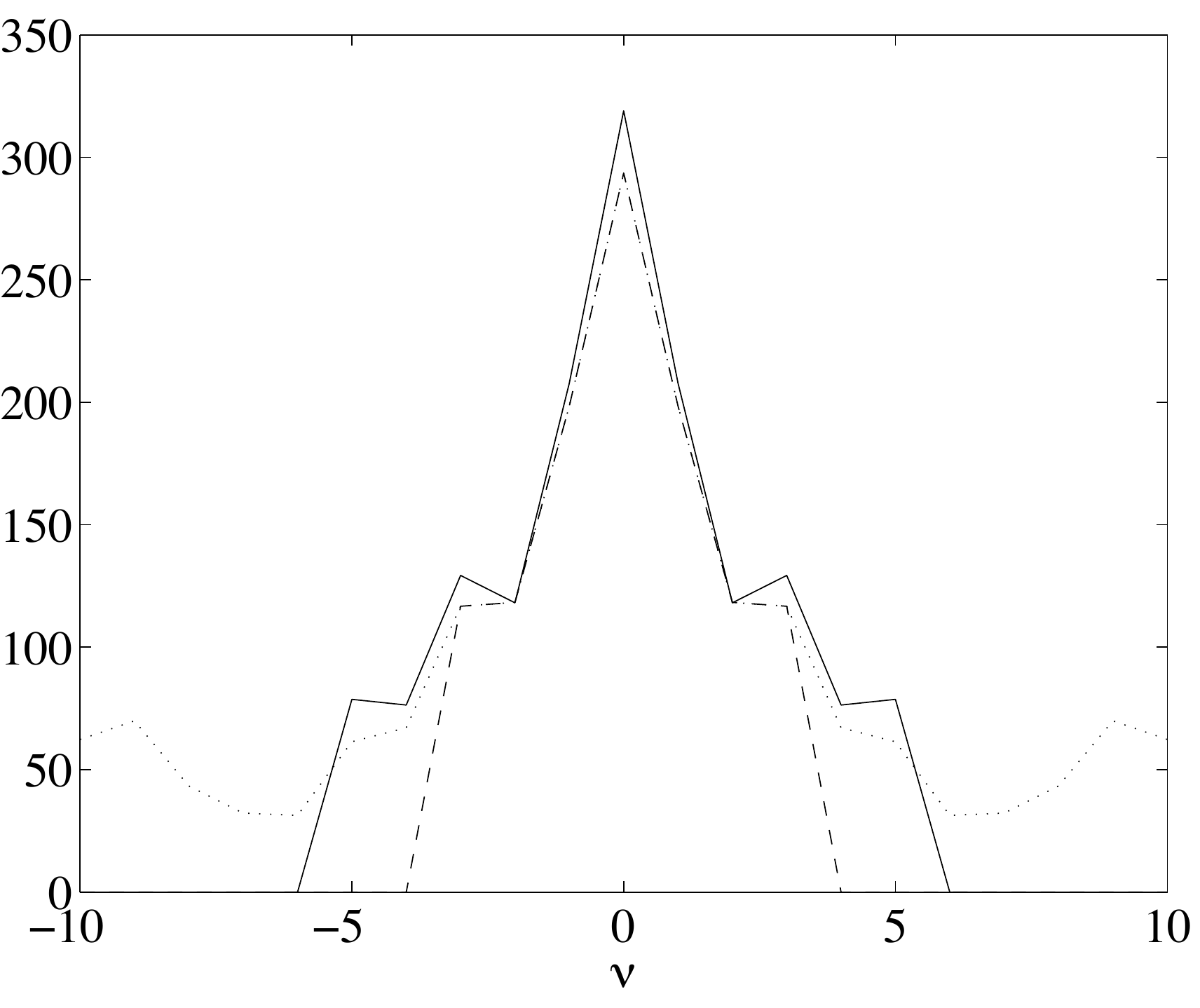}
      \hspace{0.5cm}
      \includegraphics[scale=0.3]{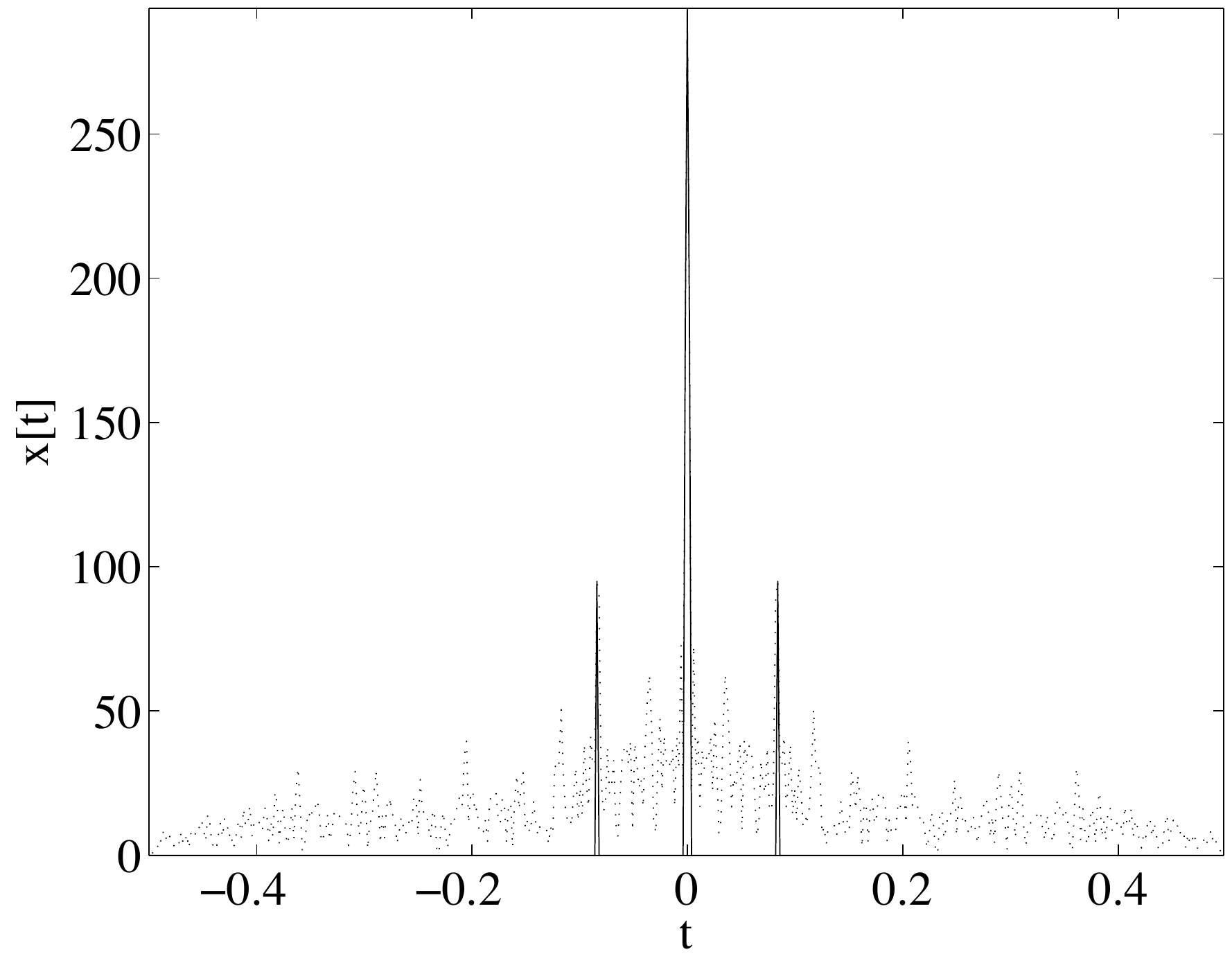}
           \end{center}
     \caption{(a) TEAF (dashed), EMAF (dotted) and $N$-AF (solid) for $\nu=0$ for one realisation of the MA process. (b) TEAF (solid) and EMAF (dotted) for one realisation of the UM process for $\tau=0$. (c) LTEAF (dashed), EMAF (dotted) and $N$-AF (solid) for $\nu=0$ for one realisation of the TVMA process. (d) LTEAF (solid) and EMAF (dotted) for one realisation of the TVMA process for $\tau=0$.
}
     \label{fig:exstat788}
\end{figure}

We consider the analytic process $X[t]$ corresponding to the real-valued stationary Moving Average (MA) process $R[t]=\sum_{i=0}^{L}w_i \xi[t-i]$, where $\xi\sim \mathcal{N}(0,\sigma_{\xi}^2)$ and $\E\left\{\xi[t]\xi[t-\tau]\right\}=\sigma^2_{\xi}\delta[\tau]$. The autocorrelation of this process is
\begin{equation}
\label{statauto}
\widetilde{M}_{RR^*}[\tau]=\E\left\{R[t]R[t-\tau]\right\}=\sigma^2_{\xi}\sum_{i=0}^{L-|\tau|} w_i w_{i+|\tau|} \qquad \text{for}\quad |\tau|\leq L.
\end{equation}
We generate realisations of the process with ${\bf{w}}=\begin{bmatrix} 1& 0.33& 0.266& 0.2& 0.133& 0.066\end{bmatrix}^T$, length $N=256$, $L=5$,  and $\sigma^2_{\xi}=1$. Fig.~\ref{fig:exstat78}(a)--(c) show the EMAF of one realisation of the process and the result of thresholding the  EMAF. Both thresholding procedures have zeroed out most of the points for $\nu\neq0$, and correspond to substantial improvements to the EMAF. Fig.~\ref{fig:exstat788}(a) shows the EMAF and the thresholded EMAF for one realisation with the $N$-AF for $\nu=0$ and $\tau\in[-10,10]$. We see that the thresholded EMAF and the $N$-AF are very similar, and only when the variance of the EMAF become comparable to the modulus square of the AF is the process thresholded. This exactly corresponds to the support of the real-valued process' autocorrelation. 

The results of the MSE estimation for these estimators are shown in Fig.~\ref{fig:exstat78}(d)--(f). Again, in most regions of the ambiguity plane we have a satisfying small value of the MSE, but at the ends, due to the low value of the threshold, suffer from the previously noted effects. Thresholding has reduced the MSE, and we see from Table~\ref{table:tabl1} that the MSE is actually reduced by more than a factor of 100, which is a substantial reduction. Note also that the LTEAF has lower MSE than the TEAF. We estimate the total spread of this process, see Table \ref{table:tabl2}. The theoretical spread is in this case equal to $11$ cells out of $512\times 511$ which is $ 4.2044\times 10^{-5}$. Our estimated spread is larger, due to the analytic signal spreading energy in $\tau$ and the points at the ends that have not been successfully thresholded, but still reflects the geometry of the
support. Using the discrete analytic signal and problems with edge effects
makes the inflated spread inevitable. Whilst propositions~\ref{edet}--\ref{eunif} give the decay of the variance of the EMAF, at the rim of the ambiguity plane such expressions are not accurate as the O(1) error term will be of comparable magnitude.
\subsection{Uniformly modulated process}
\begin{figure}
    \begin{center}
    \subfigure[]{\includegraphics[scale=0.27]{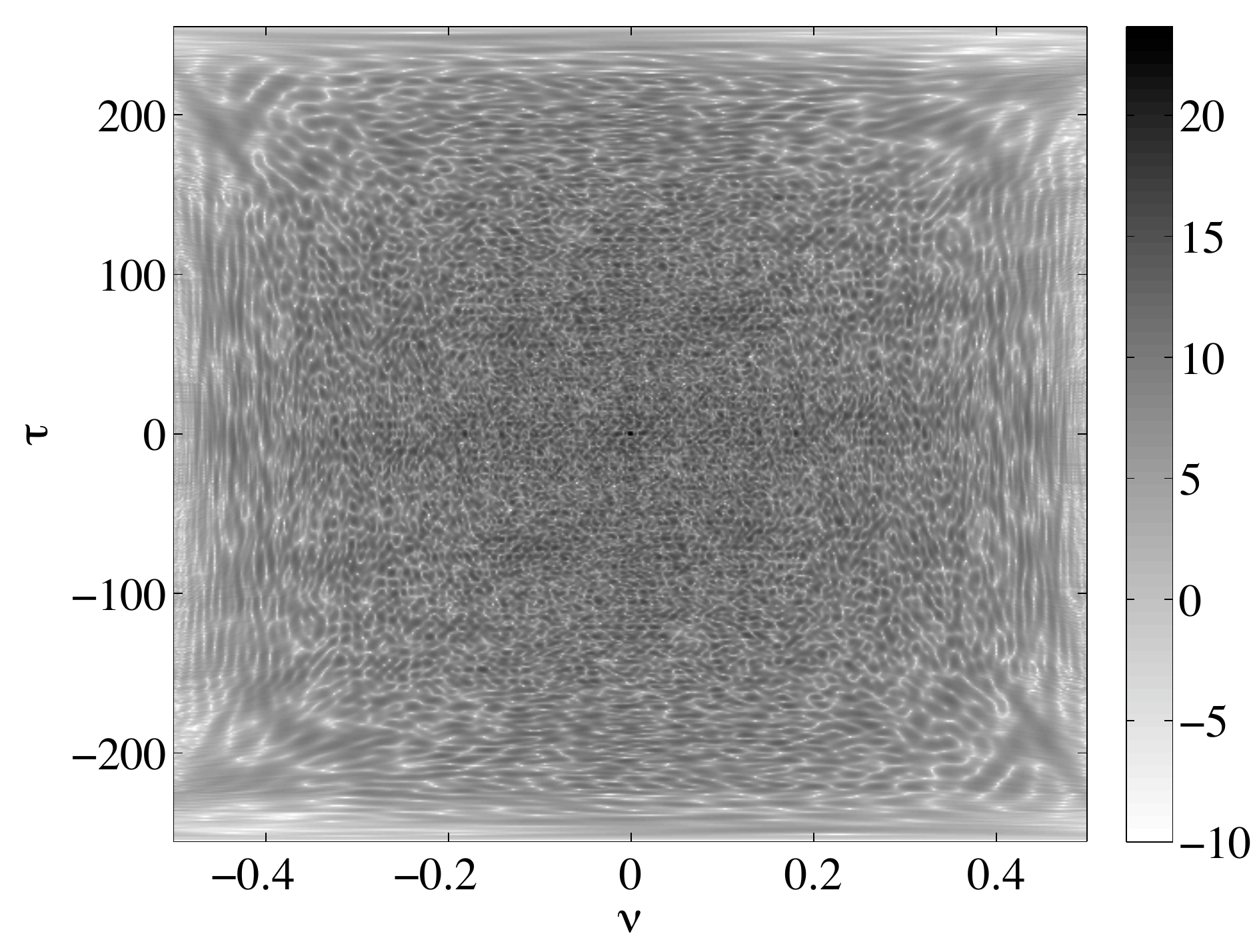}}
    \hspace{0.5cm}
     \subfigure[]{\includegraphics[scale=0.27]{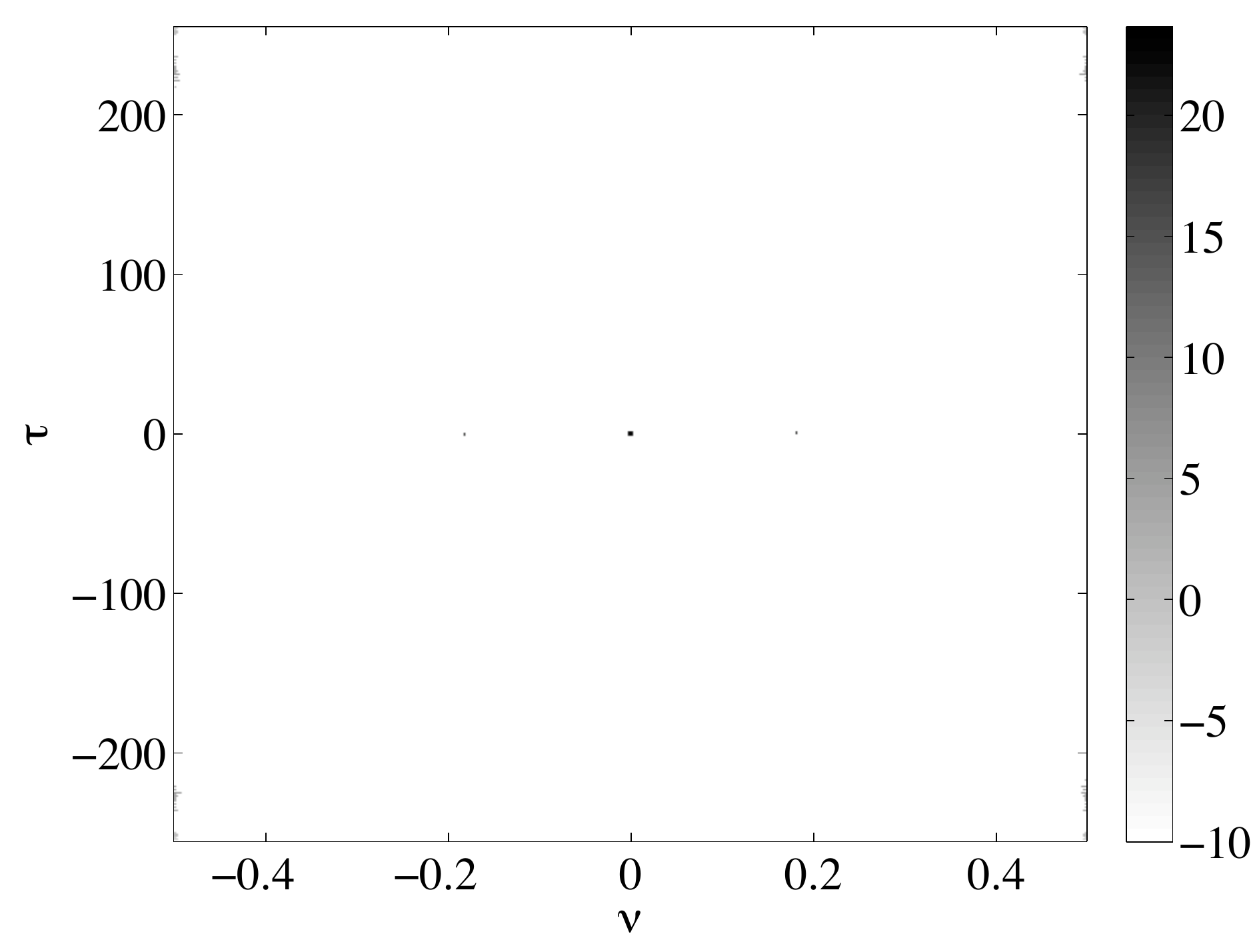}}\\
      \subfigure[]{\includegraphics[scale=0.27]{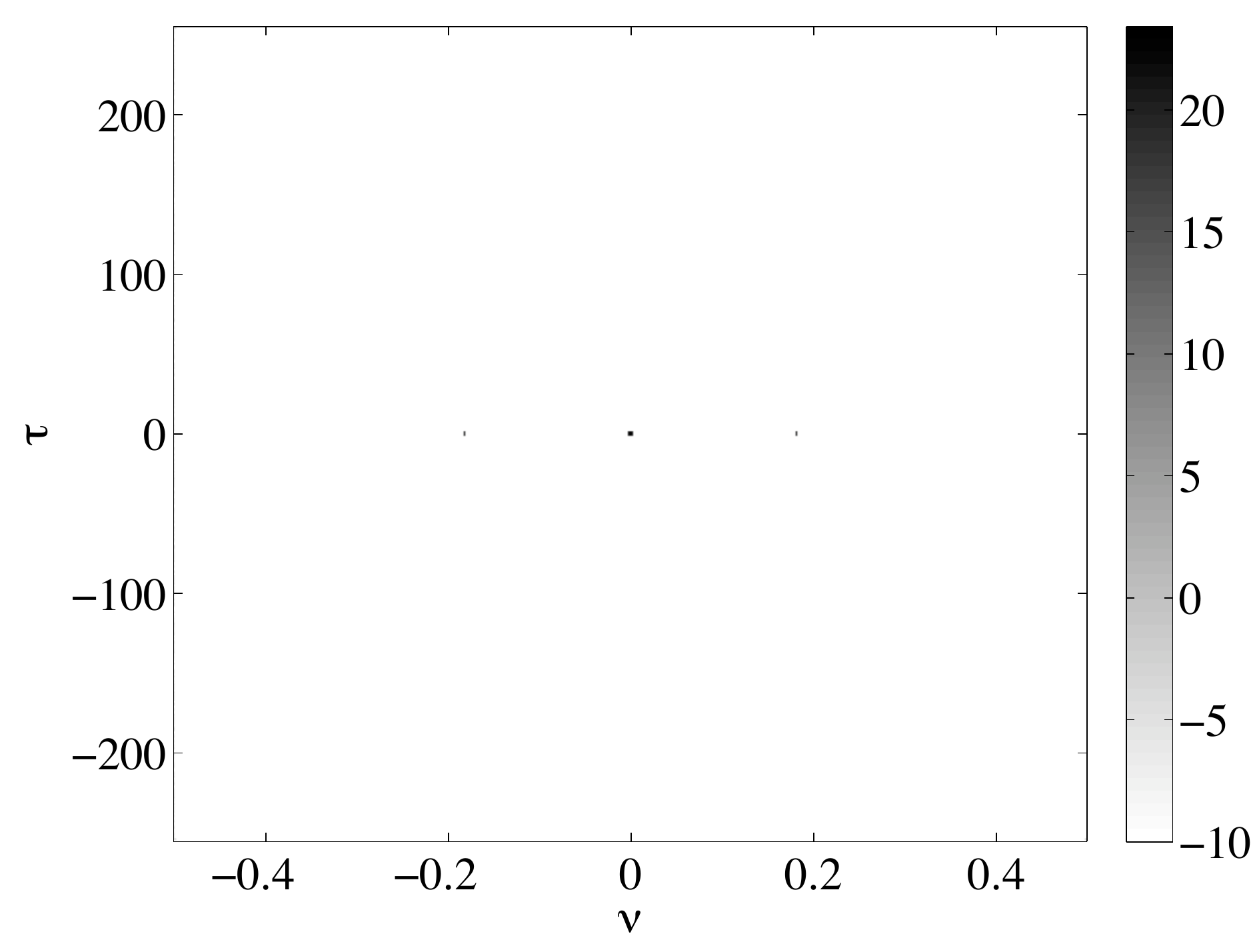}}
          \hspace{0.5cm}
           \subfigure[]{\includegraphics[scale=0.27]{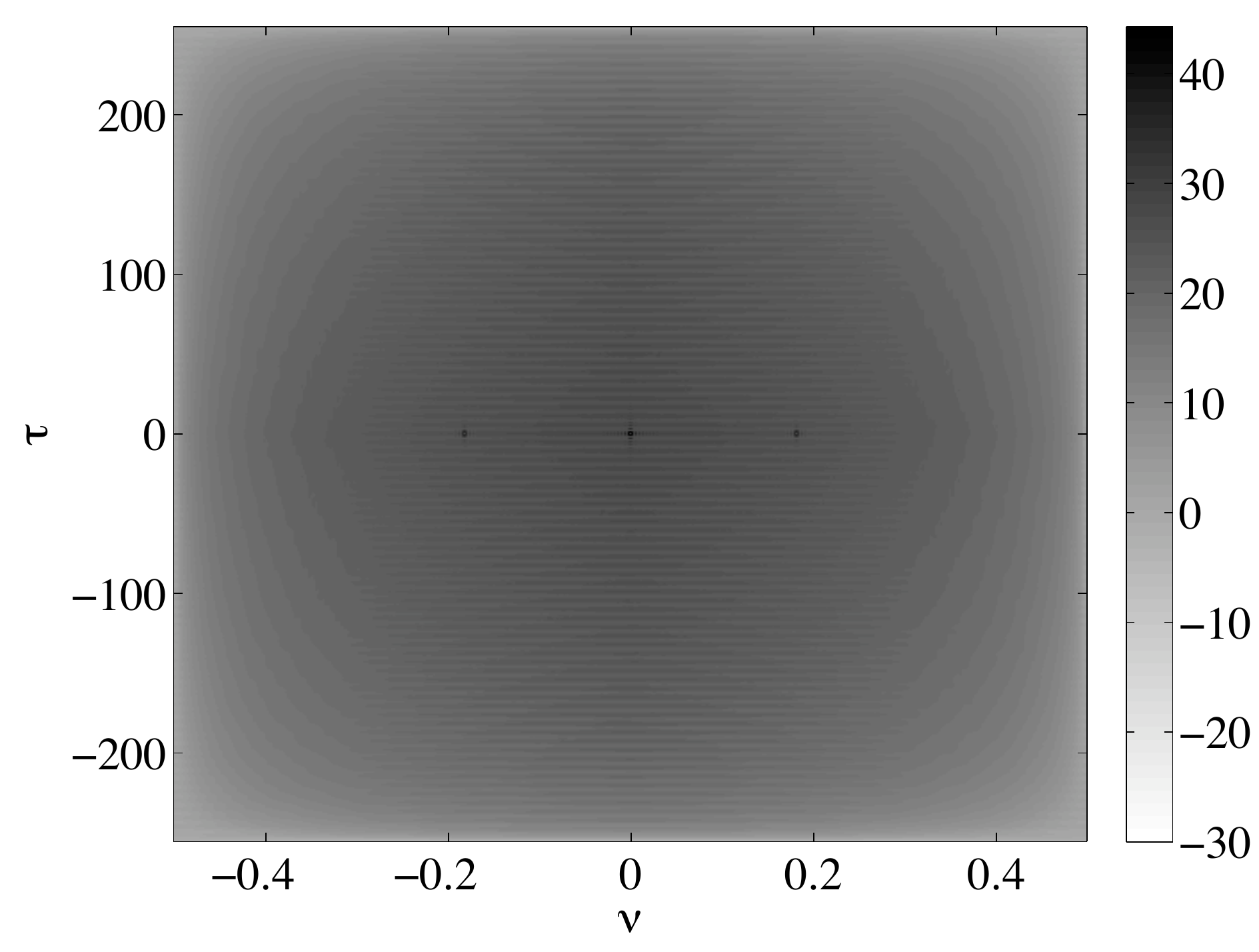}}\\
     \subfigure[]{\includegraphics[scale=0.27]{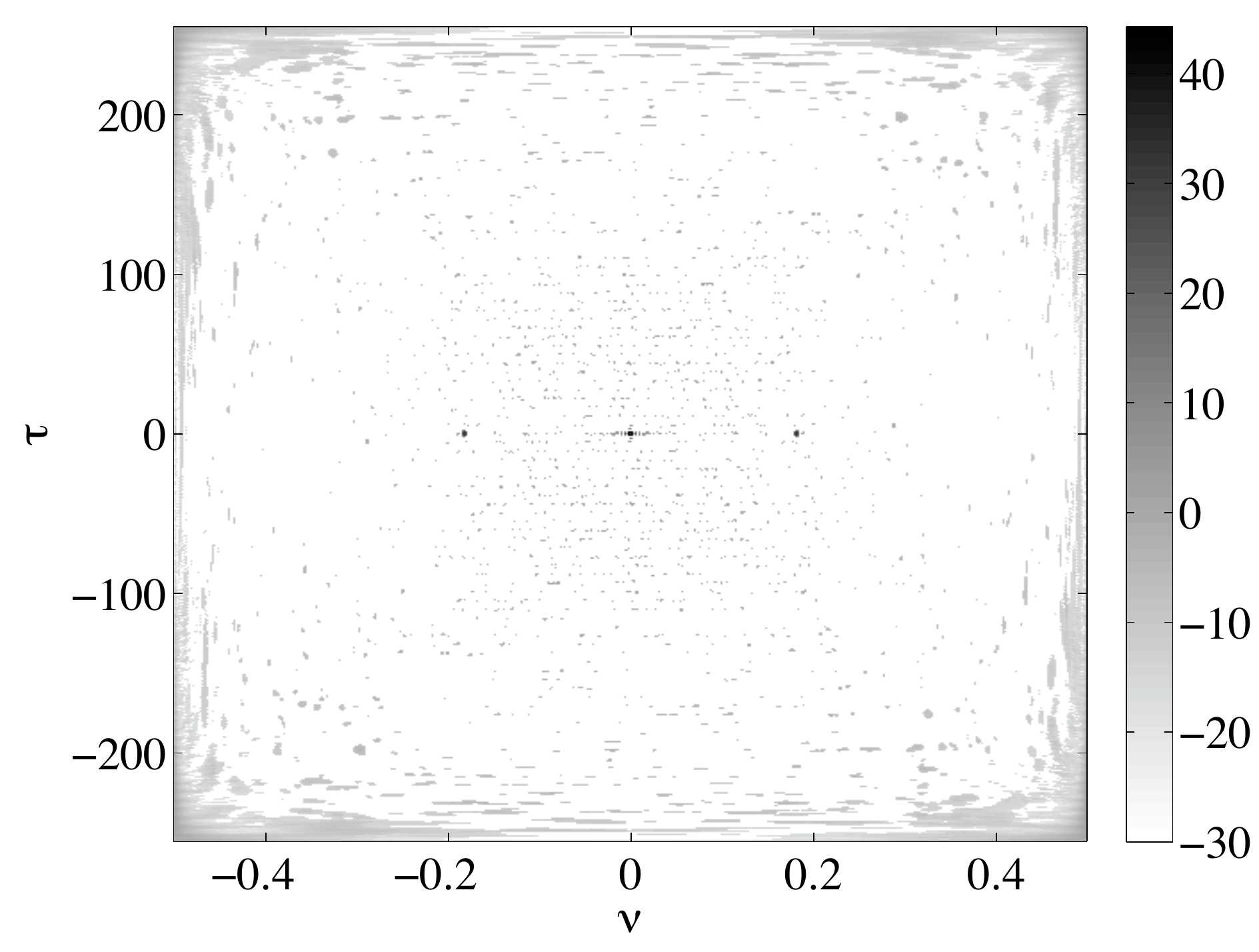}}
         \hspace{0.5cm}
     \subfigure[]{\includegraphics[scale=0.27]{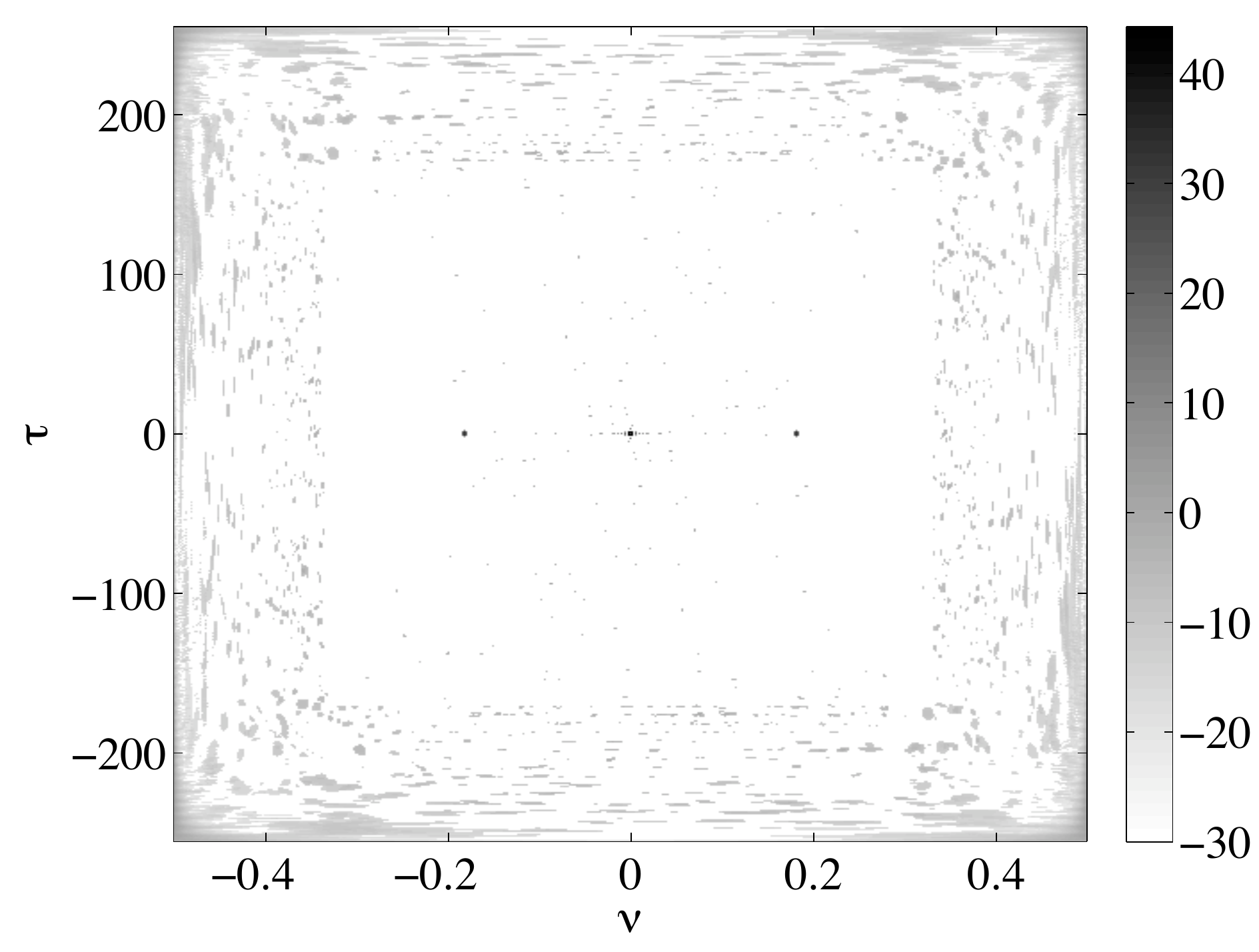}}
          \hspace{0.5cm}
           \end{center}
     \caption{UM process. (a) EMAF (b) TEAF (c) LTEAF for one realisation. Estimation of the MSE for (d) EMAF (e) TEAF (f) LTEAF. All are on a dB scale.}
     \label{fig:exunimod78}
\end{figure}

Next, we consider the analytic process $X[t]$ corresponding to the real-valued, non-stationary, Uniformly Modulated (UM) process,
$R[t]=\sigma_X[t]\xi[t]$, where $\xi\sim \mathcal{N}(0,1)$ and $\E\left\{\xi[t]\xi[t-\tau]\right\}=\delta[\tau]$. The time-varying variance is defined as $\sigma^2_X[t]=\sin^2(2\pi f_0t)$, which has an FT of $\Sigma_{XX^*}(\nu)=\frac{1}{4}\left(2\delta(\nu)-\delta(\nu-2f_0)-\delta(\nu+2f_0)\right).
$ The $N$-AF is
$
{\cal A}_{X X^*}^{(N)}(\nu,\tau]=
N $ for $\nu=0$ and $\tau=0$, ${\cal A}_{X X^*}^{(N)}(\nu,\tau]=-N(1/2-|2f_0|)$ for $\nu=\pm 2f_0, \tau=0$, and zero otherwise.
We generate a realisation of the process of length $N=256$ with $f_0=0.09$. The EMAF and the results of the two thresholding procedures based on this realisation are shown in~\ref{fig:exunimod78}(a) -- (c). The EMAF is very noisy, but both the TEAF and the LTEAF are substantially cleaner. In Fig.~\ref{fig:exstat788}(b) we see the EMAF and TEAF for $\tau=0$, and we observe that the thresholding has kept only the three points specified by the $N$-AF. Estimates of the MSE of these methods are shown in Fig.~\ref{fig:exunimod78}(d)--(f). Again we see that the thresholding has greatly reduced the MSE.  From Table~\ref{table:tabl1} we see that also for this process the MSE is reduced with a factor over one hundred from the EMAF to the TEAF and LTEAF, with the LTEAF doing a bit better than the TEAF.  The theoretical spread of this process is $1.1466\times 10^{-5}$, and our estimated spread is given in Table~\ref{table:tabl2}. 
\subsection{Time-varying MA}
\begin{figure}
    \begin{center}
    \subfigure[]{\includegraphics[scale=0.27]{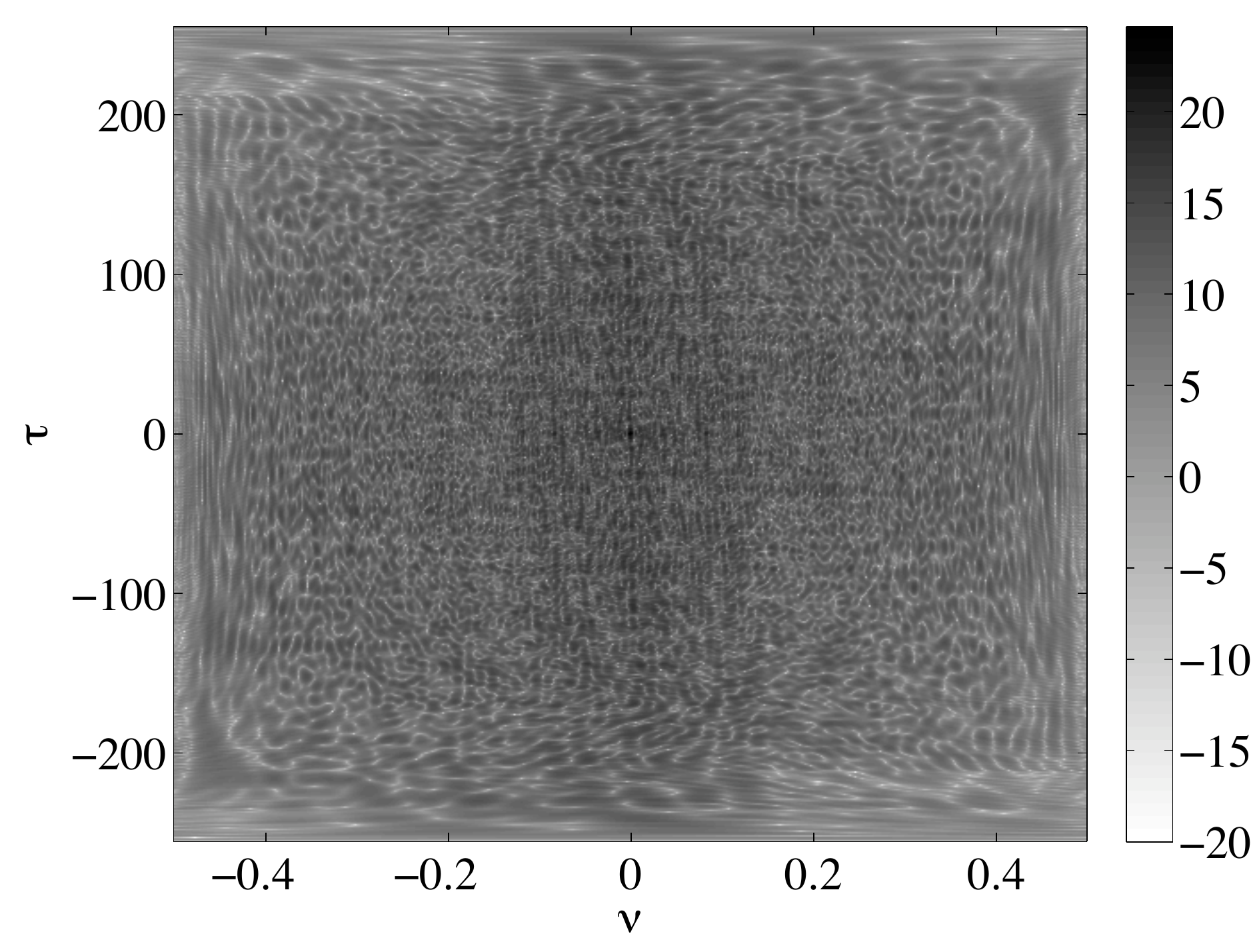}}
    \hspace{0.5cm}
     \subfigure[]{\includegraphics[scale=0.27]{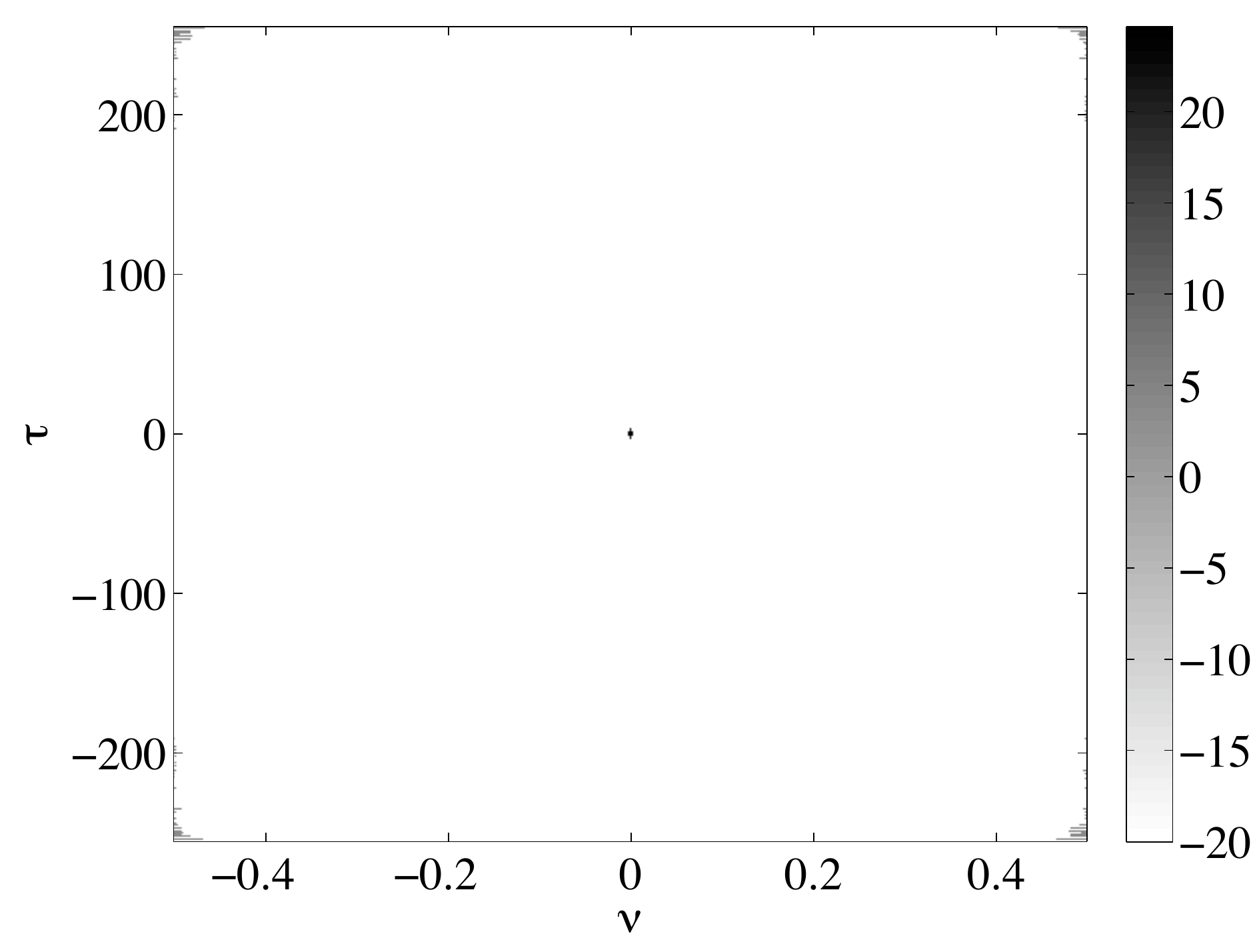}}\\
      \subfigure[]{\includegraphics[scale=0.27]{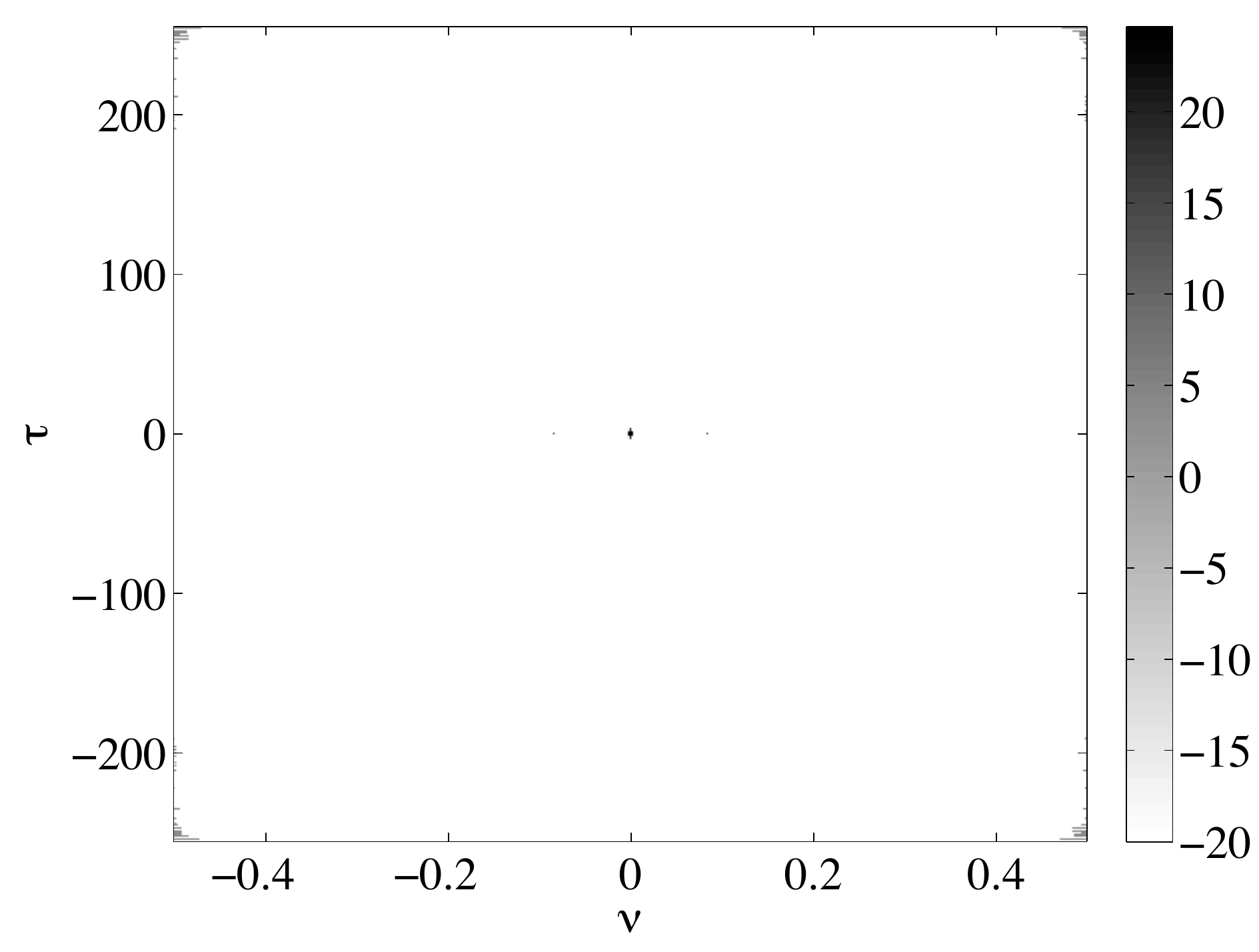}}
          \hspace{0.5cm}
           \subfigure[]{\includegraphics[scale=0.27]{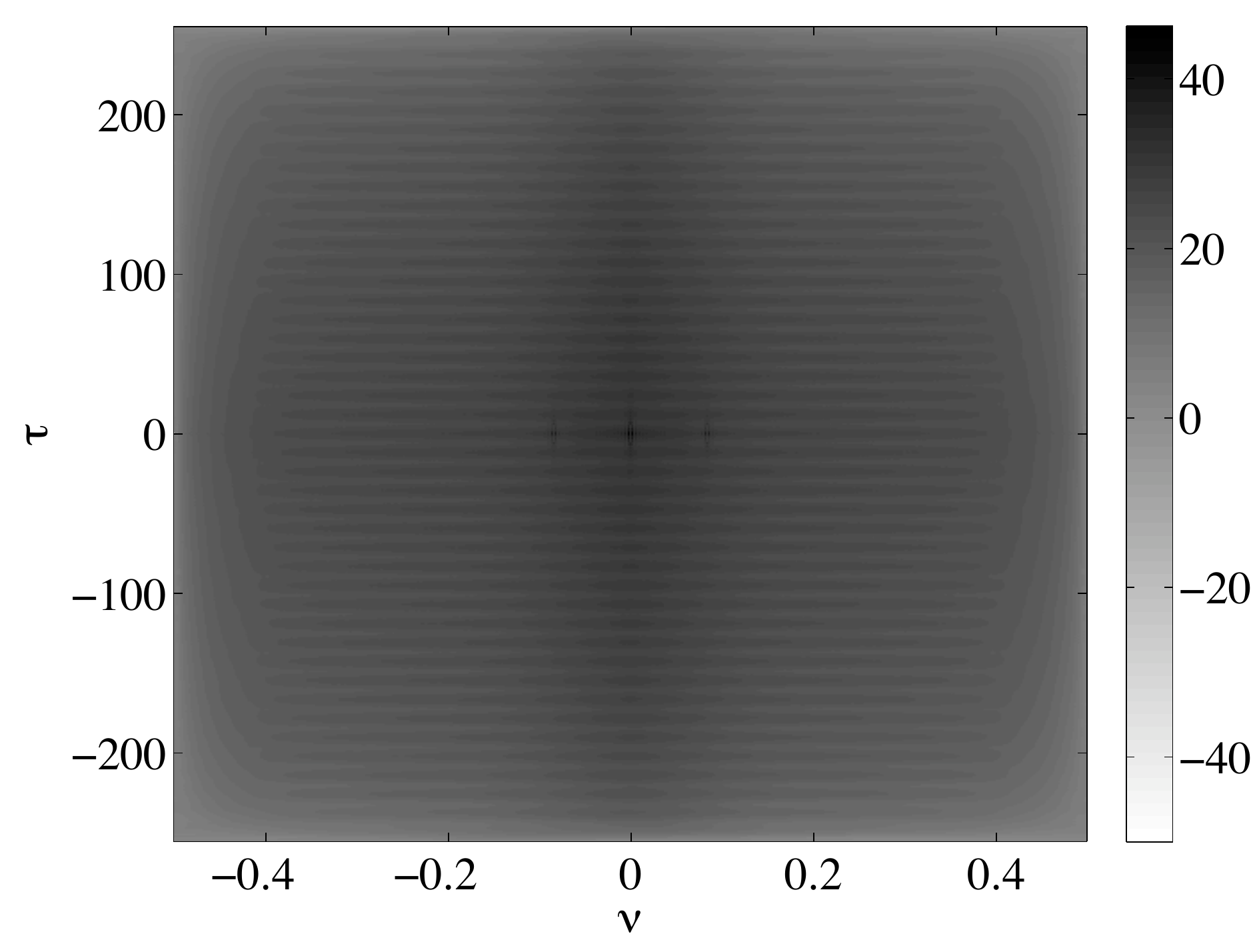}}\\
     \subfigure[]{\includegraphics[scale=0.27]{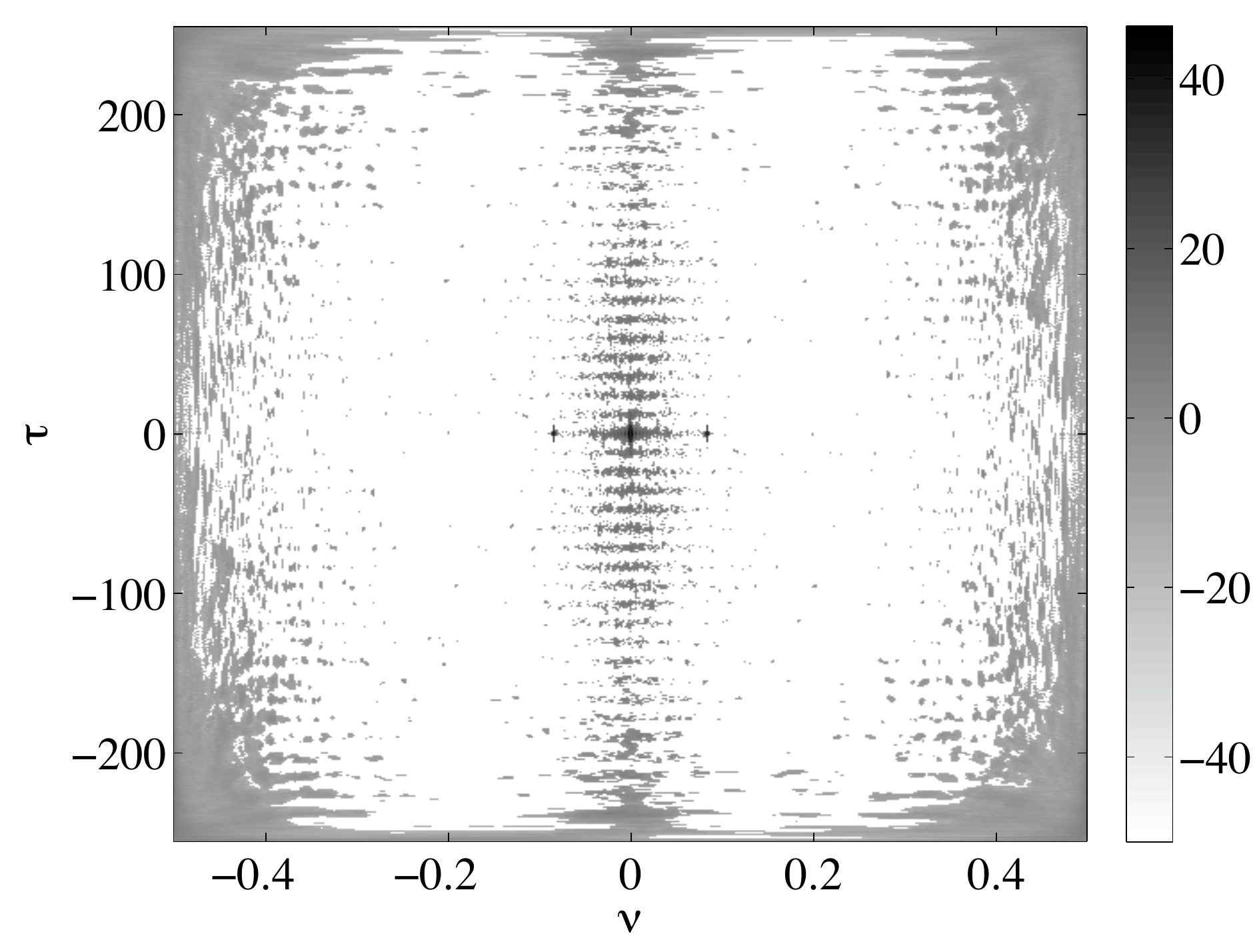}}
               \hspace{0.5cm}
     \subfigure[]{\includegraphics[scale=0.27]{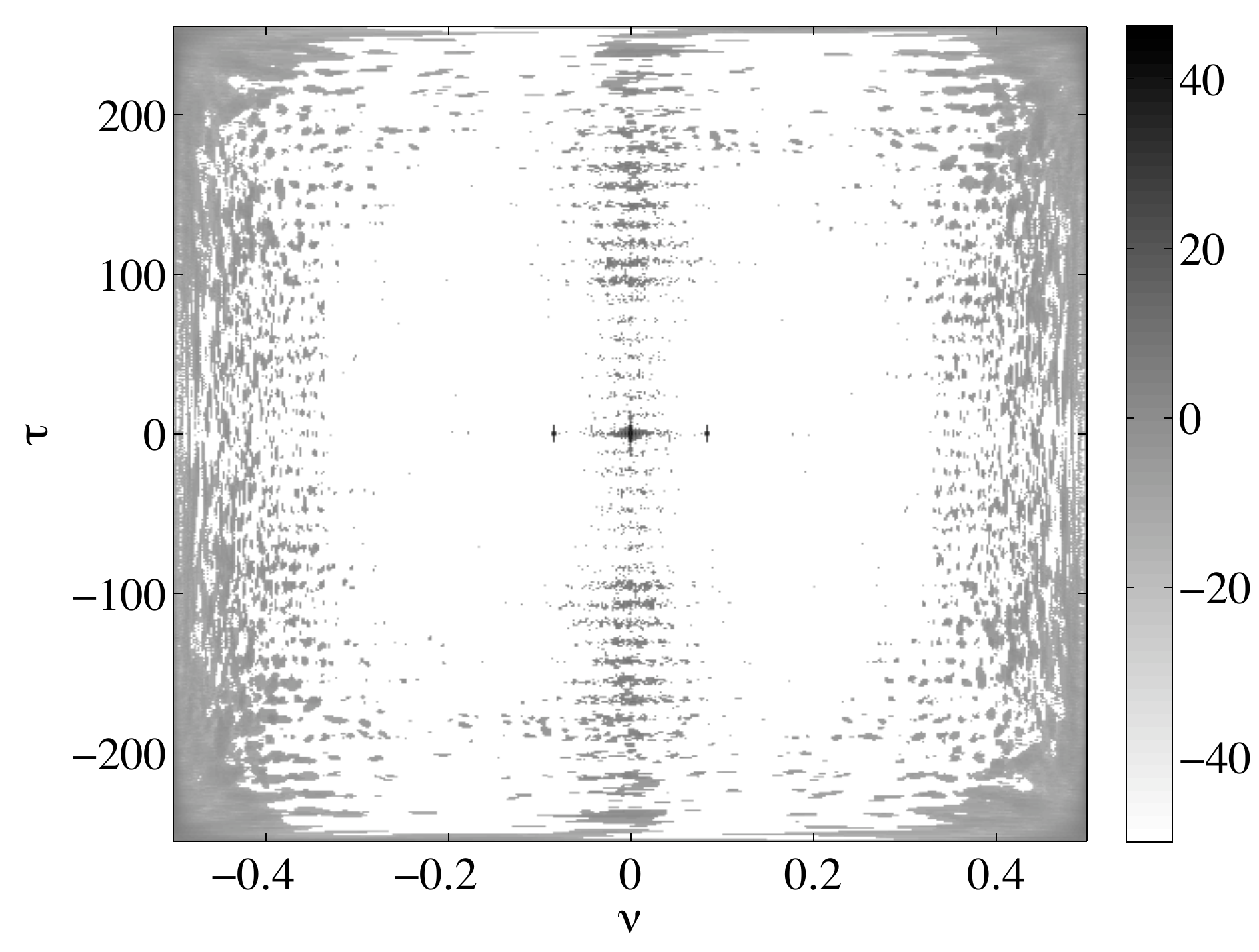}}
     \end{center}
     \caption{TVMA process. (a) EMAF (b) TEAF (c) LTEAF for one realisation. Estimation of the MSE for (d) EMAF (e) TEAF (f) LTEAF. All are on a dB scale. }
     \label{fig:extvma}
\end{figure}

Finally, we will combine the MA with the uniformly  modulated process, thus defining a real-valued Time-Varying MA (TVMA) as
$R[t]=\sigma_X[t]\sum_{i=0}^{L}w_i \xi[t-i]$, where the $w$'s and $\xi[t]$ is as defined in Section~\ref{subsecma}. We use $\sigma_X[t]=\sin(2\pi f_0 t)$ with $f_0=0.042$, and we generate samples of length $N=256$. As always, we work with the analytic process corresponding to the real-valued process. We find the $N$-AF of this process as
\begin{equation}
\widehat{\cal{A}}_{XX^*}^{(N)}(\nu,\tau]=
\begin{cases}
(N-|\tau|)\int_{0}^{1/2}\left[\widetilde{S}(f-f_0)+\widetilde{S}(f+f_0)\right]e^{j2\pi f\tau} df \qquad &\text{for }\nu=0,|\tau|\leq L\\
-(N-|\tau|)\int_{0}^{1/2-2f_0}\widetilde{S}(f+f_0)e^{j2\pi f\tau} df   \qquad &\text{for }\nu=2f_0,|\tau|\leq L\\
-(N-|\tau|)\int_{2f_0}^{1/2}\widetilde{S}(f-f_0)e^{j2\pi f\tau} df   \qquad &\text{for }\nu=-2f_0,|\tau|\leq L\\
\end{cases}
\end{equation}
where $\widetilde{S}(f)$ is the FT of the autocorrelation in~(\ref{statauto}).
We show the EMAF, TEAF and LTEAF of one realisation in Fig.~\ref{fig:extvma}(a)--(c), and we see that the thresholding has worked well, retaining only
a few points. Fig.~\ref{fig:exstat788}(c) shows the $N$-AF, the EMAF and the LTEAF for $\nu=0$. Likewise,  Fig.~\ref{fig:exstat788}(d) shows the EMAF and LTEAF at $\tau=0$.  These two figures demonstrate that the geometry of
the non-stationary process is retained by the thresholding estimator, even
if the spread of the EMAF is cut short when the variance of the EMAF becomes too large.
The estimated MSEs are shown in~\ref{fig:extvma}(d) -- (f), and the total MSE in Table~\ref{table:tabl1}. 
The MSE for the LTEAF shows a distinct improvement, especially near the region $(0,0)$.
The thresholding methods have again reduced the MSE with a factor over one hundred, and the total estimated spread demonstrates that the AF of this process is extremely sparse.
\section{Conclusions \label{conclusion}}
In this paper we have introduced a new class of estimators for the AF of a
non-stationary process that exhibits sparsity in the ambiguity plane. The
AF is a fundamental characterisation of a non-stationary process, and many
important properties of a process can be deduced from its AF. The inherent
resolution of the process in global time and frequency is an example of
such.
The AF has been used in estimating the generating mechanism of a
non-stationary process~\cite{Jachan07}. It is also a popular tool in radar
and
sonar~\cite{Blahut1991ed}. Despite this fact little efforts have been
focused on the estimation of the AF, especially to produce estimators
amenable to the determination of spread
(size of the support). Characterisation
of limited support is vital in designing the best analysis tools for
generic second order non-stationary processes, a problem that remains open
\cite{Donoho98}.

Based on the assumption of compression of the AF (small
support), we proposed different threshold procedures for estimating the
AF. The advantage of using the threshold estimator is that the size of the
magnitude of the AF is compared with the magnitude of the AF at other
time-frequency cells. Only if the local magnitude is sufficiently large is
the value not thresholded.
This should enforce a strict support from for example a process whose AF
is only extended underspread,
see \cite{Matz06}[p.~1072]. Unlike Matz and Hlawatsch \cite{Matz06}  we do
not calculate the spread of the process by fitting the support of the AF
into a box centered at $(0,0)$, but simply count the number of non-zero AF
coefficients. Conceptually this can be thought of as determining a finite
number of cells that would represent most of the structure of the observed
process. If a finite number of cells represent
a full process then estimation of its generating mechanism is possible.
Hence whilst it clearly is not necessary to constrain the process to be
underspread, i.e. be concentrated in support around the origin in the AF
domain \cite{Pfander}, it is necessary to constrain the possible degree of
dependency in the process to ensure that the generating mechanism of the
process could be estimated.

Pivotal to thresholding procedures is determining a suitable threshold. We
implemented the thresholding with a global variance estimate as well as a
local variance estimate, both adjusted for each point in the ambiguity
plane. For deterministic signals in analytic white noise we also proposed a
procedure which removes bias in the estimator caused by the noise. We
demonstrated the superior properties of the threshold estimator in finite
sample sizes for a variety of common types of non-stationary and
stationary processes. The reductions in MSE compared to a previously used,
if naive, estimator for the AF were considerable, up to factors of over a
hundred. Formally our proposed threshold procedure for zero-mean processes
is only valid if the process is underspread in which case we determine the
asymptotic distribution of the EMAF. We conjecture that asymptotic
normality can be shown for a larger class of processes, and that in the
case of any process with compressed AF, thresholding is an appropriate
estimation procedure.
The estimated spread
does not fully reflect the degree of sparsity of the AF: an inflation is
accrued due to the usage of the discrete analytic signal and edge effects.
Resolving such effects fully remains an outstanding issue. The definition
of spread in this paper has the clear advantage of interpretation,  corresponding to the fraction of
points where the mean of the EMAF dominates the variance of the EMAF. 

The AF remains the least studied of time-frequency representations of
non-stationary signals, perhaps because its arguments lack direct global
interpretability. A large class of processes that can be estimated exhibit
compression in this domain. We anticipate that further study of the AF of
non-stationary processes will yield understanding into what generating
mechanisms can be determined from a given sample with fixed sampling rates
and sample size.


\renewcommand{\theequation}{\thesection.\arabic{equation}}
\renewcommand{\theequation}{A-\arabic{equation}}
\setcounter{equation}{0}  
\appendices

\section{Proof of Proposition \ref{edet} \label{prop1}}
The Empirical Cross Ambiguity Function for two finite equal length samples of signals
$\left\{g_1[t]\right\}_t$ and $\left\{g_2[t]\right\}_t$ of length $N$
is given by
$\widehat{A}_{g_1g_2^*}(\nu,\tau]=\sum_{t=\max(0,\tau)}^{N-1+\min(\tau,0)} g_1[t] g_2^*[t-\tau]e^{-j2\pi\nu t}.$ Then
\begin{eqnarray}
\nonumber
\widehat{A}_{X X^*}(\nu,\tau]&=&
\sum\limits_{t=\max(0,\tau)}^{N-1+\min(0,\tau)}\widehat{M}_{XX^*}[t,\tau]  e^{-j2\pi \nu t}
=\sum\limits_{t=\max(0,\tau)}^{N-1+\min(0,\tau)}\left(g[t]+W[t]\right)\left( g^*[t-\tau] +W^*[t-\tau]\right)+ e^{-j2\pi \nu t}\nonumber\\
&=&\widehat{A}_{gg^*}(\nu,\tau]+\widehat{A}_{W g^*}(\nu,\tau]+\widehat{A}_{g W^*}(\nu,\tau]+\widehat{A}_{WW^*}(\nu,\tau].
\label{sum}
\end{eqnarray}
As $g[t]$ is deterministic, and $W[t]$ is zero-mean, it follows $\E\left\{\widehat{A}_{W g^*}(\nu,\tau]\right\}=0,$ 
and $\E\left\{\widehat{A}_{gW^*}(\nu,\tau]\right\}=0$. Using~(\ref{repofEAF}) for $\tau\ge 0$ and the fact that $S_{X X^*}(\nu,f)=\sigma_W^2\delta(\nu)$ for $f\geq 0$ and zero otherwise for the noise process, we find for $\tau > 0$
\begin{eqnarray}
\nonumber
\E\left\{\widehat{A}_{WW^*}(\nu,\tau] \right\}&=&\int_{0}^{1/2}\int_{0}^{1/2} e^{j\pi (\nu') (N-1-\tau)}\E\left\{d\widetilde{X}(\nu'+f)d\widetilde{X}^*(f)\right\}
e^{j2\pi (\nu'+f)\tau}D_{N-\tau}(\nu'-\nu)e^{-j\pi\nu(N+\tau-1)}\\
&=&\int_{0}^{1/2}\int_{0}^{1/2} e^{j\pi\nu' (N-1-\tau)}\sigma^2_W\delta(\nu') e^{j2\pi (\nu'+f)\tau}D_{N-\tau}(\nu'-\nu)¨
e^{-j\pi\nu(N+\tau-1)}\;df_1\;df_2\nonumber \\ \nonumber
&=&e^{-j\pi\nu(N+\tau-1)}D_{N-\tau}(\nu)\int_{0}^{1/2}\sigma^2_W e^{j2\pi f\tau}\;df\nonumber\\
&=&\frac{\sigma^2_W}{2}e^{-j\pi\nu(N+\tau-1)}D_{N-\tau}(\nu)e^{j\pi\tau/2}\sinc(\pi\tau/2),
\nonumber
\label{expect}
\end{eqnarray}
where (as usual) $\sinc(x)=\sin(\pi x)/\pi x$. Thus taking expectations of~(\ref{sum}) and using the linearity of $\E\left\{\cdot
\right\}$, the result follows. {\em Mutatis mutandis} the calculations are
implemented for $\tau<0$. 

We start from~(\ref{sum}) and note that $\var\left\{\widehat{A}_{X X^*}(\nu,\tau] \right\}$ can be written in terms of covariances of $A_{gg^*}(\nu,\tau]$, $A_{gW^*}(\nu,\tau]$ {\em etc.}
Here, $g[t]$ is deterministic and $W[t]$ is Gaussian proper and so
\begin{eqnarray}
\nonumber
\var\left\{\widehat{A}_{gg^*}(\nu,\tau]\right\}&=&0,\quad \cov\left\{\widehat{A}_{gg^*}(\nu,\tau],A_{Wg^*}(\nu,\tau]\right\}=0\\
\cov\left\{\widehat{A}_{gg^*}(\nu,\tau],\widehat{A}_{gW^*}(\nu,\tau]\right\}&=&0,\quad
\cov\left\{\widehat{A}_{gg}(\nu,\tau],\widehat{A}_{WW^*}(\nu,\tau]\right\}=0
\nonumber\\
\nonumber
\cov\left\{\widehat{A}_{W g^*}(\nu,\tau],A_{gW^*}(\nu,\tau]\right\}&=&0,\quad
\cov\left\{\widehat{A}_{Wg^*}(\nu,\tau],A_{WW^*}(\nu,\tau]\right\}=0\\
\cov\left\{\widehat{A}_{gW^*}(\nu,\tau],A_{WW^*}(\nu,\tau]\right\}&=&0.
\label{ae4}
\end{eqnarray}
We next define $G(f,\tau]=\sum\limits_{t=0}^{N-1-|\tau|} g[t-|\tau|I(\tau < 0)]e^{-j2\pi ft}$, where $I(\cdot)$ is the indicator function, and note that $G(f,\tau]$ is supported on negative frequencies as well even if $g[t]$ is analytic. If $N-|\tau|\rightarrow\infty$, $G(f,\tau]$ will tend toward only being supported on positive frequencies. 
Then, for $\tau>0$,
{\small\begin{eqnarray}
\nonumber
\var\left\{\widehat{A}_{Wg^*}(\nu,\tau]\right\}&=&\var\left\{\sum_{t=0}^{N-1-\tau} W[t+\tau] g^*[t]e^{-j2\pi   \nu (t+\tau)}\right\}\\
\nonumber
&=&\var\left\{\int_{0}^{1/2} d\widetilde{X}(f_1)\left( \sum_{t=0}^{N-1-\tau} g[t]e^{-j2\pi (f_1-\nu)t}\right)^*
e^{j2\pi\tau f_1}\,df_2\right\}\\
\nonumber
&=&\var\left\{\int_{0}^{1/2} d\widetilde{X}(f_1)G^*(f_1-\nu,\tau]e^{j2\pi\tau f_1}\,df_2\right\}\\
\nonumber
&=&\int_{0}^{1/2}\int_{0}^{1/2} E\left\{d\widetilde{X}(f_1)d\widetilde{X}^*(f_1)\right\} G^*(f_1-\nu,\tau]G(f_2-\nu,\tau]e^{j2\pi\tau (f_1-f_2)}\\
&=&\sigma^2_W\int_{0}^{1/2} \left| G(f-\nu,\tau]\right|^2 df
\label{grej}
\end{eqnarray}}
After a change of variables $f'=f-\nu$, we are given an outer integral of $\int_{-\nu}^{1/2-\nu}$
over $f'$, rather than $\int_0^{1/2}$.
For $\nu>0$ we can rewrite this as $\int_0^{1/2-\nu}$ plus a contribution
that will have the inner integrals integrating to negligible contributions.
For $\nu<0$ we can rewrite this as $\int_{-\nu}^{1/2}$ plus a contribution
that will have the inner integrals  integrating to negligible contributions. 
We denote $h(\nu,\tau]=\int_{\max(-\nu,0)}^{1/2-\max(0,\nu)} \left| G(f,\tau]\right|^2 df$. 
For a generic harmonizable process $X[t]$, $\tau>0$ gives
\begin{equation*}
\begin{split}
\var\left\{\widehat{A}_{XX^*}(\nu,\tau]\right\}
&=\int_{0}^{1/2}\int_{0}^{1/2}\int_{0}^{1/2}\int_{0}^{1/2}e^{j2\pi(f_1-\alpha_1)\tau}e^{j\pi (f_1-f_2-\alpha_1+\alpha_2) (N-1-\tau)}D_{N-\tau}(f_1-f_2-\nu)\\
&\,D_{N-\tau}(\alpha_1-\alpha_2-\nu)\E\left\{ d\widetilde{X}(f_1)d\widetilde{X}^*(f_2)d\widetilde{X}^*(\alpha_1)d\widetilde{X}(\alpha_2) \right\}-\left|\E\left\{\widehat{A}_{XX^*}(\nu,\tau]\right\}\right|^2.
\end{split}
\end{equation*}
This follows by direct calculation starting from~(\ref{repofEAF}).
Using Isserlis' theorem \cite{Isserlis1918} we write the variance as
{\small \begin{alignat}{1}
\nonumber
&\var\left\{\widehat{ A}_{X X^*}(\nu,\tau]\right\}
=\int_{0}^{1/2}\int_{0}^{1/2}\int_{0}^{1/2}\int_{0}^{1/2} S_{XX^*}(f_1-\alpha_1,\alpha_1) S_{X X^*}(\alpha_2-f_2,f_2)\\
&e^{j\pi (f_1-f_2-\alpha_1+\alpha_2) (N-1-\tau)}e^{j2\pi(f_1-\alpha_1)\tau}D_{N-\tau}(f_1-f_2-\nu)D_{N-\tau}(\alpha_1-\alpha_2-\nu) df_1 df_2 d\alpha_1 d\alpha_2,
\label{varharm}
\end{alignat}} 
which for analytic white noise reduces to
\begin{alignat}{1}
\nonumber
\var\left\{\widehat{ A}_{X X^*}(\nu,\tau]\right\}
&=\sigma^4_W\int_{0}^{1/2}\int_{0}^{1/2} D^2_{N-\tau}(f_1-f_2-\nu)df_1 df_2\\
\nonumber&=\sigma^4_W(N-\tau)\int_{\max(0,-\nu)}^{1/2+\min(0,-\nu)}\int_{-(1/2-g_2)(N-\tau)}^{g_2(N-\tau)} \sinc^2\left(\xi\right)d\xi dg_2\\
&=\sigma^4_W(N-\tau)(1/2-|\nu|)+O\left(1\right).
\nonumber\end{alignat}
Implementing the same calculations for $\tau<0$ derives analogous expressions with $\tau$ replaced by $|\tau|$.
Putting~(\ref{ae4}), (\ref{grej}) and $\var\left\{A_{gW^*}(\nu,\tau]\right\}=\var\left\{A_{Wg^*}(-\nu,-\tau]\right\}$ together, and {\em mutatis mutandis} implementing the calculations for $\tau<0$, thus yields
\begin{eqnarray*}
\var\left\{\widehat{A}_{X X^*}(\nu,\tau] \right\}&=&0+\sigma^2_{W}h(\nu,\tau]+\sigma^2_{W}h(-\nu,-\tau]
+\sigma^4_{W}(N-|\tau|)(1/2-|\nu|)+O\left(1\right).
\end{eqnarray*}

We start from~(\ref{sum}) and note that $\rel\left\{\widehat{A}_{X X^*}(\nu,\tau] \right\}$ can be written as an aggregation of relations.
As $g[t]$ is deterministic and $W[t]$ is zero-mean Gaussian,
\begin{eqnarray}
\label{eq:3er}
\rel\left\{\widehat{A}_{gg^*}(\nu,\tau]\right\}&=&0,\quad \rel\left\{\widehat{A}_{gg^*}(\nu,\tau],A_{Wg^*}(\nu,\tau]\right\}=0\\
\rel\left\{\widehat{A}_{gg^*}(\nu,\tau],\widehat{A}_{gW^*}(\nu,\tau]\right\}&=&0,\quad
\rel\left\{\widehat{A}_{gg^*}(\nu,\tau],\widehat{A}_{WW^*}(\nu,\tau]\right\}=0
\nonumber\\
\nonumber
\rel\left\{\widehat{A}_{Wg^*}(\nu,\tau],\widehat{A}_{WW^*}(\nu,\tau]\right\}&=&0,\quad
\rel\left\{\widehat{A}_{gW^*}(\nu,\tau],\widehat{A}_{WW^*}(\nu,\tau]\right\}=0.
\label{ae42}
\end{eqnarray}
For $\tau \ge 0$ 
\begin{eqnarray}
&&\rel\left\{\widehat{A}_{Wg^*}(\nu,\tau],\widehat{A}_{gW^*}(\nu,\tau]\right\}=\E\left\{\widehat{A}_{Wg^*}(\nu,\tau]\widehat{A}_{gW^*}(\nu,\tau]\right\}\nonumber\\
&&=\E\left\{\int_{0}^{1/2}\int_{0}^{1/2} d\widetilde{X}(f_1)d\widetilde{X}^*(f_2)e^{j2\pi(f_1-\nu)\tau}
G^*(f_1-\nu,\tau]G(f_2+\nu,-\tau]e^{j2\pi\nu\tau}\right\}\nonumber \\
&&=\sigma^2_W \int_{0}^{1/2} G^*(f-\nu,\tau]G(f+\nu,-\tau]e^{j2\pi(f-2\nu)\tau}df.
\label{nyekva1}
\end{eqnarray}
The factor $e^{-j4\pi\nu\tau}$ is not present for $\tau < 0$. We define $h'(\nu,\tau]=\int_{\max(0,-\nu)}^{1/2+\min(0,-\nu)} G^*(f,\tau]G(f+2\nu,-\tau]e^{j2\pi(f-\text{sign}(\tau)\nu)\tau}df$, where $\text{sign}(\tau)=1$ for $\tau \geq 0$ and 
$\text{sign}(\tau)=-1$ for $\tau < 0$. 
If $G(f,\tau]$ is mainly supported only over a range of frequencies we would
assume with increasing $\nu$ that $h'(\nu,\tau]\rightarrow 0$. Furthermore we note that
$\rel\left\{\widehat{A}_{Wg^*}(\nu,\tau]\right\}=0$, and $\rel\left\{\widehat{A}_{gW^*}(\nu,\tau]\right\}=0.$
We can note that for a generic harmonizable process $X[t]$,
\begin{equation*}
\begin{split}
&\rel\left\{\widehat{A}_{XX^*}(\nu,\tau]\right\}
=\int_{0}^{1/2}\int_{0}^{1/2}\int_{0}^{1/2}\int_{0}^{1/2}e^{j\pi (f_1-f_2+\alpha_1-\alpha_2) (N-1-\tau)}e^{j2\pi(f_1+\alpha_1)\tau}
e^{-2j\pi \nu(N+\tau-1)}\\
&D_{N-\tau}(f_1-f_2-\nu)\,D_{N-\tau}(\alpha_1-\alpha_2-\nu)\E\left\{ d\widetilde{X}(f_1)
d\widetilde{X}^*(f_2)d\widetilde{X}(\alpha_1)d\widetilde{X}^*(\alpha_2)  \right\}-\E\left\{\widehat{A}_{XX^*}(\nu,\tau]\right\}^2.
\end{split}
\end{equation*}
This follows by direct calculation starting from~(\ref{repofEAF}). Using Isserlis' theorem and duplicating the calculations for the variance, we write the relation as
{\small\begin{alignat}{1}
\nonumber
&\rel\left\{\widehat{A}_{X X^*}(\nu,\tau]\right\}
=\int_{0}^{1/2}\int_{0}^{1/2}\int_{0}^{1/2}\int_{0}^{1/2} S_{X X^*}(f_1-\alpha_2,\alpha_2) S_{X X^*}(\alpha_1-f_2,f_2)\\
&e^{j\pi (f_1-f_2+\alpha_1-\alpha_2) (N-1-\tau)}e^{j2\pi(f_1+\alpha_1)\tau}D_{N-\tau}(f_1-f_2-\nu) D_{N-\tau}(\alpha_1-\alpha_2-\nu)e^{-2j\pi \nu(N+\tau-1)} df_1 df_2 d\alpha_1 d\alpha_2.
\label{relofthing}
\end{alignat}}
For an analytic white-noise process this corresponds to
\begin{equation*}
\begin{split}
&\rel\left\{\widehat{A}_{X X^*}(\nu,\tau]\right\}
=\int_{0}^{1/2}\int_{0}^{1/2}\int_{0}^{1/2}\int_{0}^{1/2} \sigma^2_{W}\delta(f_1-\alpha_2) \sigma^2_{W}\delta(\alpha_1-f_2)e^{j\pi (f_1-f_2+\alpha_1-\alpha_2) (N-1-\tau)}
\\
& e^{j2\pi(f_1+\alpha_1)\tau}D_{N-\tau}(f_1-f_2-\nu)D_{N-\tau}(\alpha_1-\alpha_2-\nu) e^{-2j\pi \nu(N+\tau-1)}df_1 df_2 d\alpha_1 d\alpha_2\\
&=\int_{0}^{1/2}\int_{0}^{1/2} \sigma^4_{W} 
e^{j\pi (f_1-f_2+f_2-f_1) (N-1-\tau)}e^{j2\pi(f_1+f_2)\tau}D_{N-\tau}(f_1-f_2-\nu)\\
& D_{N-\tau}(f_2-f_1-\nu) e^{-2j\pi \nu(N+\tau-1)}df_1 df_2.
\end{split}
\end{equation*}
If $\nu=O\left(1/(N-\tau)\right)$ then this corresponds to (up to order $(N-\tau)/N$)
{\tiny
\begin{equation*}
\begin{split}
&\rel\left\{\widehat{A}_{X X^*}(\nu,\tau]\right\}=
\int_{0}^{1/2}\int_{(f-1/2)(N-\tau)}^{f(N-\tau)} \sigma^4_{W} 
e^{j4\pi f\tau}D_{N-\tau}\left(\frac{f'}{(N-\tau)}-\nu\right)D_{N-\tau}\left(-\frac{f'}{(N-\tau)}-\nu\right)  \frac{e^{-2j\pi \nu(N+\tau-1)}}{N-\tau}df\,df'\\
&=\int_{0}^{1/2}\int_{(f-1/2-\nu)(N-\tau)}^{(f-\nu)(N-\tau)} \sigma^4_{W}  \frac{e^{j4\pi f\tau}D_{N-\tau}\left(\frac{f'}{(N-\tau)}\right) D_{N-\tau}\left(\frac{-f'}{(N-\tau)}-2\nu\right)e^{-2j\pi \nu(N+\tau-1)}df\,df'}{N-\tau}.
\end{split}
\end{equation*}}
If $\nu$ is not $O\left(1/(N-\tau)\right)$, then $\rel\left\{\widehat{A}_{X X^*}(\nu,\tau]\right\}
$ is negligible. If $\nu=O\left(1/(N-\tau)\right)$
then with $\xi(\nu)=(N-\tau)\nu=O\left(1\right)$ we obtain that
{\small\begin{equation*}
\begin{split}
\rel&\left\{\widehat{A}_{X X^*}(\nu,\tau]\right\}=
(N-\tau)\left(\sigma^4_{W}\int_{-\infty}^{\infty} 
\sinc(f')\sinc( f'+2\xi(\nu))e^{-2j\pi \nu(N+\tau-1)}\,df' \int_{\max(0,\nu)}^{1/2+\min(0,\nu)}e^{j4\pi f\tau}\,df\right)\\
&=(N-\tau)\sigma^4_{W} L\left(N-\tau,\nu\right)e^{-2j\pi \nu(N+\tau-1)}
\frac{e^{j4\pi (1/2+\min(0,\nu))\tau}-e^{j4\pi \max(0,\nu)\tau}}{j4\pi \tau}+O\left(1 \right)\\
&=-(N-\tau)\sigma^4_{W} L\left(N-\tau,\nu\right)e^{-2j\pi \nu(N-1)}
\left[\frac{\sin(2\pi |\nu| \tau)}{2\pi \tau}I\left(\tau\neq 0
\right)-\frac{1}{2}I\left(\tau= 0\right)\right]+O\left(1 \right).
\end{split}
\end{equation*}}
This defines the quantity $L\left(N-\tau,\nu\right)$, which decays like $O\left(\frac{1}{\xi(\nu)^2}\right)$. Thus if $\tau \neq 0$
{\small\begin{eqnarray*}
\rel\left\{\widehat{A}_{X X^*}(\nu,\tau] \right\}&=&-(N-\tau)|\nu|\sigma^4_{W} L\left(N-\tau,\nu\right)e^{-2j\pi \nu(N-1)}
\sinc(2|\nu| \tau)+2\sigma^2_{W}h'(\nu,\tau]+O\left(1 \right).
\end{eqnarray*}}
{\em Mutatis mutandis} calculations can be replicated for $\tau<0$.

\section{Proof of Proposition \ref{estat} \label{prop4}}
For $\tau\ge 0$ we note that
{\small\begin{alignat}{1}
\nonumber
\E\left\{\widehat{A}_{X X^*}(\nu,\tau]\right\}&=\int_{0}^{1/2}\int_{0}^{1/2} e^{j\pi (f_1-f_2) (N-1-\tau)}\E\left\{d\widetilde{X}(f_1)d\widetilde{X}^*(f_2)\right\}
e^{j2\pi f_1\tau}D_{N-\tau}(f_1-f_2-\nu) e^{-j\pi\nu(N+\tau-1)}\\
&= D_{N-\tau}(\nu)e^{-j\pi\nu(N+\tau-1)}\widetilde{M}_{X X^*}[\tau]
\nonumber.
\end{alignat}}
Note that $D_{N-\tau}(\nu)$ is an even function, and the expression is valid for $\nu<0$. Calculations can {\em mutatis mutandis} be replicated for negative $\tau$.
%
We note that with $\tau\ge 0$
{\small \begin{eqnarray*}
\var\left\{\widehat{A}_{X X^*}(\nu,\tau]\right\}
&=&\int_{0}^{1/2}\int_{0}^{1/2} \widetilde{S}_{X X^*}(f_1) \widetilde{S}_{X X^*}(f_2)D^2_{N-\tau}(f_1-f_2-\nu)df_1 df_2\\
&=&(N-\tau)(1/2-|\nu|)\overline{A}_{X X^*}(-\nu,0]+O\left(1\right),\end{eqnarray*}}
with
$\overline{A}_{X X^*}(\nu,\tau]=\int_{\max(0,-\nu)}^{1/2+\min(\nu,0)}\widetilde{S}_{X X^*}(f-\nu) \widetilde{S}_{X X^*}(f)e^{j4\pi f\tau}\;df/(1/2-|\nu|).$
{\em Mutatis mutandis} the calculations can be replicated for $\tau<0$.
Finally we start from~(\ref{relofthing}) and note that for stationary
processes
{\small\begin{eqnarray*}
\rel\left\{\widehat{A}_{X X^*}(\nu,\tau]\right\}
&=&\int_{0}^{1/2}\int_{0}^{1/2} \widetilde{S}_{X X^*}(f_1) \widetilde{S}_{X X^*}(f_2) e^{j2\pi(f_1+f_2)\tau}\\
&& D_{N-\tau}(f_1-f_2-\nu) e^{-2j\pi \nu (N+\tau-1)} D_{N-\tau}(f_1-f_2-\nu) df_1df_2\\ 
&=&e^{-2j\pi \nu (N+\tau-1)}\int_{0}^{1/2}\int_{(N-\tau)(g_2-1/2-\nu)}^{(N-\tau)(g_2-\nu)} \widetilde{S}_{X X^*}(g_2) \widetilde{S}_{X X^*}\left(g_2-\frac{g_1}{N-\tau}-\nu\right)e^{j2\pi (2g_2-\frac{g_1}{N-\tau}-\nu)\tau} \\
\nonumber
&& D_{N-\tau}\left( \frac{g_1}{N-\tau}\right)
D_{N-\tau}\left(\frac{g_1+2(N-|\tau|)\nu)}{N-\tau}\right) dg_1 dg_2/(N-\tau) \\
&=& e^{-2j\pi \nu (N+\tau-1)}(N-\tau)(1/2-|\nu|)L(N-\tau,\nu)\overline{A}_{X X^*}(\nu,\tau]+O\left(1 \right).
\end{eqnarray*}}
{\em mutatis mutandis} results may be derived for $\tau<0$.

\section{Proof of Proposition \ref{eunif}  \label{prop7}}
Using~(\ref{repofEAF}) and by direct calculation we note that
{\small
\begin{eqnarray*}
\nonumber
&&\E\left\{\widehat{A}_{X X^*}(\nu,\tau]\right\}=\int_{0}^{1/2}\int_{0}^{1/2} e^{j\pi (f_1-f_2) (N-1-\tau)}\E\left\{d\widetilde{X}(f_1)d\widetilde{X}^*(f_2)\right\} e^{j2\pi f_1\tau}D_{N-\tau}(f_1-f_2-\nu)e^{-j\pi\nu(N+\tau-1)}\\
&&=\int_{0}^{1/2}\int_{f-1/2}^{f} e^{j\pi \nu' (N-1-\tau)}\sum_{\tau'}A_{X X^*}\left(\nu',\tau'
\right]e^{-j2\pi (f-\nu')\tau'} e^{j2\pi f\tau}D_{N-\tau}(\nu'-\nu)e^{-j\pi\nu(N+\tau-1)}\;d\nu'\;df
\\
&&=\int_{\max(0,-\nu)}^{1/2-\max(\nu,0)}e^{j2\pi f\tau}
\int_{(f-1/2)(N-\tau)}^{f(N-\tau)} e^{j\pi \nu' (N-1-\tau)/(N-\tau)}\Sigma_{X X^*}\left(\frac{\nu'}{N-\tau}+\nu\right)D_{N-\tau}\left(\frac{\nu'}{N-\tau}\right)\;d\nu'/(N-\tau)\;df\\
&&=\int_{\max(0,-\nu)}^{1/2-\max(\nu,0)}e^{j2\pi f\tau}\Sigma_{X X^*}(\nu)
\int_{-\infty}^{\infty} e^{j\pi \nu' (N-1-\tau)/(N-\tau)}
\sinc(\nu')\;d\nu'\;df+O\left(1\right)\\
&&=\Sigma_{X X^*}(\nu)
\frac{e^{j2\pi (1/2-\max(\nu,0))\tau}-e^{j2\pi \max(-\nu,0))\tau}}{j2\pi \tau}+O\left(1\right)\nonumber\\
&&=(1/2-|\nu|)\Sigma_{X X^*}(\nu) e^{j\pi (1/2-\nu)\tau}\sinc((1/2-|\nu|)\tau)+O\left(1\right).
\end{eqnarray*}}
{\em Mutatis mutandis} we may derive the results for $\tau<0$.
We start from~(\ref{varharm}), and again by direct calculation
{\small \begin{eqnarray*}
&&\var\left\{\widehat{A}_{X X^*}(\nu,\tau]\right\}
=\int_{0}^{1/2}\int_{0}^{1/2}\int_{0}^{1/2}\int_{0}^{1/2} S_{X X^*}(f_1-\alpha_1,\alpha_1) S_{X X^*}(f_2-\alpha_2,\alpha_2) e^{j2\pi(f_1-\alpha_1)\tau}  \\
&& e^{j\pi (f_1-f_2-\alpha_1+\alpha_2) (N-1-\tau)}D_{N-\tau}(f_1-f_2-\nu) D_{N-\tau}(\alpha_1-\alpha_2-\nu) df_1 df_2 d\alpha_1 d\alpha_2\\
&&=\int_{0}^{1/2}\int_{0}^{1/2}\int^{f}_{f-1/2}\int^{\alpha}_{\alpha-1/2} \Sigma_{X X^*}(f-\alpha) \Sigma_{X X^*}^*(f-\nu'-\alpha+\alpha')e^{j\pi (\nu'-\alpha') (N-1-\tau)}
\\
&& e^{j2\pi(f-\alpha)\tau}D_{N-\tau}(\nu'-\nu) D_{N-\tau}(\alpha'-\nu) d\nu' d\alpha' df d\alpha\\
&&=\int_{-\nu}^{1/2-\nu}\int_{-\nu}^{1/2-\nu}
\int^{f(N-\tau)}_{(f-1/2)(N-\tau)}
\int^{\alpha(N-\tau)}_{(\alpha-1/2)(N-\tau)} \Sigma_{X X^*}(f-\alpha) 
\Sigma_{X X^*}^*\left(f-\frac{\nu'-\alpha'}{N-\tau}-\alpha\right)
\\
&& e^{j\pi (\nu'-\alpha') (N-1-\tau)/(N-\tau)}e^{j2\pi(f-\alpha)\tau}D_{N-\tau}(\nu'/(N-\tau)) D_{N-\tau}( \alpha'/(N-\tau)) d\nu' d\alpha' df d\alpha/(N-\tau)^2\\
&& \nonumber\\
&&=\int_{\max(0,-\nu)}^{1/2-\max(\nu,0)}\int_{\max(0,-\nu)}^{1/2-\max(\nu,0)} \Sigma_{X X^*}(f-\alpha) \Sigma_{X X^*}^*(f-\alpha)
 e^{j2\pi(f-\alpha)\tau} d\alpha df+O\left(1 \right) \\
&&=\int_{-1/2+|\nu|}^{1/2-|\nu|}\left(\frac{1}{2}-|\nu|\right) \left|\Sigma_{X X^*}(f) \right|^2
e^{j2\pi y\tau}  df+O\left(1\right)\\
&&=(1/2-|\nu|)(N-\tau)\int_{-1/2+|\nu|}^{1/2-|\nu|} \frac{\left|\Sigma_{X X^*}(f) \right|^2}{(N-\tau)}e^{j2\pi f\tau}df+O\left(1\right).
\end{eqnarray*}}
We start from~(\ref{relofthing})
{\small \begin{eqnarray*}
&&\rel\left\{\widehat{A}_{X X^*}(\nu,\tau]\right\}
=\int_{0}^{1/2}\int_{0}^{1/2}\int_{0}^{1/2}\int_{0}^{1/2} S_{X X^*}(f_1-\alpha_2,\alpha_2) S_{X X^*}(\alpha_1-f_2,f_2)   
 e^{j\pi (f_1-f_2+\alpha_1-\alpha_2) (N-1-\tau)}\\
&&\times e^{j2\pi(f_1+\alpha_1)\tau}D_{N-\tau}(f_1-f_2-\nu) D_{N-\tau}(\alpha_1-\alpha_2-\nu)e^{-2j\pi\nu(N+\tau-1)} df_1 df_2 d\alpha_1 d\alpha_2\\
&&=
\int_{0}^{1/2}\int_{0}^{1/2}\int_{0}^{1/2}\int_{0}^{1/2} 
\Sigma_{X X^*}(f_1-\alpha_2)\Sigma_{X X^*}^*(f_2-\alpha_1)
e^{j\pi (f_1-f_2+\alpha_1-\alpha_2) (N-1-\tau)}\\
&& e^{j2\pi(f_1+\alpha_1)\tau}D_{N-\tau}(f_1-f_2-\nu)D_{N-\tau}(\alpha_1-\alpha_2-\nu)e^{-2j\pi\nu(N+\tau-1)} df_1 df_2 d\alpha_1 d\alpha_2
\end{eqnarray*}}
We implement the change of variables $\nu''=(\nu'-\nu)(N-\tau)$ and $\alpha''=(\alpha'-\nu)(N-\tau)$.
{\small \begin{eqnarray*}
&&\rel\left\{\widehat{A}_{X X^*}(\nu,\tau]\right\}
=\int_{0}^{1/2}\int_{0}^{1/2}\int_{(f-1/2-\nu)(N-\tau)}^{(f-\nu)(N-\tau)}
\int_{(\alpha-1/2-\nu)(N-\tau)}^{(\alpha-\nu)(N-\tau)}
\Sigma_{X X^*}\left(f-\alpha+\frac{\alpha''}{N-\tau}+\nu\right)\\&&
\times\Sigma_{X X^*}^*\left(f-\frac{\nu''}{N-\tau}-\nu-\alpha\right)
 e^{j\pi (\nu''+\alpha'') (N-1-\tau)/(N-\tau)+j\pi(N-1-\tau)2\nu}e^{j2\pi(f+\alpha)\tau}D_{N-\tau}\left(
 \frac{\nu''}{N-\tau}\right)\\
&& D_{N-\tau}\left(\frac{\alpha''}{N-\tau}\right)e^{-2j\pi\nu(N+\tau-1)} \frac{1}{(N-\tau)^2}d\alpha'' d\nu'' df d\alpha\\
&&=e^{-4j\pi\nu\tau}\int_{\max(0,\nu)}^{1/2+\min(0,\nu)}\int_{\max(0,\nu)}^{1/2+\min(0,\nu)}
\Sigma_{X X^*}(f-\alpha+\nu)\Sigma_{X X^*}^*(f-\nu-\alpha) e^{j2\pi(f+\alpha)\tau}
\,df d\alpha+O\left(1\right).
\end{eqnarray*}}
{\em Mutatis mutandis} the expressions may be derived for $\tau<0$.

\section{Proof of Theorem  \label{theorem13} \ref{distEAF}}
We outline the proof for $\tau\ge 0$, the same arguments hold {\em mutatis
mutandis} for $\tau<0$. We let $\{X[t]\}$ be the analytic extension of a real-valued process $\{Y[t]\}$ or equivalently $X[t]=Y[t]+jU[t]$.
As $\left\{Y[t]\right\}$ is a real-valued underspread process
there exists a constant $T=O\left(1\right)$ such that
$\forall\;|\tau|\ge T$, $\cov\left\{Y[t],Y[t+\tau]\right\}=0$. This implies that
$\forall\; |\tau|\ge T$ $\cov\left\{Y[t],U[t+\tau]\right\}=O\left(1/(\tau-T)\right)$, where $\left\{U[t]\right\}$ is the discrete Hilbert transform of $\left\{Y[t]\right\}$. For convenience
assume $N-\tau=Tn$. This is of course not a necessity but simplifies the proofs. Otherwise for $Tm$ such that $Tm<N-\tau$ and $T(m+1)>N-\tau$ we can split
the sum in the EAF into two parts, and show the latter part has negligible contributions
to the total sum. The first part, summing from zero over $t$ up to $Tm$, can be treated as if $N-\tau=Tm$ with $m=n$.
We now take a full length observation $\mathbf{Y}=\left[Y_0,\dots,Y_{N-1}\right]$
and construct $n$ subvectors by the following mechanism
\begin{equation*}
\mathbf{Y}_1=\left[Y[0],Y[T],\dots,Y[(n-1)T]\right],\quad 
\mathbf{Y}_2=\left[Y[1],Y[T+1],\dots,Y[(n-1)T+1]\right],\quad \dots .
\end{equation*}
The vectors are constructed so that the elements of $\mathbf{Y}_k$ are uncorrelated
and if we define $\mathbf{U}_k$ and $\mathbf{X}_{k}$, as the obvious
extensions to $\mathbf{Y}_k$, then $\cov\left(Y_k[t],U_l[s]\right)=O\left(1/|k-l+T(t-s)|\right)$,
$\cov\left(X_{k}[t],X_{l}[s]\right)=O\left(1/|k-l+T(t-s)|\right)$ if $k\neq
l$. 
Before constructing the CLT type of argument let us study how we shall represent $\widehat{A}_{X X^*}(\nu,\tau]$. We write
\begin{eqnarray*}
\widehat{A}_{XX^*}(\nu,\tau]&=&\sum_{u=0}^{T-1}\sum_{v=0}^{n-1} 
x[vT+u+\tau]x^*[vT+u] e^{-j2\pi \nu (vT+u)}e^{-j2\pi \nu\tau}.
\end{eqnarray*}
The expectation of $\widehat{A}_{X X^*}(\nu,\tau]$ follows directly from this expression, and we additionally note that $\E\left\{\widehat{M}_{XX^*}[t,\tau]\right\}=M_{XX^*}[t,\tau]$.

Using Isserlis' formula \cite{Isserlis1918} and the assumption that the process is proper, we find the variance of $\widehat{A}_{X X^*}(\nu,\tau]$ as
{\small \begin{eqnarray*}
\var\left\{\widehat{A}_{X X^*}(\nu,\tau]\right\}&=& \sum_{v_1=0}^{n-1} \sum_{v_2=0}^{n-1}\sum_{u_1=0}^{T-1} \sum_{u_2=0}^{T-1} c_N(\nu,\tau;u_1,u_2,v_1,v_2],
\end{eqnarray*}}
where 
\begin{eqnarray*}
c_N(\nu,\tau;u_1,u_2,v_1,v_2]&=&e^{-j2\pi \nu (T(v_1-v_2)+u_1-u_2)}M_{XX^*}[v_1T+u_1+\tau,T(v_1-v_2)+u_1-u_2]\\
&&
M_{XX^*}^*[v_1T+u_1,T(v_1-v_2)+u_1-u_2].
\end{eqnarray*}

As $\{Y[t]\}$ is exactly underspread if $v_1\neq v_2$ and $v_1 \neq v_2\pm 1$ we have that with $t_l=v_l T+u_l$, 
$c_N(\nu,\tau;u_1,u_2,v_1,v_2]=O\left(1/(t_1-t_2)^2\right).$
Furthermore, $\sum_{|v_1-v_2|>1}c_N(\nu,\tau;u_1,u_2,v_1,v_2]=O\left(\log(n)\right).$ Otherwise 
\begin{equation*}
c_N(\nu,\tau;u_1,u_2,v_1,v_2]=M_{XX^*}[v_1T+u_1+\tau,u_1-u_2+(v_1-v_2)T]M_{XX^*}^*[v_1T+u_1
,u_1-u_2+(v_1-v_2)T].
\end{equation*}
Thus we find that (up to $O\left(\log(n) \right)$)
\begin{eqnarray}
\nonumber
\var\left\{\widehat{A}_{XX^*}(\nu,\tau] \right\}
&=&\sum_{u_1=0}^{T-1} \sum_{u_2=u_1-(T-1)}^{u_1+(T-1)}\sum_{v=0}^{n-1} e^{-j2\pi \nu [u_1-u_2]}
M_{XX^*}[v T+u_1+\tau,u_1-u_2]M_{XX^*}^*[v T+u_1,u_1-u_2].
\end{eqnarray}
Let $x=v T+u_1$ and $\tau'=u_1-u_2$. Then this expression is rewritten for $\tau\ge 0$ as
\begin{eqnarray*}
\var\left\{\widehat{A}_{XX^*}(\nu,\tau] \right\}&=&
\sum_{x=0}^{N-\tau-1} \sum_{\tau'=-(T-1)}^{T-1} e^{-j2\pi \nu \tau'}
M_{XX^*}[x+\tau,\tau']M_{XX^*}^*[x,\tau']+O\left(\log(n)\right).
\end{eqnarray*}
Thus $\var\left\{\widehat{A}_{XX^*}(\nu,\tau] \right\}=O\left(N\right)$.
We note
that
\begin{eqnarray*}
\rel\left\{\widehat{A}_{XX^*}(\nu,\tau] \right\}&=&\sum_{v_1=0}^{n-1} \sum_{v_2=0}^{n-1}\sum_{u_1=0}^{T-1} \sum_{u_2=0}^{T-1} r_N(\nu,\tau;u_1,u_2,v_1,v_2]
(\nu,\tau],
\end{eqnarray*}
\begin{eqnarray*}
r_N(\nu,\tau;u_1,u_2,v_1,v_2]&=&e^{-j2\pi \nu (T(v_1+v_2)+u_1+u_2+2\tau)}M_{XX^*}[v_1T+u_1+\tau,T(v_1-v_2)+u_1-u_2+\tau]\\
&&\times M_{XX^*}^*[v_1T+u_1,T(v_1-v_2)+u_1-u_2-\tau]\\
\end{eqnarray*}
If $\left| t_1-t_2+\tau\right|>T$  and $\left|  t_1-t_2-\tau\right|>T$
\begin{equation*}
r_N(\nu,\tau;u_1,u_2,v_1,v_2]=
O\left(\frac{1}{(t_1-t_2+\tau)(t_1-t_2-\tau)}\right),
\end{equation*}
whilst if $\left|  t_1-t_2+\tau\right|>T$  and $\left|
 t_1-t_2-\tau\right|<T$
\begin{equation*}
r_N(\nu,\tau;u_1,u_2,v_1,v_2]=
O\left(\frac{1}{t_1-t_2+\tau}\right).
\end{equation*}
Thus if $|\tau|>T$,
$\sum_{v_1, v_2}r_N(\nu,\tau;u_1,u_2,v_1,v_2]=
O\left(\log(n)\right).$ If $|\tau|<T$
$
\sum_{|v_1-v_2|>1}r_N(\nu,\tau;u_1,u_2,v_1,v_2]=
O\left(\log(n)\right).$
If $\tau<T$ and $|v_1 -v_2|\le 1$  we have that
$
r_N(\nu,\tau;u_1,u_2,v_1,v_2]=
e^{-j2\pi \nu (T(v_1+v_2)+u_1+u_2+2\tau)}M_{XX^*}[v_1T+u_1+\tau,T(v_1-v_2)+u_1-u_2+\tau]M_{XX^*}^*[v_1T+u_1,T(v_1-v_2)+u_1-u_2-\tau]
.$
We deduce that
with $x=v T+u_1$ and $\tau'=u_1-u_2$ that the relation of
$\widehat{A}_{XX^*}(\nu,\tau]$ is
\begin{alignat}{1}
&=\left\{\begin{array}{lcr}
O\left(\log(n)\right) &{\mathrm{if}} & \left|\tau\right|>T\\
\sum_{x=0}^{N-\tau-1} \sum_{\tau'=-(T-1)}^{T-1} e^{-j2\pi \nu [2x+2\tau-\tau']}
M_{XX^*}[x+\tau,\tau'+\tau]M_{XX^*}^*[x,\tau'-\tau]&{\mathrm{if}} & \left|\tau\right|<T
\end{array} \right. .
\label{rely2}
\end{alignat}
This completes the proof of the order structure of the EAF of an underspread
process.

To prove a CLT for the EAF we add the extra assumptions stated in the theorem.
We need to use results for random variables that are both non-stationary
and correlated to determine large same results. We note that for $\tau>0$
\begin{equation*}
\widehat{A}_{X X^*}(\nu,\tau]=\sum_{t=0}^{N-\tau-1} Q_{N-|\tau|}[t]+
\sum_{t=0}^{N-\tau-1} e^{-j2\pi\nu(t+\tau)}\widehat{M}_{XX^*}[t+\tau,\tau].
\end{equation*}
We note that $\left\{ Q_{N-|\tau|}[t],\quad t=0,\dots,N-\tau-1\right\}$ is a triangular array of centred random variables, that these by assumption are strongly mixing, and have finite second moments. The constraint Peligrad~\cite{Peligrad} makes on the correlation is satisfied because $X[t]$ is strictly underspread, and the Hilbert transform only induces suitably decaying correlation from such a process.
We have assumed the Lindeberg condition holds, and note (\ref{haga}) from our assumptions. Hence from Theorem 2.1 of Peligrad \cite{Peligrad} the theorem follows. The condition in~(\ref{haga}) does not have to be stated for the real and imaginary part separately as the relation divided by the variance goes to zero for $(\nu,\tau)\notin {\cal D}$. We can deduce the exact form of the mean, variance and correlation from the previous part of the theorem.



\begin{thebibliography}{10}
\providecommand{\url}[1]{#1}
\csname url@samestyle\endcsname
\providecommand{\newblock}{\relax}
\providecommand{\bibinfo}[2]{#2}
\providecommand{\BIBentrySTDinterwordspacing}{\spaceskip=0pt\relax}
\providecommand{\BIBentryALTinterwordstretchfactor}{4}
\providecommand{\BIBentryALTinterwordspacing}{\spaceskip=\fontdimen2\font plus
\BIBentryALTinterwordstretchfactor\fontdimen3\font minus
  \fontdimen4\font\relax}
\providecommand{\BIBforeignlanguage}[2]{{%
\expandafter\ifx\csname l@#1\endcsname\relax
\typeout{** WARNING: IEEEtran.bst: No hyphenation pattern has been}%
\typeout{** loaded for the language `#1'. Using the pattern for}%
\typeout{** the default language instead.}%
\else
\language=\csname l@#1\endcsname
\fi
#2}}
\providecommand{\BIBdecl}{\relax}
\BIBdecl

\bibitem{Donoho98}
D.~L. Donoho, S.~Mallat, and R.~von Sachs, ``Estimating covariances of locally
  stationary processes: rates of convergence of best basis methods,''
  Statistics, Stanford University, Standford, California, USA, Tech. Rep.,
  1998.

\bibitem{Matz06}
G.~Matz and F.~Hlawatsch, ``Nonstationary spectral analysis based on
  time-frequency operator symbols and underspread approximations,'' \emph{IEEE
  Trans. Inf. Theory}, vol.~52, pp. 1067--1086, March 2006.

\bibitem{Priestly65}
M.~B. Priestley, ``Evolutionary spectra and non-stationary processes,''
  \emph{J. Roy. Stat. Soc. B}, vol.~47, no.~2, pp. 204--237, 1965.

\bibitem{Pfander}
G.~E. Pfander and D.~F. Walnut, ``Measurement of time-variant linear
  channels,'' \emph{IEEE Trans. Inf. Theory}, vol.~52, pp. 4808--4820, November
  2006.

\bibitem{Blahut1991ed}
R.~E. Blahut, W.~Miller, and C.~H. Wilcox, Eds., \emph{Radar and Sonar, Part
  I}.\hskip 1em plus 0.5em minus 0.4em\relax New York, USA: Springer Verlag,
  1991.

\bibitem{Flandrin99}
P.~Flandrin, \emph{Time-Frequency/Time-Scale Analysis}.\hskip 1em plus 0.5em
  minus 0.4em\relax San Diego, CA: Academic Press, 1999.

\bibitem{Ma2006}
N.~Ma and J.~T. Goh, ``Ambiguity-function-based techniques to estimate doa of
  broadband chirp signals,'' \emph{IEEE Trans. Signal Process.}, vol.~54, pp.
  1826--1839, May 2006.

\bibitem{Jachan07}
M.~Jachan, G.~Matz, and F.~Hlawatsch, ``Time-frequency arma models and
  parameter estimators for underspread nonstationary random processes,''
  \emph{IEEE Trans. Signal Proc.}, vol.~55, pp. 4366--4381, 2007.

\bibitem{Frylewicz06}
P.~Fryzlewicz and G.~P. Nason, ``Haar-{Fisz} estimation of evolutionary wavelet
  spectra,'' \emph{J. Roy. Stat. Soc. B}, vol.~68, pp. 611--634, 2006.

\bibitem{vSachs96}
R.~von Sachs and K.~Schneider, ``Wavelet smoothing of evolutionary spectra by
  nonlinear thresholding,'' \emph{ACHA}, vol.~3, pp. 268--282, 1996.

\bibitem{Hedges2002}
R.~A. Hedges and B.~W. Suter, ``Numerical spread: Quantifying local
  stationarity,'' \emph{Digital Signal Processing}, vol.~12, pp. 628--643,
  2002.

\bibitem{Johnstone97}
I.~M. Johnstone and B.~W. Silverman, ``Wavelet threshold estimators for data
  with correlated noise,'' \emph{J. Roy. Stat. Soc. B}, vol.~59, pp. 319--351,
  1997.

\bibitem{Cramer40}
H.~Cram\'er, ``On the theory of stationary random processes,'' \emph{Ann.
  Math.}, vol.~41, pp. 215--230, January 1940.

\bibitem{Kozek}
W.~Kozek, ``Matched {Weyl}-{Heisenberg} expansions of nonstationary
  environments,'' Ph.D. dissertation, Universit{\"a}t {Wien}, Vienna, Austria,
  1996.

\bibitem{Matz}
G.~Matz, ``A time-frequency calculus for time-varying systems and
  non-stationary processes with applications,'' Ph.D. dissertation,
  Universit{\"a}t {Wien}, Vienna, Austria, 2000.

\bibitem{Jeong92}
J.~Jeong and W.~J. Williams, ``Alias-free generalized discrete-time
  time-frequency distributions,'' \emph{IEEE Trans. Signal Process.}, vol.~40,
  pp. 2757--2765, November 1992.

\bibitem{Neeser93}
F.~D. Neeser and J.~L. Massey, ``Proper complex random processes with
  applications to information theory,'' \emph{IEEE Trans. Inf. Theory},
  vol.~39, pp. 1293--1302, July 1993.

\bibitem{Schreier03}
P.~J. Schreier and L.~L. Scharf, ``Stochastic time-frequency analysis using the
  analytic signal: Why the complementary distribution matters,'' \emph{IEEE
  Trans. Signal Process.}, vol.~51, pp. 3071--3079, December 2003.

\bibitem{Auslander1984}
L.~Auslander and R.~Tolimieri, ``Characterizing the radar ambiguity
  functions,'' \emph{IEEE Trans. Inf. Theory}, vol.~30, pp. 832--836, November
  1984.

\bibitem{Auslander1985}
------, ``Radar ambiguity functions and group-theory,'' \emph{SIAM Journal on
  Mathematical Analysis}, vol.~16, pp. 577--601, 1985.

\bibitem{Bekir1993}
E.~Bekir, ``Unaliased discrete-time ambiguity function,'' \emph{JOSA}, vol.~94,
  pp. 817--826, 1993.

\bibitem{Tolimieri1995}
R.~Tolimieri and R.~S. Orr, ``Poisson summation, the ambiguity function and the
  theory of weyl-heisenberg frames,'' \emph{The Journal of Fourier Analysis and
  Applications}, vol.~1, pp. 233--247, 1995.

\bibitem{Ferguson96}
T.~S. Ferguson, \emph{A Course in Large Sample Theory}.\hskip 1em plus 0.5em
  minus 0.4em\relax London, UK: Chapman \& Hall, 1996.

\bibitem{Olhede07}
S.~Olhede, ``Hyperanalytic denoising,'' \emph{IEEE Trans. Image Process.},
  vol.~16, pp. 1522--1537, June 2007.

\bibitem{Wasserman07}
L.~Wasserman, \emph{All of Nonparametric Statistics}.\hskip 1em plus 0.5em
  minus 0.4em\relax New York, USA: Springer Verlag, 2007.

\bibitem{Averbuch2006}
A.~Averbuch, R.~R. Coifman, D.~L. Donoho, M.~Elad, and M.~Israeli, ``Fast and
  accurate polar fourier transform,'' \emph{Applied and Computational Harmonic
  Analysis}, vol.~21, pp. 145--167, 2006.

\bibitem{Isserlis1918}
L.~Isserlis, ``On a formula for the product-moment coefficient of any order of
  a normal frequency distribution in any number of variables,''
  \emph{Biometrika}, vol.~12, pp. 134--139, 1918.

\bibitem{Peligrad}
M.~Peligrad, ``On the asymptotic normality of sequences of weak dependent
  random variables,'' \emph{J. Theo. Prob.}, vol.~9, pp. 703--715, 1996.

\end{thebibliography}
\end{document}